\documentclass{bmcart}

\usepackage{amsthm,amsmath,amssymb}
\usepackage[utf8]{inputenc} 

\usepackage{multirow}
\usepackage{booktabs}
\usepackage{graphics}
\usepackage{adjustbox}
\usepackage{subcaption}
\usepackage[font=small,labelfont=bf]{caption}
\usepackage{url}
\usepackage{graphicx}

\startlocaldefs
\endlocaldefs

\begin{document}

\begin{frontmatter}

\begin{fmbox}
\dochead{Research}


\title{Subjective performance evaluation of bitrate allocation strategies for MPEG and JPEG Pleno point cloud compression}



\author[
  addressref={aff1},                   
  corref={aff1},                       
  email={davi.nachtigalllazzarotto@epfl.ch}   
]{\inits{D.L.}\fnm{Davi} \snm{Lazzarotto}}
\author[
  addressref={aff1},
  email={michela.testolina@epfl.ch}
]{\inits{M.T.}\fnm{Michela} \snm{Testolina}}
\author[
  addressref={aff1},
  email={touradj.ebrahimi@epfl.ch}
]{\inits{T.E.}\fnm{Touradj} \snm{Ebrahimi}}






\address[id=aff1]{
  \orgdiv{Multimedia Signal Processing Group (MMSPG)},             
  \orgname{\'Ecole Polytechnique F\'ed\'erale de Lausanne (EPFL)},          
  \city{Lausanne},                              
  \cny{Switzerland}                                    
}



\end{fmbox}


\begin{abstractbox}

\begin{abstract} 

The recent rise in interest in point clouds as an imaging modality has motivated standardization groups such as JPEG and MPEG to launch activities aiming at developing compression standards for point clouds. 
Lossy compression usually introduces visual artifacts that negatively impact the perceived quality of media, which can only be reliably measured through subjective visual quality assessment experiments. 
While MPEG standards have been subjectively evaluated in previous studies on multiple occasions, no work has yet assessed the performance of the recent JPEG Pleno standard in comparison to them. 
In this study, a comprehensive performance evaluation of JPEG and MPEG standards for point cloud compression is conducted. 
The impact of different configuration parameters on the performance of the codecs is first analyzed with the help of objective quality metrics. 
The results from this analysis are used to define three rate allocation strategies for each codec, which are employed to compress a set of point clouds at four target rates. 
The set of distorted point clouds is then subjectively evaluated following two subjective quality assessment protocols. 
Finally, the obtained results are used to compare the performance of these compression standards and draw insights about best coding practices.

\end{abstract}





\end{abstractbox}
%

\end{frontmatter}




\section{Introduction}\label{sec:intro}

Imaging modalities that allow the representation of three-dimensional (3D) objects and scenes have gained momentum in the last few years. 
This increase in interest has been influenced both by new applications such as virtual and augmented reality, as well as by the popularization of devices capable of measuring depth at a high resolution. 
In the field of computer graphics, meshes have been the norm for the representation of 3D content. 
However, meshes are not the most suited solution to represent all real-world scenes, either due to the irregularity of the source data or to the computational complexity of algorithms used to create the mesh from the raw acquired points. 
In those cases, representing the data as point clouds, which are composed of a set of unconnected points with geometric coordinates and associated attributes, may be a preferable alternative. 
Due to the high amount of data needed for the representation of point clouds, effective compression algorithms are needed for practical transmission and storage.

For this reason, several approaches have been proposed in the literature for the compression of point clouds. 
While some focused on encoding the geometric coordinates, others were better suited for the compression of the associated attributes such as colors. 
Since color attributes are usually required to enable human visualization, both classes of algorithms have to be combined to allow the data to be completely represented. 
Compression methods can also be grouped according to the underlying algorithm. 
For instance, some techniques project 3D models onto multiple planes and encode their projections as regular images. 
Other methods compress geometry and attributes directly in their original spatial domain. 
Many geometry coding schemes rely on the octree representation, with handcrafted transforms as the backbone of color attribute compression methods. 
Recently, deep learning techniques have been leveraged for compression, mainly through the use of convolutional autoencoders applied to geometric coordinates.

Given these advancements in the field of point cloud compression, standardization bodies such as \textcolor{black}{JPEG and MPEG} demonstrated their interest by issuing Call\textcolor{black}{s} for Proposals, where proponents from industry and academia were able to submit their compression methods to be considered as candidates for a compression standard. 
As a result, MPEG has standardized two separate compression methods, i.e. Geometry-based Point Cloud Compression (G-PCC) and Video-based Point Cloud Compression (V-PCC). 
The first relies on the octree to encode the coordinates, with an additional module allowing to encode the leaf nodes as triangular primitives. 
Similarly, two distinct transforms can be selected by the user to encode the color attributes. 
On the other hand, V-PCC employs a projection mechanism to obtain two-dimensional maps from the point cloud geometry and color, which are encoded using state-of-the-art video codecs. 
Moreover, JPEG is currently in the process of standardizing a compression method based on deep learning techniques \textcolor{black}{called JPEG Pleno}. 
In its current version, the geometry is encoded using an autoencoder composed of sparse convolutional layers, while for color coding, the projection algorithm from V-PCC is used to produce color maps that are encoded with the JPEG AI codec.

Lossy compression methods such as those reported above are able to reduce the size of the compressed data to a higher extent by omitting information from the original point cloud during the encoding process, resulting in a loss in quality in the decoded model. 
\textcolor{black}{Higher} compression leads to elevated levels of distortion, which decreases the quality of experience of the users. 
However, the impact of this distortion is hard to quantify, and even if many objective metrics have been proposed to measure the quality of distorted point clouds, there is no consensus as to which metric leads to an accurate model of the human visual system. 
For this reason, subjective quality assessment experiments are considered the most reliable method to estimate the quality of 3D models, where observers are asked to rate the visual quality of point clouds that suffer from different kinds of distortion. 
In recent years, many subjective quality assessment experiments have been conducted and reported in the literature to assess the performance of V-PCC and G-PCC. 
In such experiments, the observed stimuli are sets of point clouds compressed with these codecs at many compression levels, usually employing the parameters described in their Common Test Conditions (CTC) documents \cite{MPEG-GPCC-CTC, MPEG-VPCC-CTC, JPEG-Pleno-PC-CTTC}.  
However, to date, the JPEG Pleno compression model has never been evaluated subjectively but only compared to the MPEG standards through objective metrics. 

Since the compression performance of a codec can be affected by the features of the reference point cloud model, subjective experiments usually employ a diversified set of point clouds to evaluate the codecs at different conditions. 
While some experiments selected point clouds representing content of different nature, i.e. human figures, inanimate objects, or landscapes, others grouped them according to their low-level features, such as the curvature of the underlying surface. 
Recent studies found that point density is a determinant factor in the performance of the JPEG Pleno standard \textcolor{black}{\cite{lazzarotto2023evaluating}}, and therefore subjective experiments evaluating it should take this factor into account when deciding the models to be included in the dataset. 

Unlike the implementation of image coding algorithms such as legacy JPEG, which offers one quality parameter that can be used to control the rate-distortion trade-off, the reference software of the \textcolor{black}{JPEG Pleno and MPEG} compression methods usually offer at least two main configuration parameters that can independently control the quality factor used to encode geometry and color. 
With the correct \textcolor{black}{tuning} of these parameters, it is possible to obtain decoded point clouds with different characteristics at the same bitrate by allowing different proportions of the bitstream for color or geometry. 
However, this effect has not been evaluated in detail in previous studies, where the parameters described in the CTC are usually employed as is. 

Considering the importance of a meaningful evaluation of the recent standards that account for these factors, this paper describes a comprehensive subjective evaluation of G-PCC, V-PCC, and JPEG Pleno point cloud coding solutions. 
A set of six point clouds was selected from the JPEG Pleno test dataset containing point clouds with different sparsity levels, which were compressed with each codec at four compression levels. 
For each level, three different strategies that maintained approximately the same bitrate but allocated different proportions of the bitstream to geometry and color were employed. 
The entire dataset was evaluated in two experiments following different protocols, namely Double Stimulus Impairment Scale (DSIS) and Pairwise Comparison (PWC). 
While the first intended to obtain an overall quality value determined by the mean opinion score (MOS), the second allowed to achieve precise comparison between models compressed with the same codec and bitrate but different rate allocation strategies. 
The obtained results are used to derive conclusions regarding the performance of these three codecs, as well as the ability of popular objective quality metrics to correctly predict the ranking between codecs. 


The main contributions of this paper can be summarised as follows: 

\begin{itemize}
    \item The JPEG Pleno point cloud codec is assessed subjectively for the first time in comparison to the MPEG standards G-PCC and V-PCC, by adopting a diverse dataset with different sparsity levels.
    \item An analysis is conducted regarding the rate allocation for the three codecs, including a search for parameters that provide improved results when compared to the CTC for each rate both according to objective quality metrics and subjective scores. 
    \item The subjectively annotated dataset is publicly released to foster research on subjective and objective point cloud quality assessment \footnote{https://www.epfl.ch/labs/mmspg/downloads/mj-pccd/}.

\end{itemize}

\section{Related work}

\subsection{Point cloud compression}

The recent \textcolor{black}{JPEG Pleno and MPEG} point cloud compression standards are evaluated in the experiment described in this paper, and for this reason, these algorithms, as well as related techniques, are described in this section. 
Octrees have been used as a data structure for the representation of 3D geometry in early compression algorithms \cite{Schnabel2006a, huang2008a}, being able to represent voxelized point clouds with a given precision. 
Other algorithms used octrees pruned at earlier levels using triangular primitives to represent the leaf nodes \cite{Pavez2018a}. 
Algorithms based on these two techniques were included in the G-PCC standard \cite{MPEG-GPCC-standard} as two alternatives for geometry encoding, which are here referred to as the \emph{octree} module and the \emph{trisoup} module, respectively. 
Moreover, two alternative color coding modules are included in the same standard, namely the Region Adaptive Hierarchical Transform (RAHT) \cite{Queiroz2016a} and nearest-neighbor prediction algorithm with an update step, which is denominated the \emph{predlift} module. 
Draco \cite{Draco}, a library developed by Google for 3D data compression, also employs octree for the representation of point clouds. 
The V-PCC standard \cite{MPEG-VPCC-standard} employs a different mechanism relying on the compression of point cloud data in the 2D domain. 
For this purpose, the points are grouped into patches, and each patch is projected onto a different view and packed together into one image frame. 
For dynamic point clouds including motions, color and depth maps from different frames are assembled and compressed with a video codec. 
The latest MPEG video coding standard, i.e. VVC~\cite{bross2021overview}, has been recently used for this purpose. 

While G-PCC and V-PCC have been explored and evaluated in numerous research papers, the JPEG Pleno codec was developed only recently and therefore it is still not yet known how it compares to these standards in terms of subjective visual quality. 
Following recent advances in learning-based point cloud coding \cite{Guarda2020b, Guarda2020d, Quach2020a, Quach2020b, Wang2020b, frank2022latent}, the JPEG Pleno Point Cloud Call for Proposals \cite{JPEG-Pleno-PC-CfP} was launched with the goal of standardizing a point cloud codec based on deep learning techniques.  
As a result, a solution with joint coding of geometry and color \cite{guarda2022ipleiria} in a learning-based architecture based on a convolutional autoencoder was selected as the first version of the Verification Model (VM). 
The model has however been updated, and geometry and color are currently coded separately \cite{guarda2023point}, with the autoencoder being used to encode the coordinates of the point cloud and JPEG AI \cite{ascenso2023jpeg} compressing color maps projected into 2D views using the same mechanism as V-PCC. 
\textcolor{black}{The JPEG AI image encoder is also based on an autoencoder architecture composed of convolutional and attention layers along with a hyperprior to learn spatial dependencies in the latent representation and has demonstrated higher rate-distortion performance when compared to the intraframe mode of the state-of-the-art video coding VVC.} 
An architecture based on sparse convolutions has also been integrated into the model \cite{lazzarotto2023evaluating} for geometry coding.

\subsection{Point cloud subjective quality assessment}

A large number of studies have been conducted to evaluate the quality of point clouds in recent years. 
\textcolor{black}{Mekuria et al.} \cite{mekuria2016design} assessed the quality of compressed point cloud sequences, while \textcolor{black}{Alexiou et al.} \cite{alexiou2017towards} compared the effects of different degradation types on geometry-only point clouds using a systematic approach. 
In both cases, the subjects were able to interact with the content, either in a virtual environment through flat monitors \cite{mekuria2016design} or using augmented reality glasses \cite{alexiou2017towards}. 
A later study compared two compression methods in a controlled environment \cite{da2019point}, adopting a passive approach where the subjects would visualize videos generated from a pre-defined camera path moving around the object depicted by the point clouds. 
A larger subjectively annotated dataset was produced by \cite{yang2020predicting}, where 10 point clouds selected from the MPEG dataset suffered 7 types of degradation with different levels, producing a total of 420 distorted point clouds with associated subjective scores. 

Later works started to take the distortions generated by recent MPEG compression standards into the evaluation. 
Still aiming at a large set of point clouds to properly benchmark objective quality metrics, the WPC dataset \cite{su2019perceptual, liu2022perceptual} contained point clouds compressed with the MPEG standards V-PCC and G-PCC, the latter being used both in \emph{octree} and \emph{trisoup} mode for geometry compression. 
The IRPC dataset \cite{javaheri2020point} also evaluated both compression methods, studying the impact of the rendering method on the subjective quality, while \textcolor{black}{Perry et al.}\cite{perry2020quality} conducted an experiment in multiple labs in order to compare both codecs. 
A comprehensive study was also conducted with the two standards \cite{alexiou2019comprehensive}, evaluating the performance of both codecs following the DSIS protocol and defining the best-performing configurations for G-PCC using PWC.
Instead of employing static point clouds, the V-SENSE dataset \cite{zerman2019subjective} focused on dynamic sequences, using DSIS and PWC to evaluate point clouds compressed with V-PCC. 
The same authors later included G-PCC and Draco in the evaluation \cite{zerman2020textured}, additionally comparing the subjective quality of point clouds against meshes in similar bitrates. 
Other works concentrated instead on the impact of using head-mounted displays for the subjective inspection, either with point clouds compressed with G-PCC \cite{alexiou2020pointxr} or V-PCC \cite{viola2022impact}. 
In another study, the SIAT-PCQD \cite{wu2021subjective} was produced where V-PCC was used to compress a set of 20 point clouds with different configuration parameters, studying the effect of rate allocations outside of the CTC.

Due to the rise of learning-based methods for point cloud coding, recent studies have included such solutions in subjective experiments to assess how this specific type of distortion affects human perception. 
A first study conducted a crowdsourced evaluation \cite{lazzarotto2021benchmarking} of G-PCC, V-PCC, and two learning-based methods \cite{Quach2020a, Quach2020b}. 
The same author also used a learning-based coding tool \cite{frank2022latent} and G-PCC in an evaluation \cite{lazzarotto2022impact} with both a flat screen and a light field monitor. 
Recently, a large-scale study \cite{ak2023basics} produced subjective scores for a set of more than 1200 distorted point clouds using G-PCC, V-PCC, and a learning-based algorithm for geometry compression \cite{Quach2020a}, with the goal of fostering research on learning-based objective quality metrics. 
Draco was also compared against G-PCC, V-PCC, and a learning-based technique \cite{Guarda2020b}, at separate studies that used different visualization devices such as a flat screen \cite{prazeres2022quality}, a stereoscopic monitor \cite{prazeres2022subjective}, and head-mounted device \cite{prazeres2023subjective}. 
The same authors also conducted another evaluation \cite{prazeres2022quality2} including three different learning-based solutions \cite{Guarda2020d, Quach2020a, Wang2020b}. 
In the initial phases of the standardization process of the JPEG Pleno codec, an experiment was first conducted only with G-PCC and V-PCC \cite{perry2022subjective}, and at a later stage different learning-based proposals were evaluated and compared with these anchors \cite{prazeres2023jpeg}, resulting in the adoption of one of them as the starting point for the development of the standard. 

The work presented in this paper has therefore significant differences when compared to the previous works on the field. 
First of all, this is the first experiment to evaluate the performance of JPEG Pleno and compare it to G-PCC and V-PCC. 
Even if a previous work assessed the proposals to the JPEG Pleno Point Cloud CfP \cite{prazeres2023jpeg}, one of which was selected as the starting point to the development of the standard, the architecture of the current VM has been significantly modified when compared to the selected proposal. 
A new subjective experiment is therefore needed to consider the impact of these updates on the performance of the model. 
Moreover, the majority of the experiments previously conducted with G-PCC and V-PCC have been constrained to the CTC, not exploring the impact of different trade-offs between color and geometry quality. 
To the best of the authors' knowledge, only \textcolor{black}{SIAT-PCQD} \cite{wu2021subjective} evaluated V-PCC with different sets of configurations. 
However, G-PCC is not included in the evaluation, and in the approach adopted in the current paper, the different employed configurations are forced to have similar bitrates as the CTC, allowing to derive the conclusion as to whether other sets of parameters can allow to achieve superior performance without affecting the rate. 
Finally, a similar evaluation is conducted for JPEG Pleno, providing insights on how configuration parameters for this codec should be set for future experiments.

\section{Dataset construction}
\label{sec:dataset_construction}

\begin{table*}[]
\centering
\begin{tabular}{c c c c c c}
Point cloud & Points  &  GP & DC & DF & CGV \\ 
\toprule
\emph{Bouquet} & 3150249 & 10 & Solid & 0.418  & 41\% \\
\emph{StMichael} & 1871158 & 10 & Solid & 0.418 & 21\% \\
\emph{Soldier} & 1089091 & 10 & Solid & 0.418 & 1\% \\
\emph{Thaidancer} & 3130215 & 12 & Solid & 0.328 & 22\% \\
\emph{House\_without\_roof} & 4848745 & 12 & Dense & 0.036  & 13\% \\
\emph{Boxer} & 3493085 & 12 & Dense & 0.048 & 3\% \\

\end{tabular}

\caption{Features of the point clouds of the evaluated dataset. GP=Geometry Precision, DC=Density Class, DF=Density Factor, CGV=Color Gamut Volume.}
\label{tab:dataset_stats}

\end{table*}

\begin{figure}
    \centering
    \subfloat[\emph{Bouquet}]{
    \begin{minipage}[b]{0.32\linewidth}
    \centering
    \includegraphics[width=\linewidth]{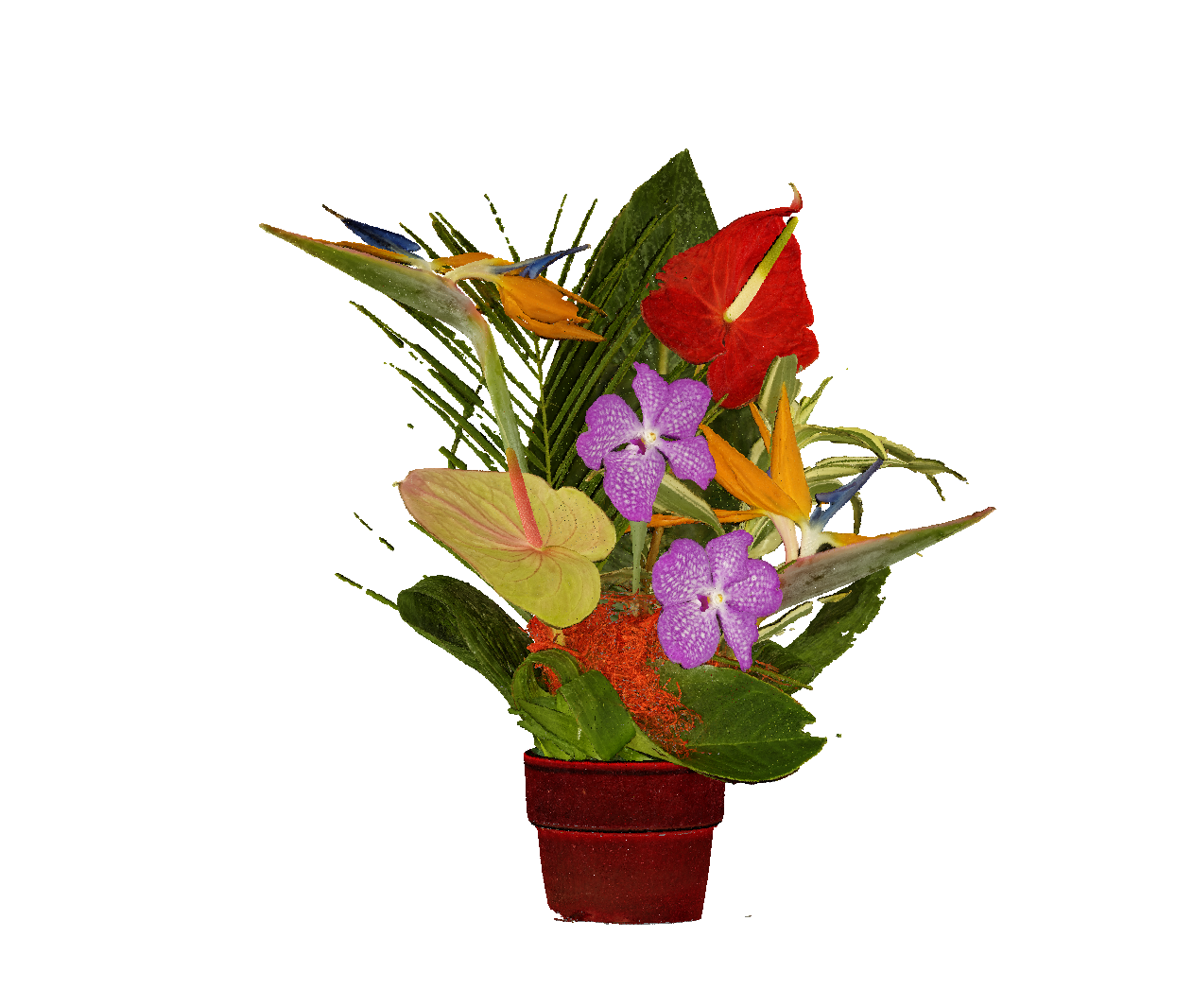}
    \end{minipage}}
    \subfloat[\emph{Soldier}]{
    \begin{minipage}[b]{0.32\linewidth}
    \centering
    \includegraphics[width=\linewidth]{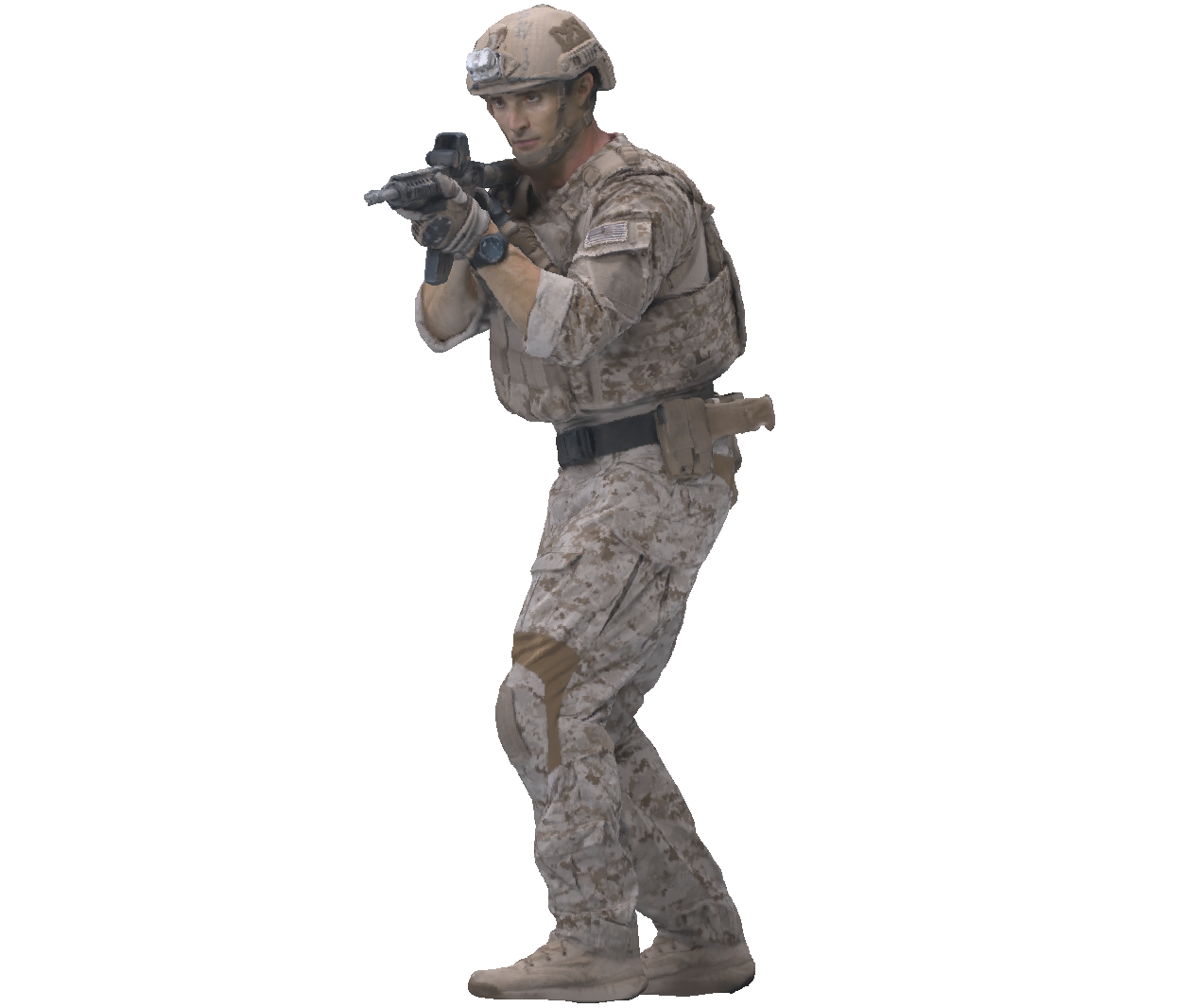}
    \end{minipage}}
    \subfloat[\emph{Boxer}]{
    \begin{minipage}[b]{0.32\linewidth}
    \centering
    \includegraphics[width=\linewidth]{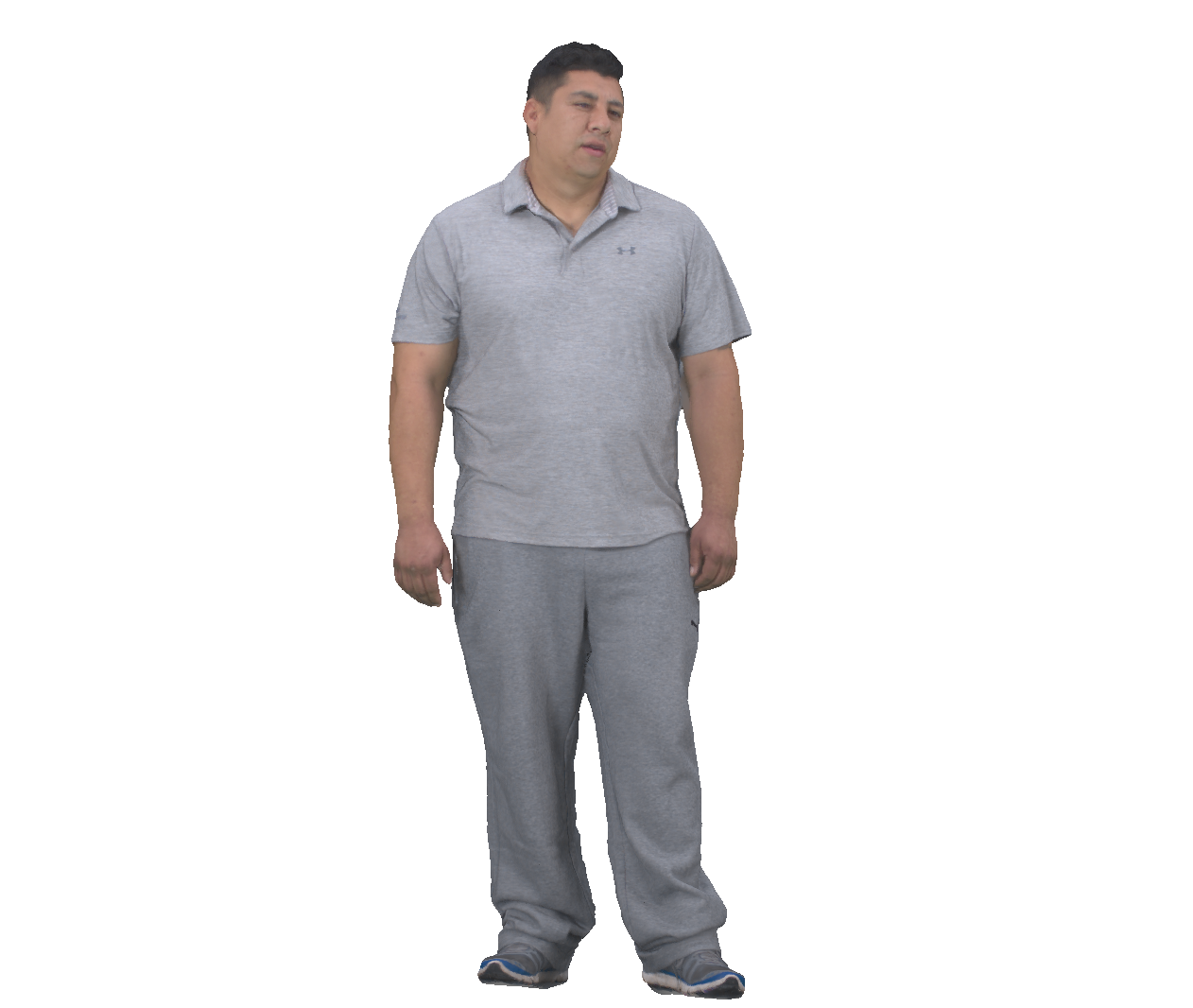}
    \end{minipage}}

    \subfloat[\emph{StMichael}]{
    \begin{minipage}[b]{0.32\linewidth}
    \centering
    \includegraphics[width=\linewidth]{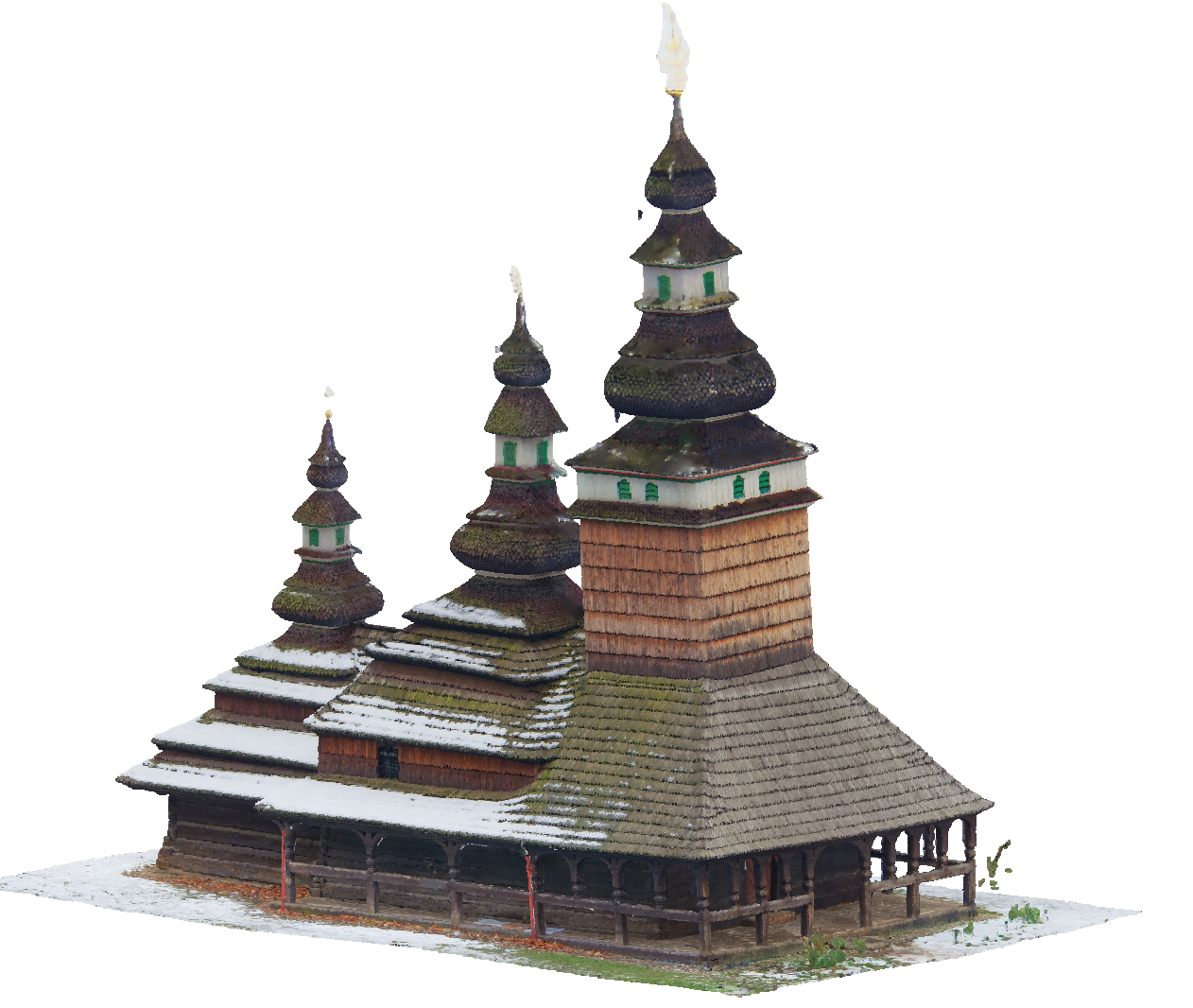}
    \end{minipage}}
    \subfloat[\emph{Thaidancer}]{
    \begin{minipage}[b]{0.32\linewidth}
    \centering
    \includegraphics[width=\linewidth]{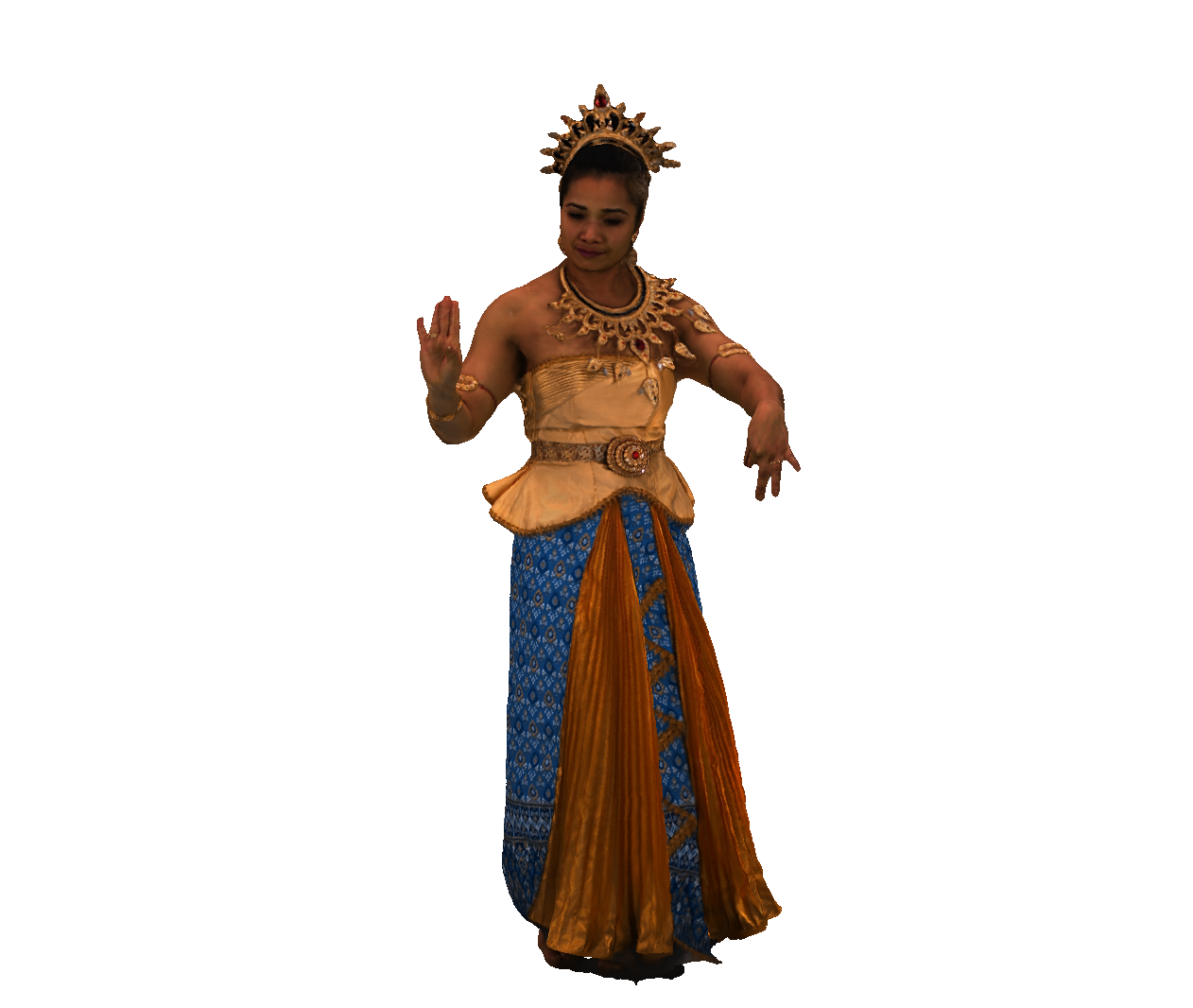}
    \end{minipage}}
    \subfloat[\emph{House\_without\_roof}]{
    \begin{minipage}[b]{0.32\linewidth}
    \centering
    \includegraphics[width=\linewidth]{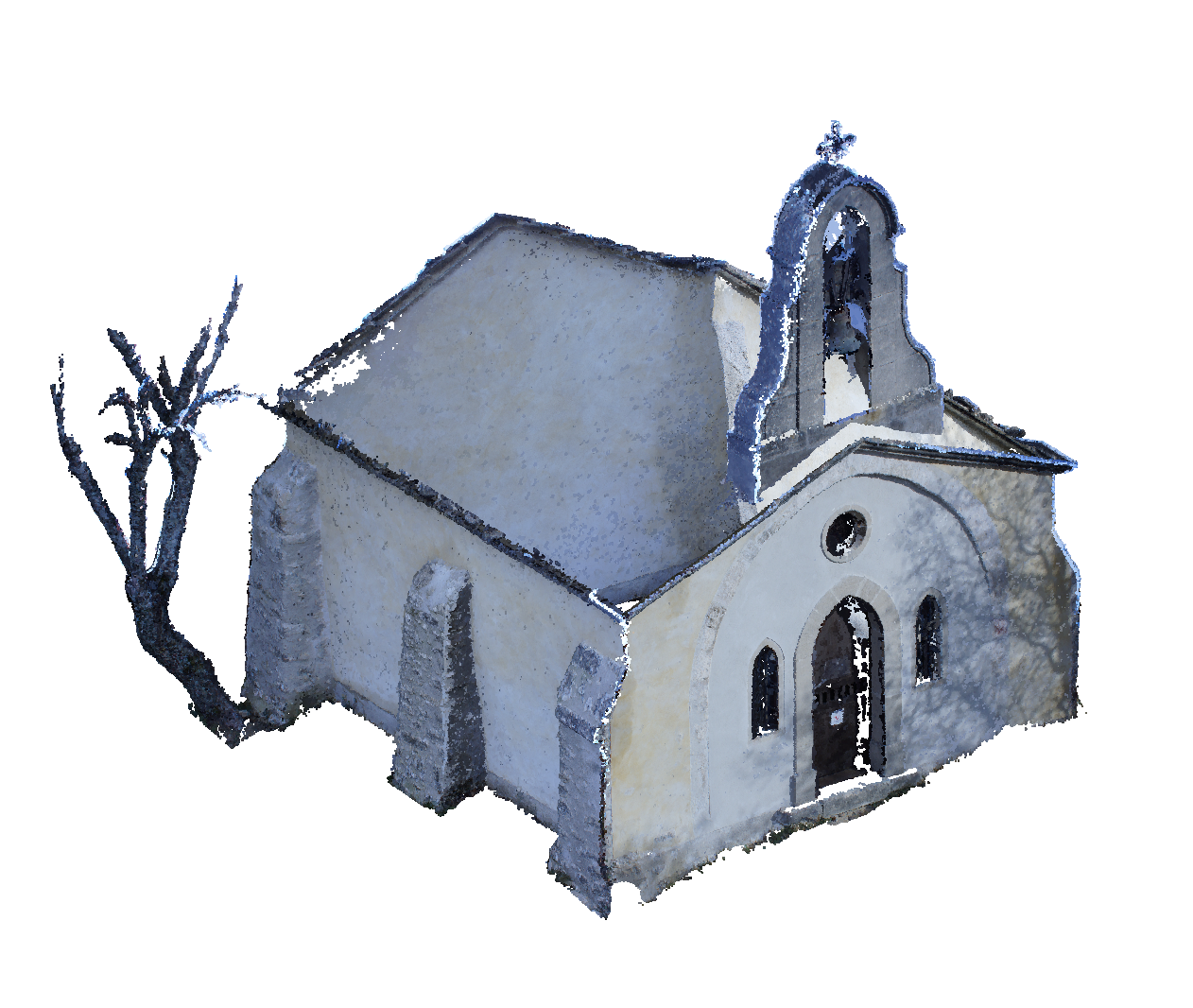}
    \end{minipage}}
    
    \caption{Point clouds in the evaluated dataset.}
    \label{fig:dataset}
\end{figure}

The design of the subjective experiments described in this paper started with the selection of the evaluated dataset. 
The test dataset described in the JPEG Pleno Point Cloud Common Training and Test Conditions (CTTC) \cite{JPEG-Pleno-PC-CfP} was designated as a starting point. 
This set contains twelve point clouds from three different density classes: solid, dense, and sparse. 
While the initial goal was to include models from the three different classes in the experiment, compression of the models of the sparse class was observed to be too time-consuming for some of the selected algorithms, especially V-PCC, and these point clouds were therefore excluded from consideration.
As for the remaining models from the test set, the aerial view from \emph{CITIUSP\_vox13} was considered to have too many elements needing visual attention for an adequate inspection in a short time frame, and \emph{Facade\_00009\_vox12} depicts a sculpture on a wall that can only be properly examined at a short angular range, not being suited for the evaluation protocols currently in use. 
The remaining six point clouds, displayed in Figure \ref{fig:dataset}, were therefore adopted for this subjective experiment. 
Table \ref{tab:dataset_stats} displays the value of different metrics computed on these point clouds extracted from the \textcolor{black}{JPEG Pleno Point Cloud CTTC} \cite{JPEG-Pleno-PC-CTTC}, showing that not only does it include models from different density classes, but also with different voxelization bit depth and color variation measured by the color gamut volume. 

The selected dataset was then compressed with G-PCC, V-PCC, and JPEG Pleno codecs. 
Four rate points were selected for each codec as a subset of their CTC, which for V-PCC and G-PCC define a set of precise parameters that result in a specific rate allocation between geometry and color. 
However, to the best of the authors' knowledge, there are no experiments in the literature that support these sets of parameters as providing the best subjective performance for a given rate. 
For this reason, two other rate allocation strategies producing the same or lower rates as the CTC were defined, with the goal of \textcolor{black}{investigating} whether they can obtain better perceptual results. 
For G-PCC and V-PCC, the baseline is denominated as P1, while the two evaluated strategies are denominated as P2 and P3. 
For JPEG Pleno, given that the CTTC does not define precise configuration parameters for each rate, the three strategies P1, P2, and P3 are proposed independently. 
The remainder of this section describes the configurations used for each codec as well as how their rate allocation was defined.

\subsection{G-PCC compression}

The reference software version 22.0 was employed to compress the dataset using G-PCC. 
Since previous experiments \cite{alexiou2019comprehensive} found that the \emph{octree} coding module outperformed \emph{trisoup} on average, it was used in this experiment for geometry coding. 
The same study also found that the performance of the color coding module \emph{predlift} is preferred by subjects when compared to RAHT and was therefore selected, similarly to other recent experiments \cite{perry2020quality, lazzarotto2022impact}. 
The configurations defined in the reference software adapt the value of two parameters to obtain different bitrates: $positionQuantizationScale$, here abbreviated to $pqs$, which determines the quantization scale applied to the geometric coordinates, and $qp$, which controls the quality of the compressed color. 
The first parameter can be set anywhere in the range from $0$ to $1$, with lower values reducing the bitrate but also producing coarser point clouds with lower quality. 
The value of $qp$ controls the compression of the color attributes, with the maximum value of $51$ producing the lowest available quality while the minimum value of $4$ corresponds to lossless compression. 
The CTC defines six rate points from $r01$ to $r06$ for the condition corresponding to lossy coding of geometry and color attributes (C2), with different values of $pqs$ and $qp$ depending on the rate point. 
The value of $pqs$ for each rate also depends on the voxelization bit depth of the input point cloud, with higher precision being associated with lower $pqs$ for the same rate. 
Table \ref{tab:gpcc_ctc_params} provides the values for $qp$ and $pqs$ defined in the CTC for all rates, with two sets of values being defined for the latter depending on the voxelization bit depth.

For the subjective experiment, only the rates from $r02$ to $r05$ were employed. 
While $r01$ resulted in very heavy visual degradation that is not likely to be useful in practical use cases, the differences between $r06$ and the reference model are almost imperceptible and would probably not be noticed in this experiment. 
In the remainder of this paper, the rates $r02$ to $r05$ defined in the G-PCC CTC are referred to as R1 to R4. 
As it can be seen in Table \ref{tab:gpcc_ctc_params}, the values of $pqs$ and $qp$ vary in a regular pattern between one rate and the other: in order to increase the rate by one level, the value of $pqs$ is doubled if it is lower than 0.5, or else its distance to 1 is halved; at the same time, the difference of the $qp$ value between two levels is always equal to 6. 
In the CTC, the variations of both parameters are coupled and they are not changed independently. 
However, even if an increase in $pqs$ and a decrease in $qp$ would result in a higher bitrate and better quality, the extent of their individual impact on both of these variables may be different. 
Moreover, given a target bitrate, many possible combinations of $pqs$ and $qp$ may result in similar rates while not necessarily keeping the visual quality constant. 

\begin{table*}[t]

\centering

\begin{tabular}{c c c c c c c c}

&  $r01$ & $r02$ & $r03$ & $r04$ & $r05$ & $r06$  \\ 
\toprule
$pqs$ 12-bit & 0.03125 & 0.0625 & 0.125 & 0.25 & 0.5 & 0.75  \\
$pqs$ 10-bit & 0.125 & 0.25 & 0.5 & 0.75 & 0.875 & 0.9375   \\
$qp$ & 51 & 46 & 40 & 34 & 28 & 22

\end{tabular}

\caption{Compression parameters suggested by G-PCC CTC for point clouds voxelized with 10 and 12 bit precision.}
\label{tab:gpcc_ctc_params}
\end{table*}

\begin{figure}
    \centering
    \begin{minipage}[b]{0.478\linewidth}
    \centering
    \includegraphics[width=\linewidth]{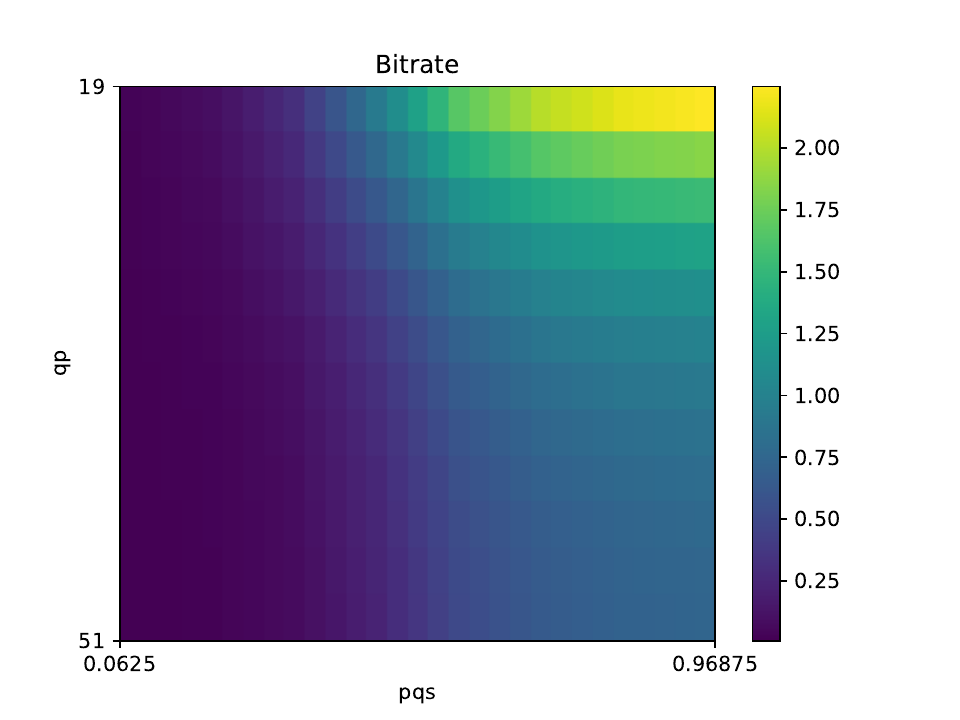}
    \end{minipage}
    \begin{minipage}[b]{0.478\linewidth}
    \centering
    \includegraphics[width=\linewidth]{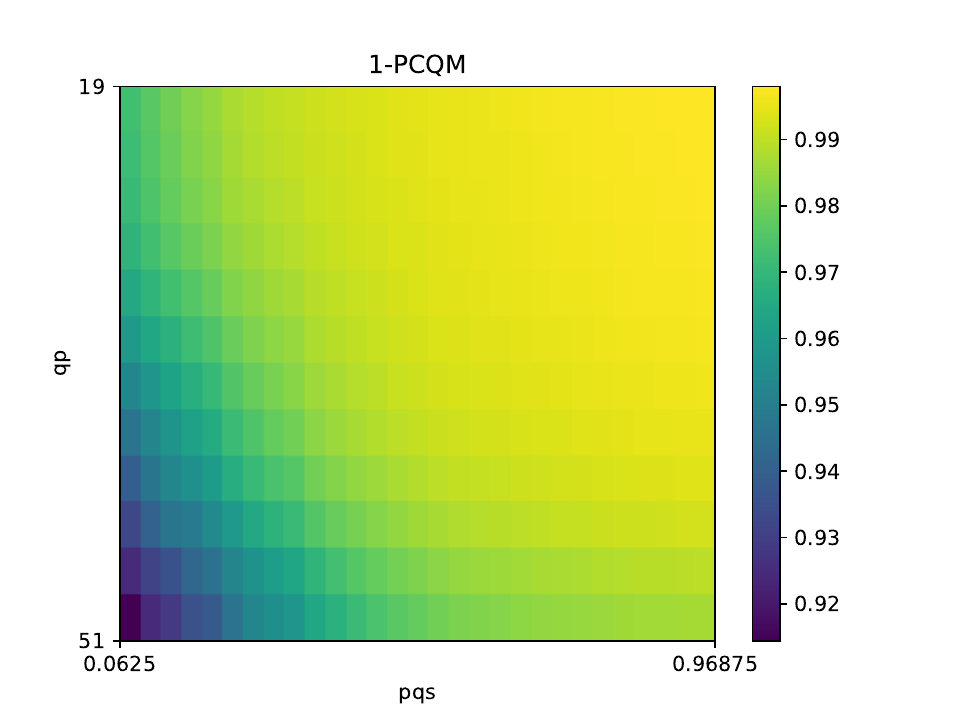}
    \end{minipage}
    
    \begin{minipage}[b]{0.478\linewidth}
    \centering
    \includegraphics[width=\linewidth]{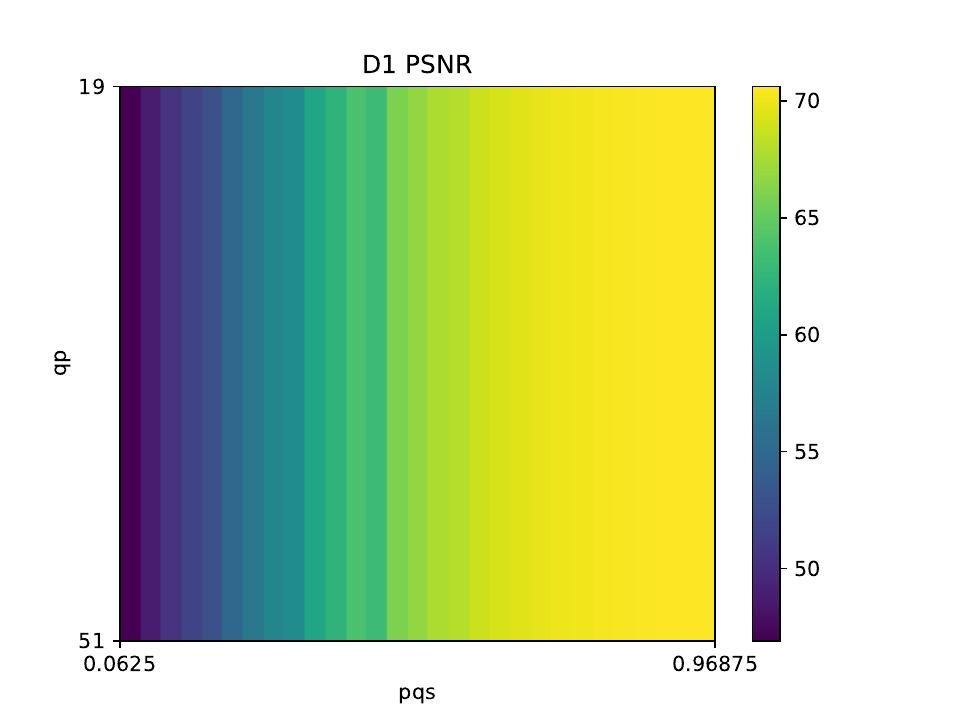}
    \end{minipage}
    \begin{minipage}[b]{0.478\linewidth}
    \centering
    \includegraphics[width=\linewidth]{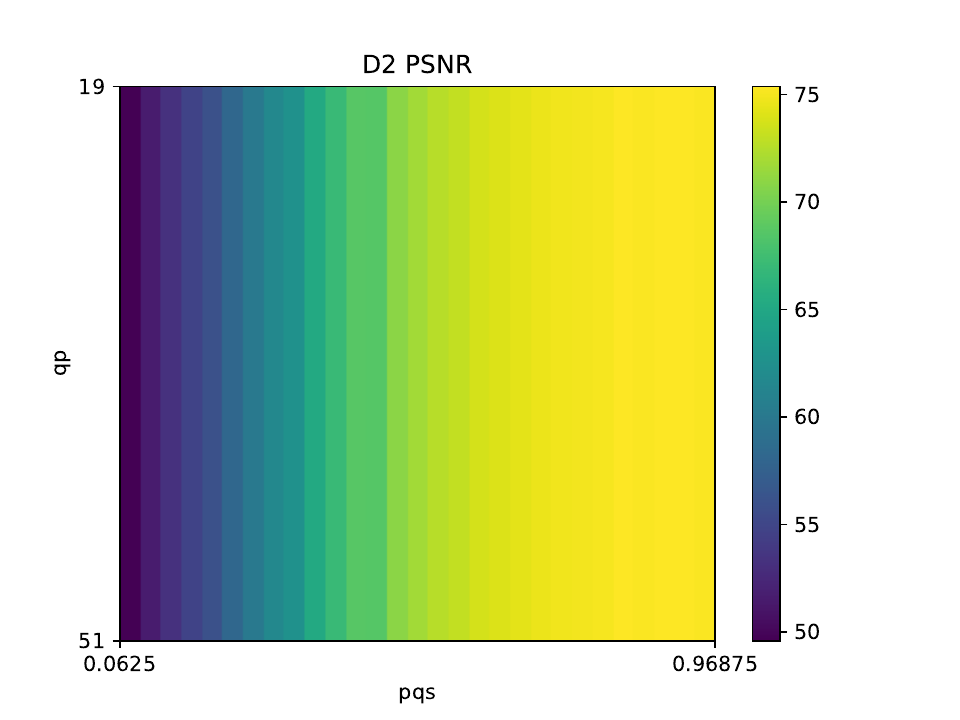}
    \end{minipage}

    \begin{minipage}[b]{0.478\linewidth}
    \centering
    \includegraphics[width=\linewidth]{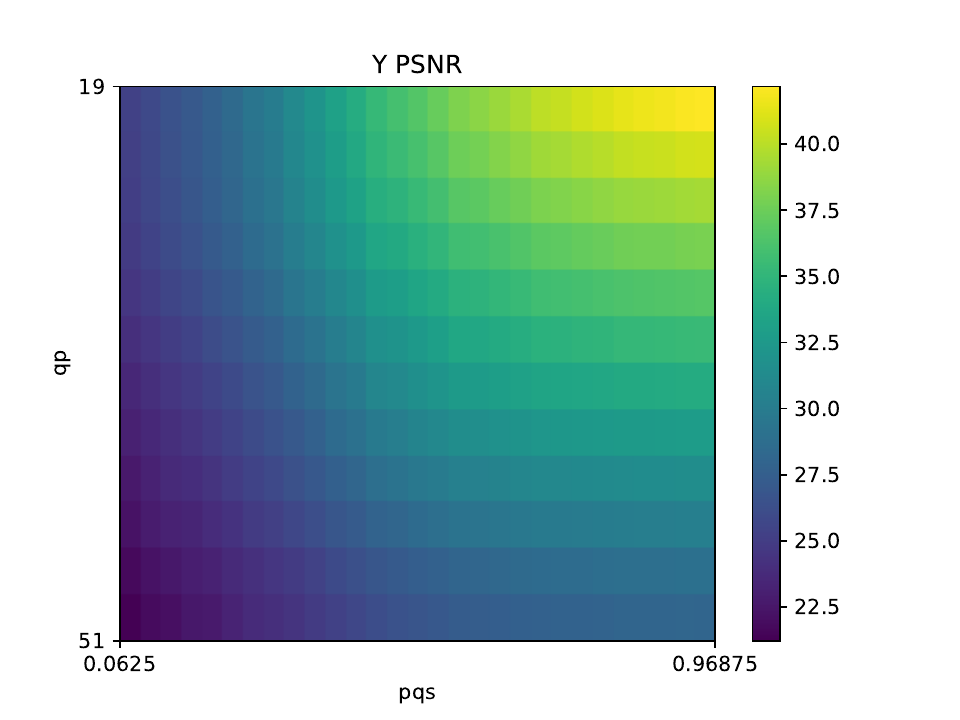}
    \end{minipage}
    \begin{minipage}[b]{0.478\linewidth}
    \centering
    \includegraphics[width=\linewidth]{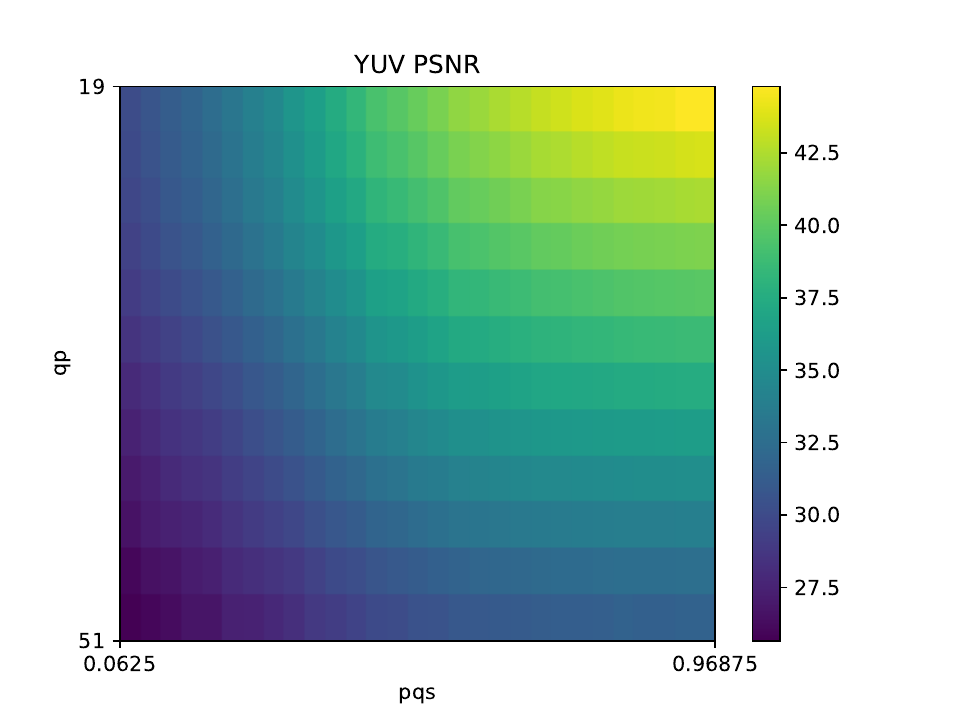}
    \end{minipage}
    
    \caption{Color maps of the bitrate and metric values for the compression of \emph{Soldier} with different values for $pqs$ and $qp$ with G-PCC.}
    \label{fig:gpcc_soldier_grid}
\end{figure}

The point clouds of the dataset were encoded with different $pqs$ and $qp$ values sampled from a uniform grid to investigate this issue, with objective quality metrics being computed between the decoded and reference models. 
The configuration files available in the reference software were used to define the other compression parameters, as recommended in the CTC. 
Moreover, for the point clouds \emph{Bouquet} and \emph{StMichael}, which are not included in the G-PCC test dataset, the configuration files from \emph{Soldier} were used since it is a point cloud from the same density class voxelized with the same bit depth. 
Point-to-point (D1) and point-to-plane (D2) \cite{tian2017geometric} PSNR were used to estimate geometric distortion, while Y PSNR and YUV PSNR were computed to estimate luminance and color distortion, respectively. 
In this study, the latter is a weighted average between the three color channels, assigning relative importance of 6 to the Y channel\textcolor{black}{, 1 to the U channel, and 1 to the V channel.} 
PCQM \cite{meynet2020pcqm} was also computed as a method aggregating geometry and color distortion in a single score. 
Even if PCQM has demonstrated a higher correlation with subjective perception in recent experiments, there is no consensus regarding the ability of any metric to accurately mimic the human visual system. 
Therefore, the values obtained in this analysis cannot be regarded as a precise predictor of quality, but rather as an estimator. 
The metric and bitrate values for \emph{Soldier} can be observed in Figure \ref{fig:gpcc_soldier_grid} as a color map indexed by $pqs$ in one axis and $qp$ in the other. 
Since a value of PCQM closer to 0 is supposed to be an estimator of higher quality, the value of 1-PCQM is used in the plot. 

Figure \ref{fig:gpcc_soldier_grid} demonstrates that the bitrate is significantly impacted both by $pqs$ and $qp$. 
Towards the highest bitrates of the grid, the color $qp$ plays the major role, with a small variation having higher effects on the final value. 
However, achieving the lowest bitrates is only possible with low $pqs$ values, with reduced impact from $qp$. 
Probably, the main reason for this observation is that low $pqs$ produces point clouds with fewer points, in which case there are a small number of attributes to encode, reducing the bitrate assigned to attributes even if low $qp$ values are used. 
The plots seem to indicate that PCQM does not suffer the same effect, with both parameters playing a similar role across the entire grid. 
Instead, the metric seems to plateau at the top right quadrant, having an exceptionally small value when both factors approach the lower range.  
If this metric correlates well with subjective perception across this range, these results suggest that there could be better strategies for rate allocation other than jointly adjusting both parameters. 
The geometry-based metrics are, as expected, only influenced by $pqs$, and the color maps from D1 PSNR and D2 PSNR are very similar. 
Therefore, these metrics cannot assist in the definition of a rate allocation strategy between color and geometry. 
However, even if Y PSNR and YUV PSNR are computed only on the color attributes, $pqs$ is observed to highly influence their values as well. 
These metrics can also be affected by geometric distortions since differences between the position of points impact the choice of nearest neighbors between reference and distorted model. 
Moreover, since a lower $pqs$ drastically reduces the number of points of the decoded point cloud, the transfer of the color attributes to the degraded geometry already incurs a loss of details by itself, even if the subsequent color compression is lossless. 
It is also observed that the added chrominance components in YUV PSNR do not change the general aspect of the metric map in comparison to Y PSNR. 

\begin{figure}
    \centering
    \subfloat[\emph{Bouquet}]{
    \begin{minipage}[b]{0.44\linewidth}
    \centering
    \includegraphics[width=\linewidth]{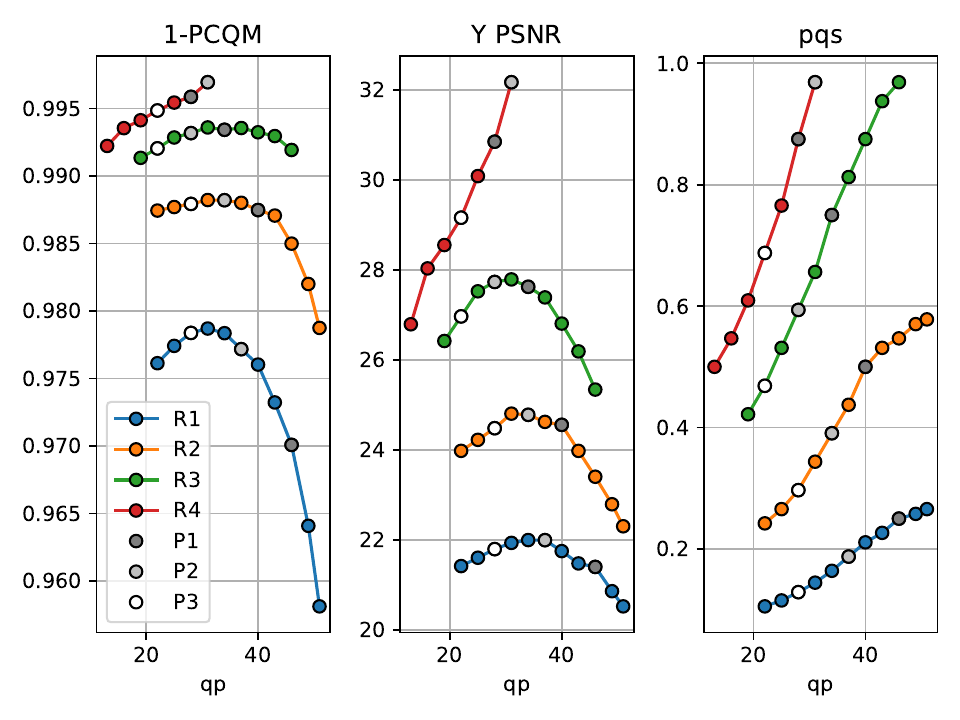}
    \end{minipage}}
    \subfloat[\emph{StMichael}]{
    \begin{minipage}[b]{0.44\linewidth}
    \centering
    \includegraphics[width=\linewidth]{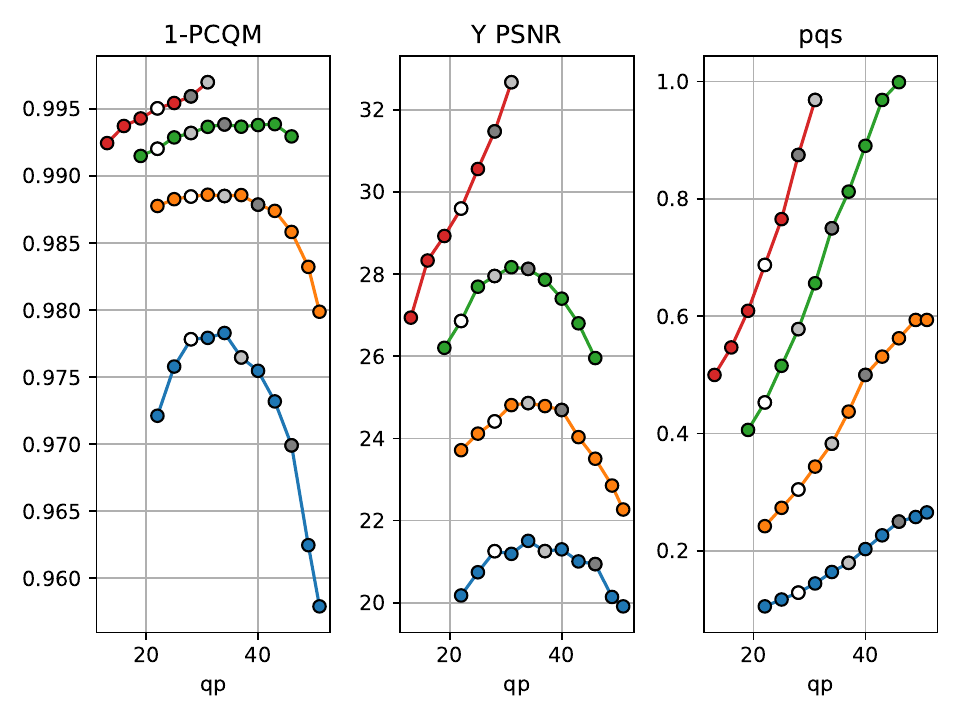}
    \end{minipage}}
    
    \subfloat[\emph{Soldier}]{
    \begin{minipage}[b]{0.44\linewidth}
    \centering
    \includegraphics[width=\linewidth]{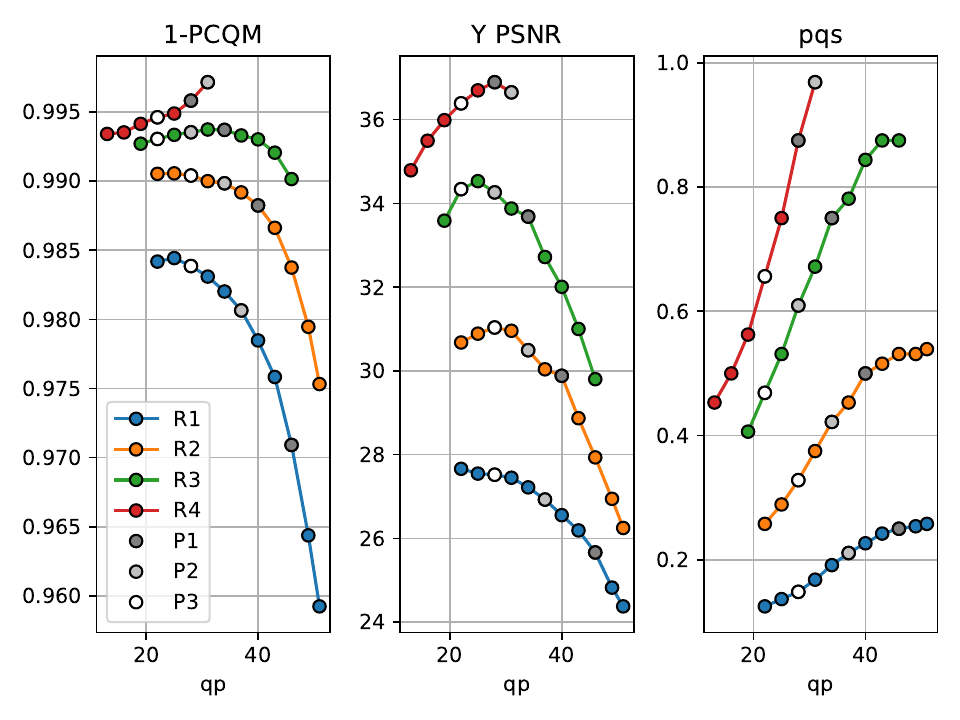}
    \end{minipage}}
    \subfloat[\emph{Thaidancer}]{
    \begin{minipage}[b]{0.44\linewidth}
    \centering
    \includegraphics[width=\linewidth]{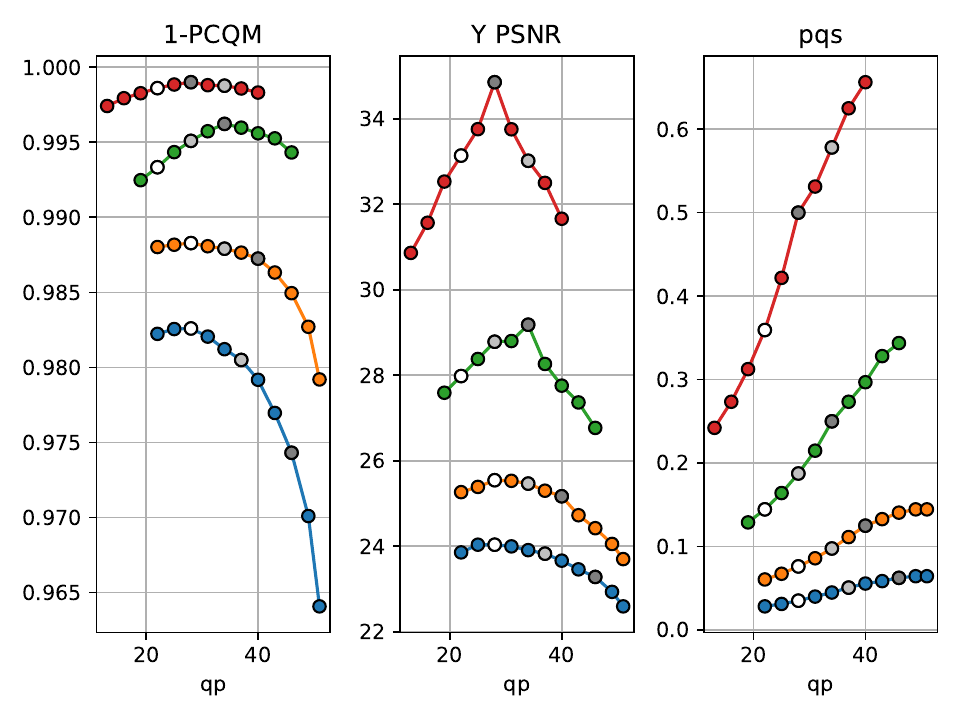}
    \end{minipage}}

    \subfloat[\emph{Boxer}]{
    \begin{minipage}[b]{0.44\linewidth}
    \centering
    \includegraphics[width=\linewidth]{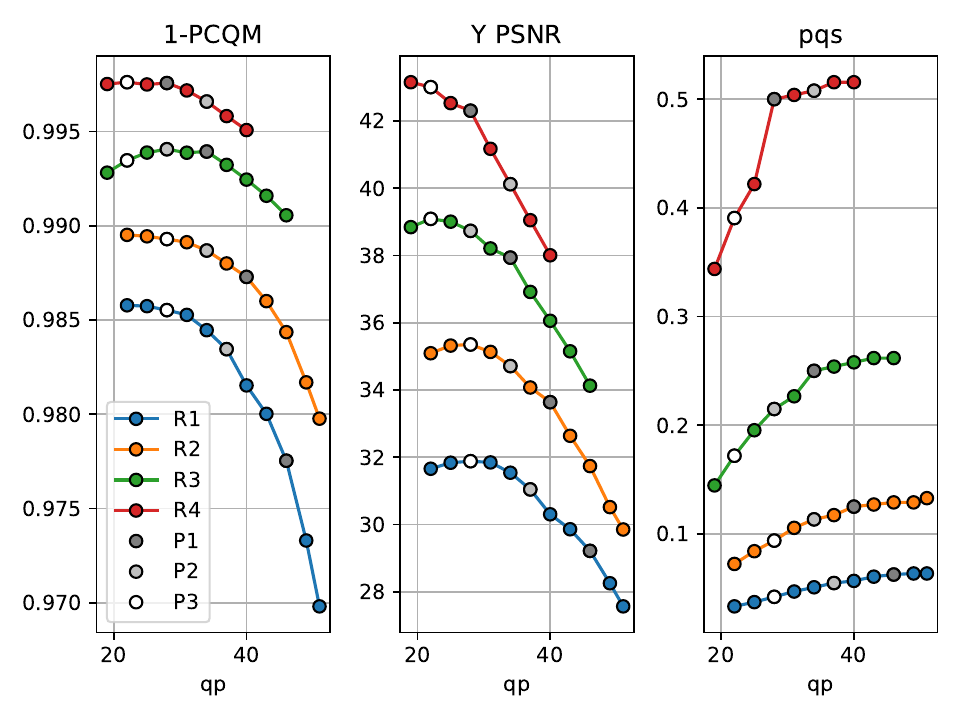}
    \end{minipage}}
    \subfloat[\emph{House\_without\_roof}]{
    \begin{minipage}[b]{0.44\linewidth}
    \centering
    \includegraphics[width=\linewidth]{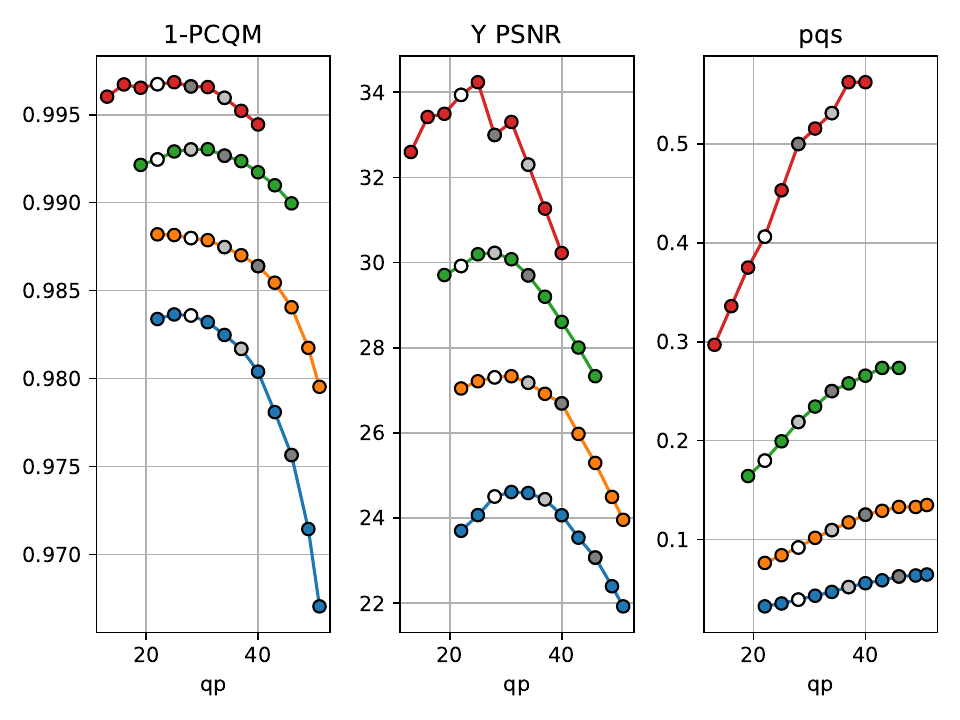}
    \end{minipage}}
    
    \caption{G-PCC isorate curves for 1-PCQM, Y PSNR and $pqs$ for each point cloud in the dataset. The points selected for the rate allocation strategies P1, P2 and P3 are highlighted.}
    \label{fig:gpcc_isorate_curves}
\end{figure}

This mismatch between the effect of the configuration parameters on the metrics and bitrate could mean that there may be configurations able to achieve better rate-distortion performance than the CTC. 
With this goal, an additional analysis was conducted where, for each rate point R1 to R4, a range of $qp$ values different from what is proposed by the CTC were tested, each one combined with a $pqs$ value needed to maintain the bitrate at the same or lower value. 
This analysis resulted in multiple isorate curves, one for each point cloud and original rate point, which can be observed in Figure \ref{fig:gpcc_isorate_curves} for PCQM and Y PSNR. 
Additionally, the $pqs$ value needed to obtain each point in the isorate curve is also illustrated. 

The majority of the obtained isorate curves show a concave shape for PCQM and Y PSNR, indicating that there might be an optimal $qp$ value that produces a point cloud with the highest quality for a given bitrate, and that a $qp$ above or below this optimal value would deliver poorer performance. 
However, this optimal $qp$ value is usually not the same for both PCQM and Y PSNR. 
For instance, for \emph{StMichael} at the rate R3, a $qp$ value of 43 produces the best point cloud according to PCQM but is far from the best-performing configuration for Y PSNR. 
As expected, Y PSNR is less affected by the geometry distortion, usually resulting in lower optimal $qp$ values than PCQM corresponding to a higher percentage of the bitrate being assigned to color. 
The rate of the point cloud is also found to affect the shape of the curves: for R4, the optimal $qp$ value for some point clouds is the highest one available in this evaluation even for Y PSNR. 
Moreover, the characteristics of the compressed point cloud heavily impacted the results as well, with lower $qp$ values being needed to optimize Y PSNR for \emph{Boxer} in comparison to \emph{Thaidancer} for example. 

Considering the available results, one option for the definition of the alternative strategies P2 and P3 could be to choose the $qp$ value that optimizes PCQM and Y PSNR, respectively. 
However, this approach would sometimes produce the same $qp$ and $pqs$ values for two or more strategies, resulting in a reduction in the number of evaluated compressed point clouds. 
For this reason, fixed values of $qp$ were selected for each allocation strategy at each rate point, being kept the same across all of the point clouds. 
The value of $pqs$ was adjusted in order to keep a constant rate across different allocation strategies, in the same way as for the computation of the values of Figure \ref{fig:gpcc_isorate_curves}. 
The values $qp$ attributed to P1, P2, and P3 are depicted in Table \ref{tab:gpcc_qp_values_for_p} and are also highlighted in Figure \ref{fig:gpcc_isorate_curves}. 
As it can be observed, lower $qp$ values are given to P2 and P3 when compared to P1 for the three lower rates, corresponding to a higher weight given to the color attributes. 
The chosen values are usually closer to the optimal PCQM and Y PSNR values than the CTC, which in most of the observed cases resulted in assigning a larger part of the bitrate to color. 
For the last rate point, P2 adopted a higher $qp$ than P1, allowing it to achieve higher metric values for many of the point clouds. 
In this case, the point clouds voxelized at 12-bit depth were compressed with $qp$ values of 34, while the remaining ones used a value of 31. 
For the latter, the associated $pqs$ value was already very close to 1, and there was thus no reason to further decrease the color quality without an increase in geometry quality in return. 
On the other hand, the value of $qp$ for P3 is always the lowest across the three rate allocation strategies, which therefore assigns the highest proportion of the bitrate to color attributes across the three strategies. 

\begin{table*}[]

\centering

\begin{tabular}{c c c c c}

&  R1 & R2 & R3 & R4  \\ 
\toprule
P1 & 46 & 40 & 34 & 28  \\
P2 & 37 & 34 & 28 & 34/31 \\
P3 & 28 & 28 & 22 & 22

\end{tabular}

\caption{Values employed for $qp$ at each rate and allocation strategy.}
\label{tab:gpcc_qp_values_for_p}
\end{table*}

\subsection{V-PCC compression}
\label{sec:vpcc_compression}

The reference software version 22.1 was used for the compression of the dataset with V-PCC. 
Configuration parameters recommended by MPEG \cite{MPEG-VPCC-Usage} were employed with V-PCC, with the VVC video codec being used to compress depth and color maps produced by the projection algorithm. 
Similarly to G-PCC, two independent parameters \emph{geometryQP} and \emph{attributeQP} are used in the configuration of the codec to define the quality of compression of the depth and color maps. 
These parameters are subsequently referred to as $gqp$ and $aqp$, respectively. 
An additional parameter \emph{occupancyPrecision} is used to control the precision of the representation of the occupancy map. 
The values given to these parameters at the rates defined in the CTC document can be observed in Table \ref{tab:vpcc_ctc_params}. 

\begin{table*}

\centering

\begin{tabular}{c c c c c c c c}

&  $r1$ & $r2$ & $r3$ & $r4$ & $r5$ \\ 
\toprule
$aqp$ & 42 & 37 & 32 & 27 & 22  \\
$gqp$  & 32 & 28 & 24 & 20 & 16   \\
\emph{occupancyPrecision}  & 4 & 4 & 4 & 4 & 2

\end{tabular}

\caption{Compression parameters suggested by V-PCC CTC.}
\label{tab:vpcc_ctc_params}

\end{table*}

The V-PCC reference software also defines a large set of parameters that are not dependent on the rate. 
This set includes parameters that are separately defined for each point cloud sequence included in the CTC test dataset and are available in the \texttt{cfg/sequence} folder of the reference software. 
However, this experiment employs point clouds that are not present in the MPEG dataset, and for this reason, do not have any official definition for such parameters. 
In this evaluation, the same sequence-specific configurations as the \emph{longdress} point cloud are employed, following the same method as the JPEG Pleno VM for the V-PCC-based projection of color maps. 
The performance of V-PCC in this evaluation may therefore not be optimal since these configuration parameters, which are not responsible for bitrate matching, were not separately tuned for each point cloud. 

\begin{figure}
    \centering
    \begin{minipage}[b]{0.478\linewidth}
    \centering
    \includegraphics[width=\linewidth]{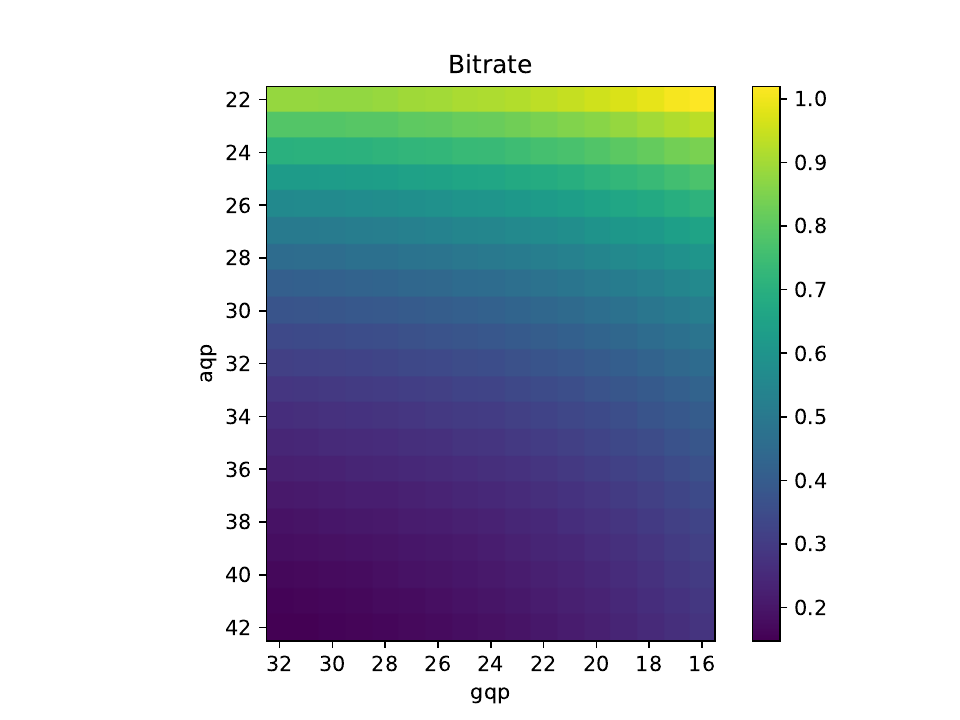}
    \end{minipage}
    \begin{minipage}[b]{0.478\linewidth}
    \centering
    \includegraphics[width=\linewidth]{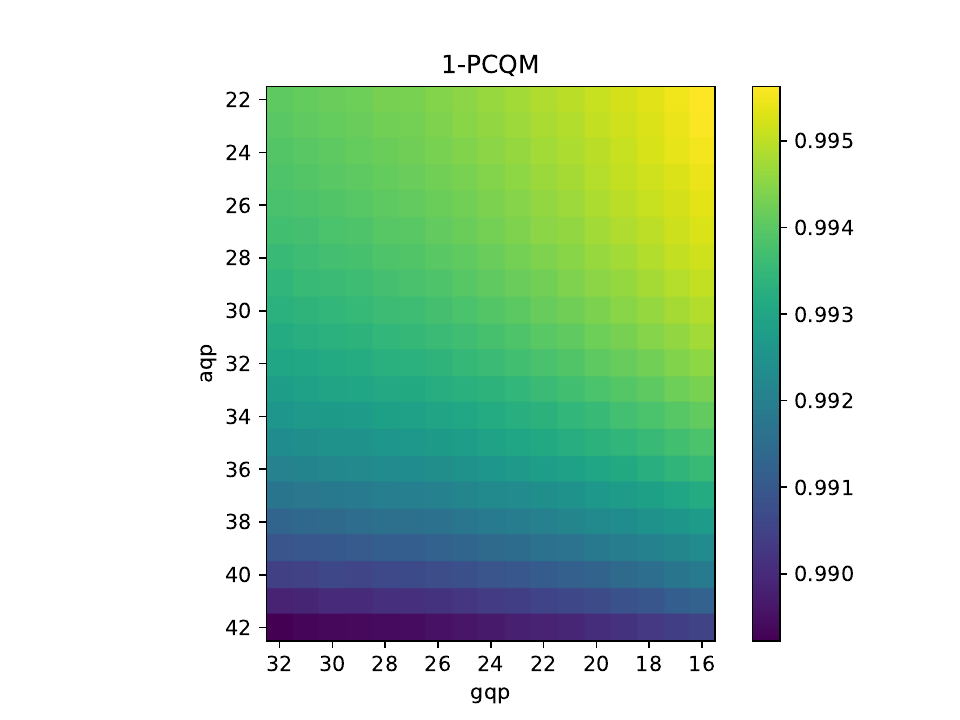}
    \end{minipage}
    
    \begin{minipage}[b]{0.478\linewidth}
    \centering
    \includegraphics[width=\linewidth]{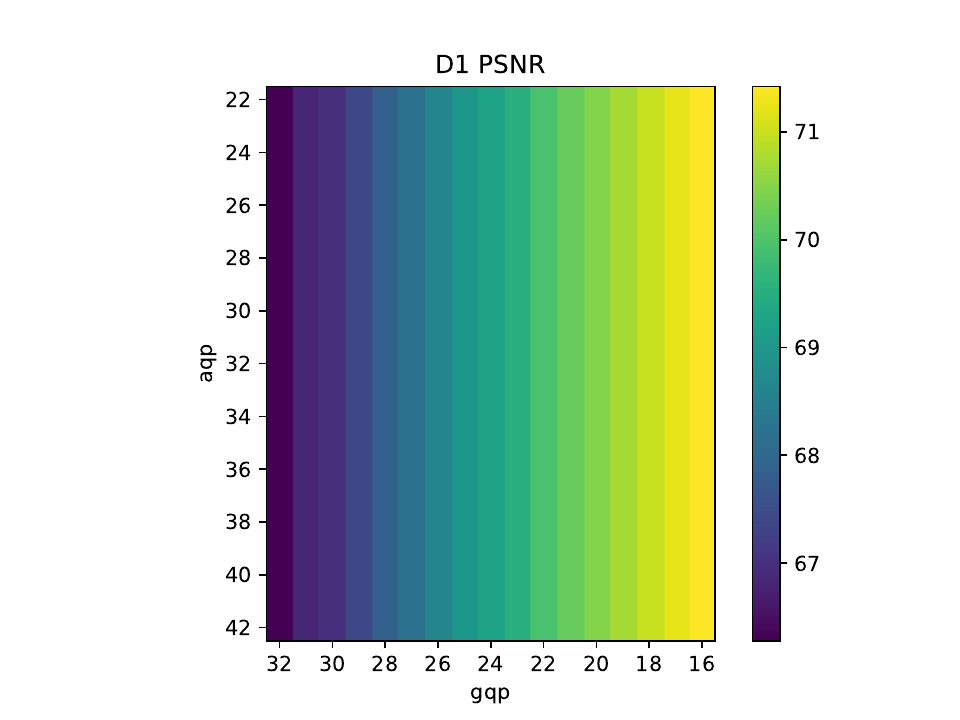}
    \end{minipage}
    \begin{minipage}[b]{0.478\linewidth}
    \centering
    \includegraphics[width=\linewidth]{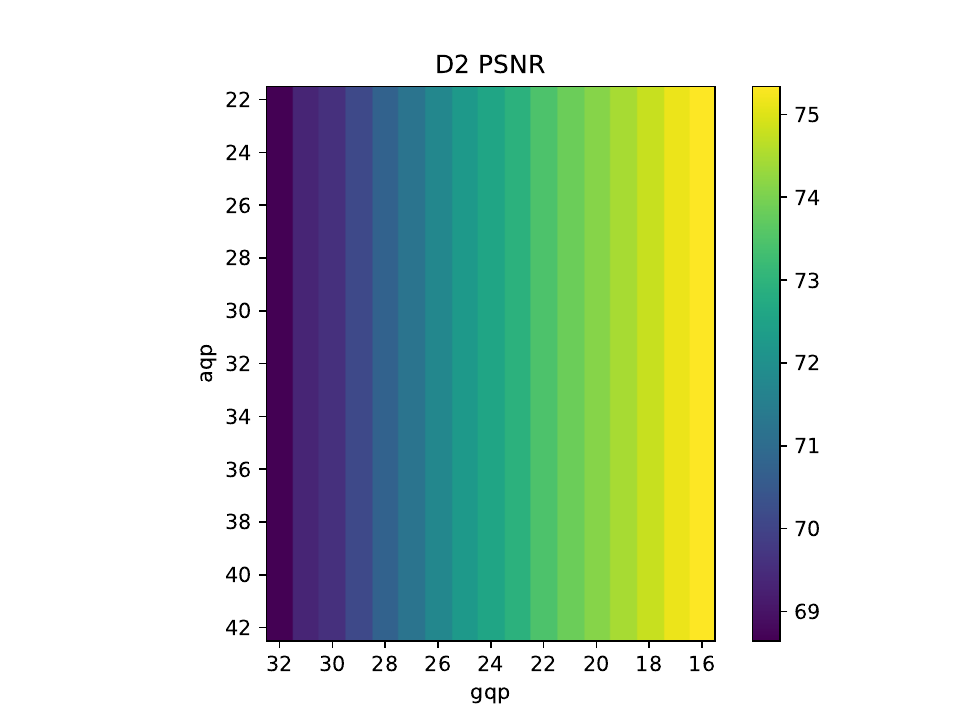}
    \end{minipage}

    \begin{minipage}[b]{0.478\linewidth}
    \centering
    \includegraphics[width=\linewidth]{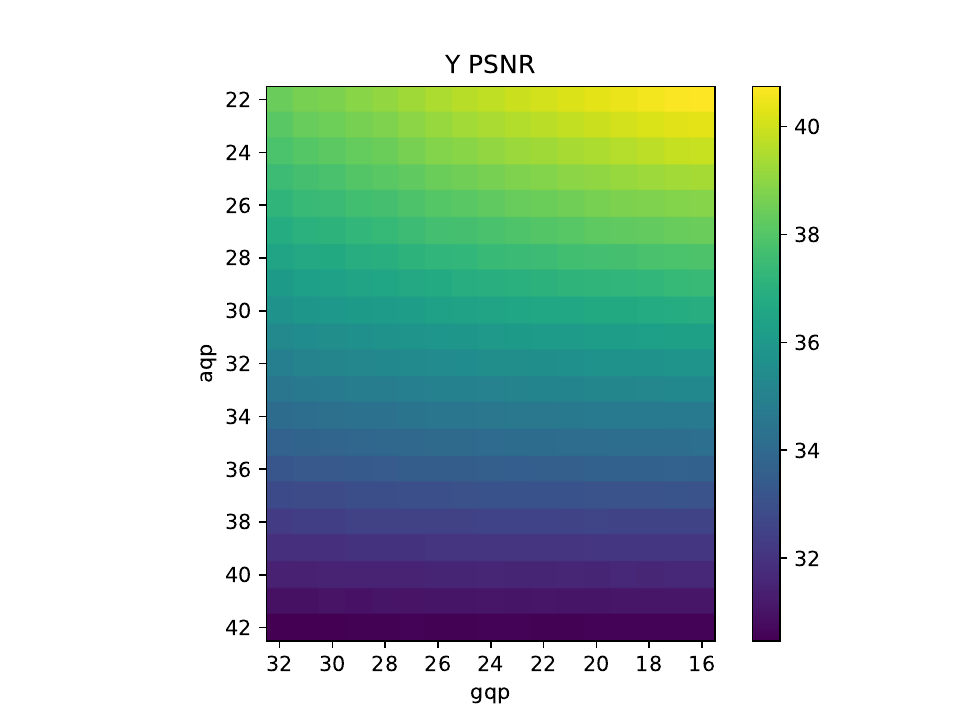}
    \end{minipage}
    \begin{minipage}[b]{0.478\linewidth}
    \centering
    \includegraphics[width=\linewidth]{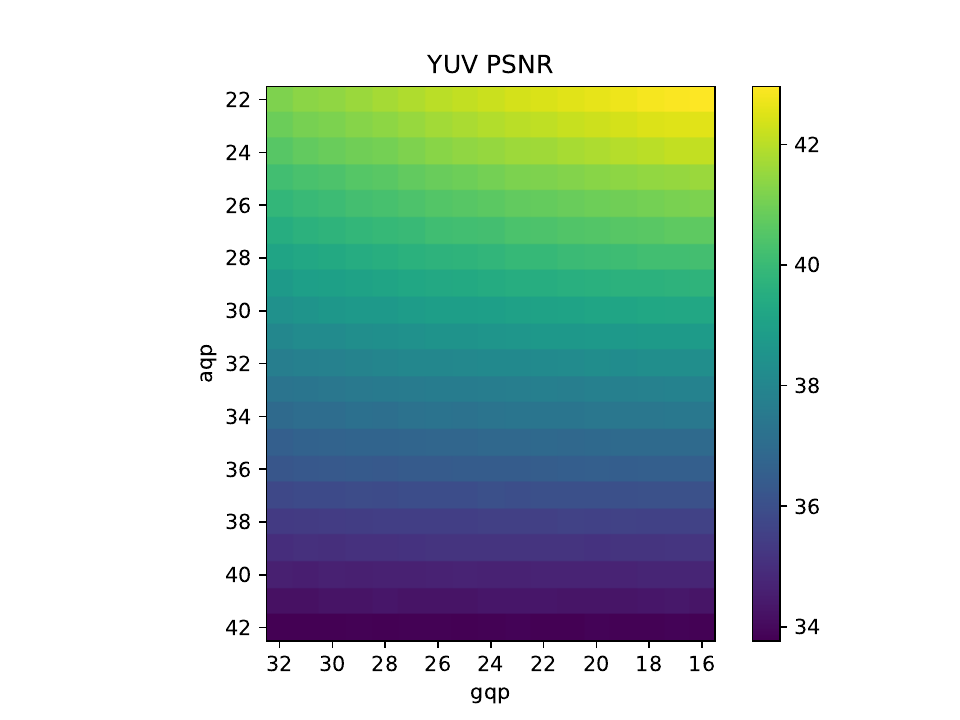}
    \end{minipage}
    
    \caption{Color maps of bitrate and metric values for the compression of \emph{Soldier} with different values for $aqp$ and $gqp$ with V-PCC.}
    \label{fig:vpcc_soldier_grid}
\end{figure}

In order to keep the bitrate and quality range as close as possible to G-PCC, the four first rate points were selected, and are similarly referred to as R1 to R4 with a capital letter in the remainder of this paper. 
Similarly to G-PCC, the parameters defined in the CTC are not guaranteed to deliver optimal visual quality for a given rate. 
For that reason, an analysis is conducted where a point cloud was encoded and decoded with $aqp$ and $gqp$ values following a grid with step size of 1, while keeping a constant \emph{occupancyPrecision} of 4. 
The color maps illustrating the corresponding bitrate and metric values can be found in Figure \ref{fig:vpcc_soldier_grid}.

\begin{figure}
    \centering
    \subfloat[\emph{Bouquet}]{
    \begin{minipage}[b]{0.42\linewidth}
    \centering
    \includegraphics[width=\linewidth]{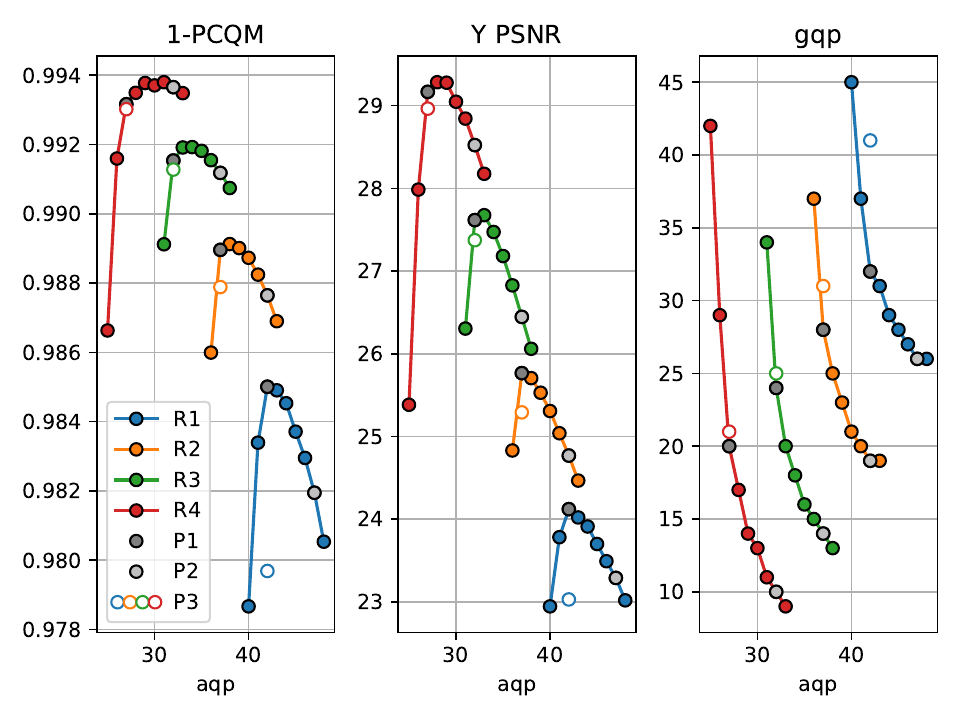}
    \end{minipage}}
    \subfloat[\emph{StMichael}]{
    \begin{minipage}[b]{0.42\linewidth}
    \centering
    \includegraphics[width=\linewidth]{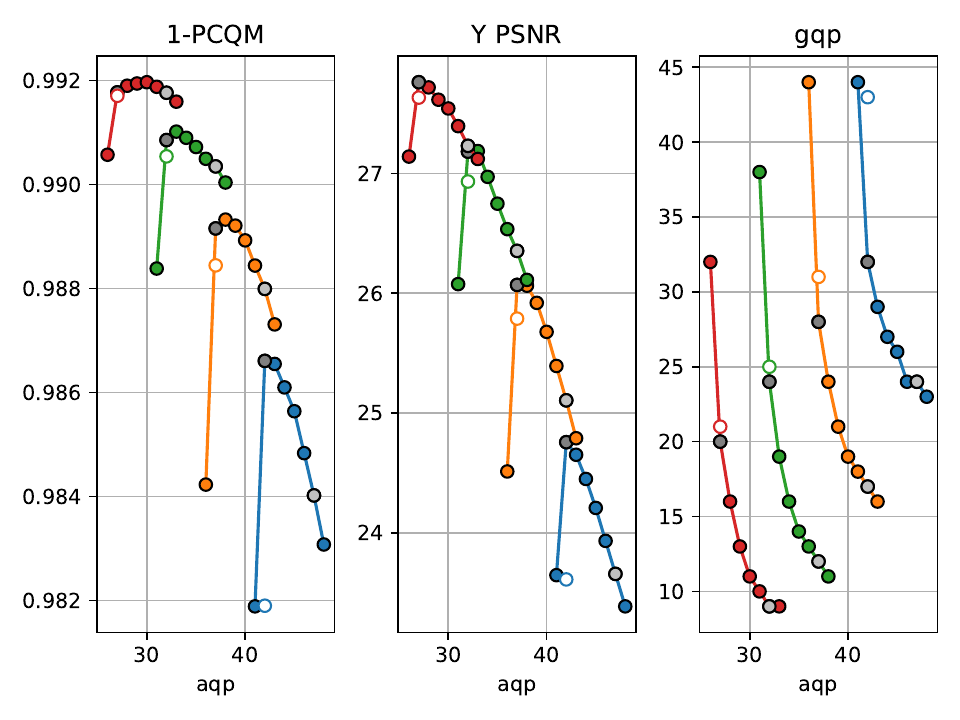}
    \end{minipage}}
    
    \subfloat[\emph{Soldier}]{
    \begin{minipage}[b]{0.42\linewidth}
    \centering
    \includegraphics[width=\linewidth]{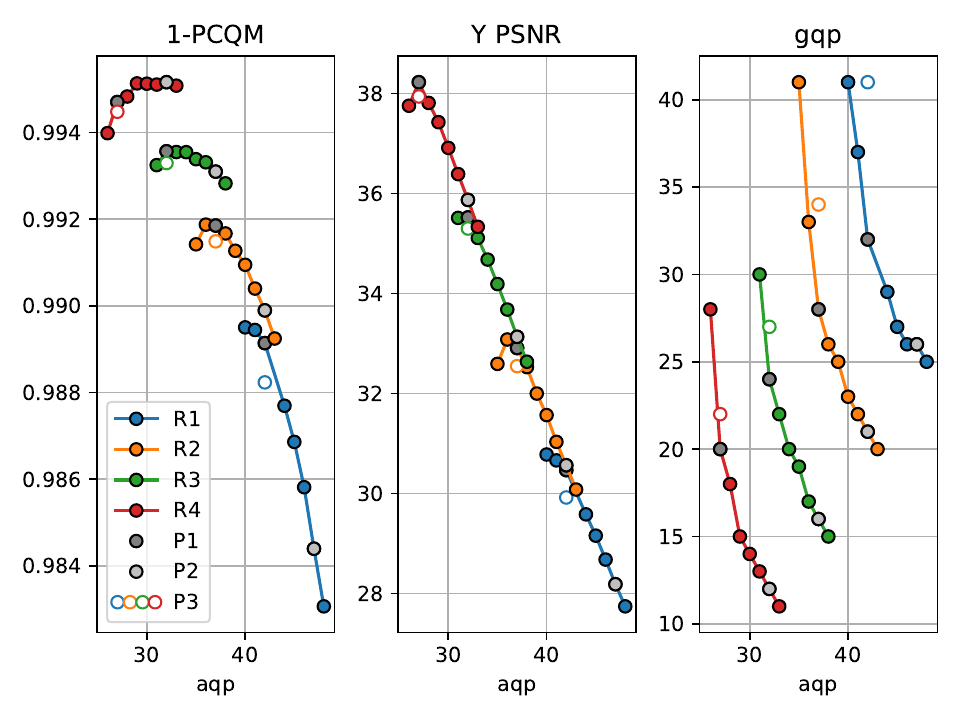}
    \end{minipage}}
    \subfloat[\emph{Thaidancer}]{
    \begin{minipage}[b]{0.42\linewidth}
    \centering
    \includegraphics[width=\linewidth]{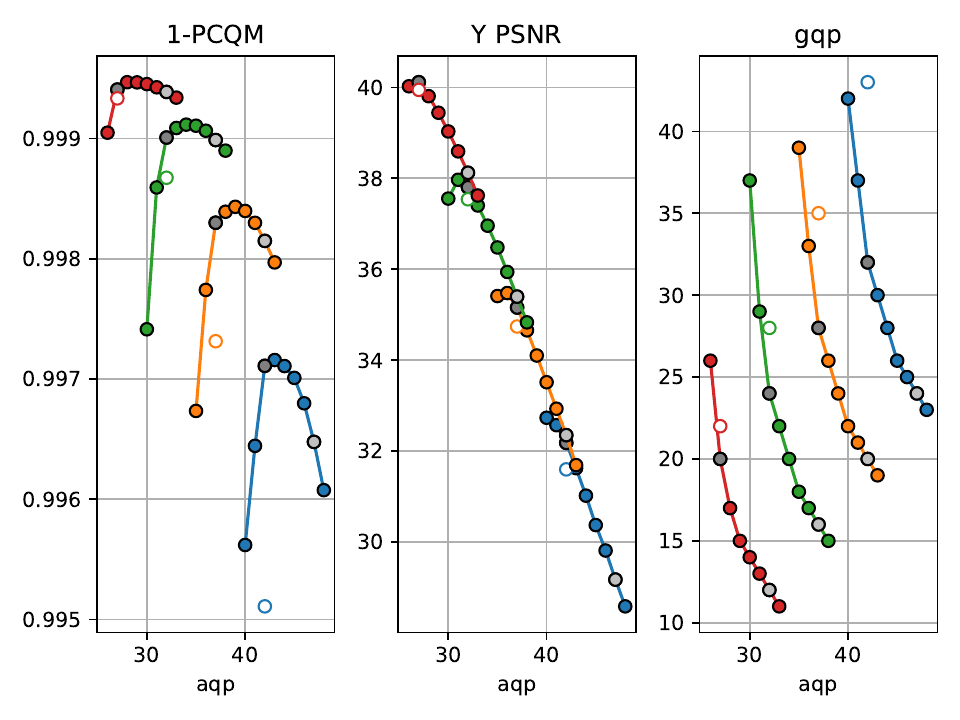}
    \end{minipage}}

    \subfloat[\emph{Boxer}]{
    \begin{minipage}[b]{0.42\linewidth}
    \centering
    \includegraphics[width=\linewidth]{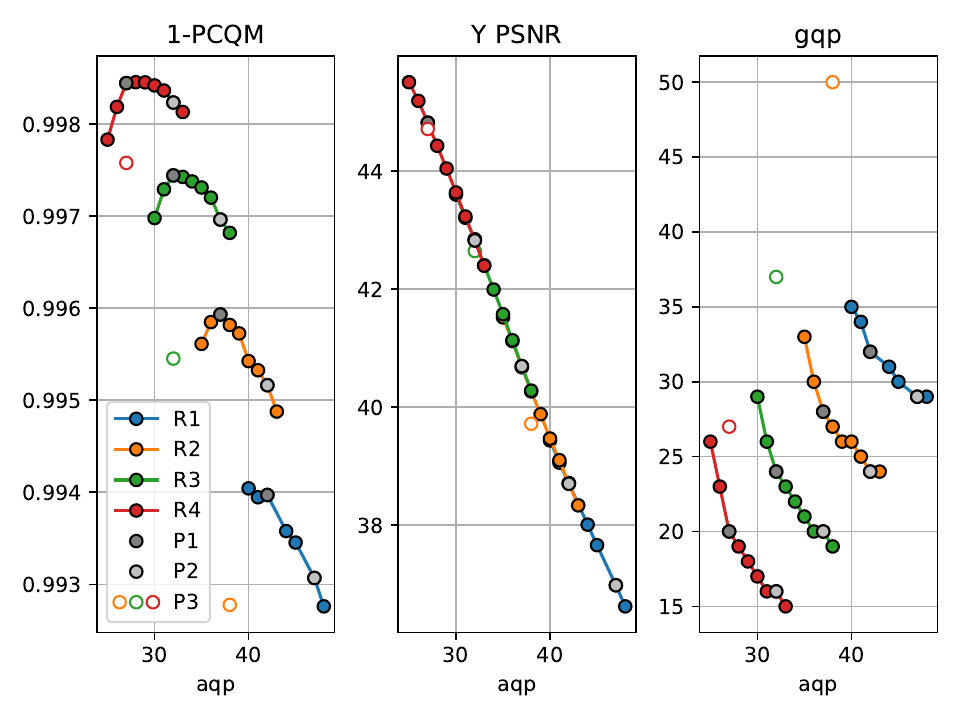}
    \end{minipage}}
    \subfloat[\emph{House\_without\_roof}]{
    \begin{minipage}[b]{0.42\linewidth}
    \centering
    \includegraphics[width=\linewidth]{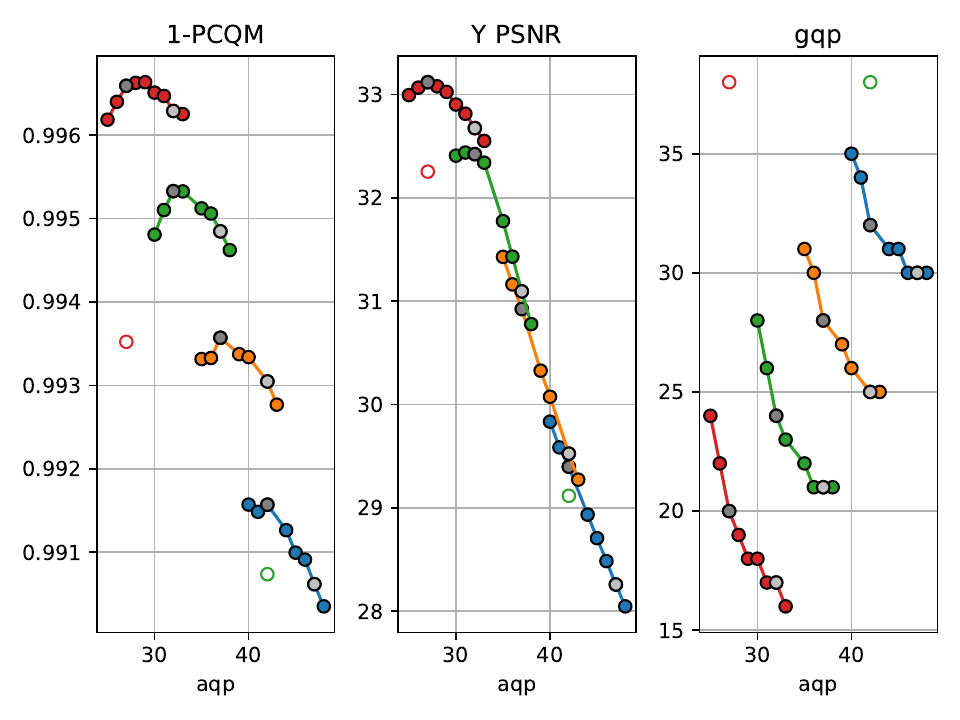}
    \end{minipage}}
    
    \caption{V-PCC isorate curves for 1-PCQM, Y PSNR and $gqp$ for each point cloud in the dataset. The points selected for the rate allocation strategies P1, P2 and P3 are highlighted. The value for \emph{occupancyPrecision} is kept at 4 except for the P3 points, where this value is set to 2.}
    \label{fig:vpcc_isorate_curves}
\end{figure}

Excluding the geometry-only metrics, the patterns presented in Figure \ref{fig:vpcc_soldier_grid} for PCQM, Y PSNR, and the bitrate share many more similarities with each other for V-PCC than for G-PCC. 
In particular, the bitrate is more impacted by $aqp$ than $gqp$, indicating a higher contribution of variations in the quality parameter for the color rather than for the geometry. 
These results are to be expected since the attribute maps are composed of three channels, while the depth maps representing the geometry are constituted of only one channel. 
Similarly, Y PSNR and YUV PSNR are also more impacted by $aqp$. 
The effect of $gqp$ on these color metrics is much lower than for G-PCC, indicating a higher level of independence between color and geometry coding. 
Since PCQM is a joint metric, its value is more impacted by $gqp$ than Y PSNR and YUV PSNR. 
However, this higher impact is more observed for lower values of $gqp$, with a predominant dependence of $aqp$ at the right half of the map where the geometry quality is worse. 
The higher apparent correlation between bitrate and PCQM may indicate that there is less opportunity for a significant increase in rate-distortion performance due to different rate allocation strategies. 
In order to further investigate this assumption, isorate curves were obtained similarly to G-PCC, i.e. searching the value for $gqp$ that keeps the same or lower bitrate as the CTC configurations for different $aqp$ values. 
The corresponding results can be observed in Figure \ref{fig:vpcc_isorate_curves}. 
In this analysis, the \emph{occupancyPrecision} was again kept at a value of 4. 

The isorate plots from Figure \ref{fig:vpcc_isorate_curves} further indicate that the configurations present in the CTC document provide good performance, always achieving the optimal or near-optimal PCQM value for its rate across the evaluated samples. 
A small increment in $aqp$ leads sometimes to improved PCQM values, but is followed by a decline when the variation is large. 
On the other hand, decreasing $aqp$ leads to a sharp decline in quality in some cases. 
The isorate curves also show that a given $aqp$ value leads to very similar Y PSNR values across different rates for the majority of point clouds, suggesting that color and geometry quality can be almost independently controlled by $gqp$ and $aqp$. 
In these cases, an analysis of the Y PSNR metric does not provide useful insights about the joint quality of color and geometry. 

The obtained results for the objective quality metrics indicate that the selection of alternative rate allocation strategies for V-PCC with superior performance than the CTC at equivalent bitrates is challenging. 
Since lower values of $aqp$ often lead to abrupt losses according to PCQM, no alternatives assigning a higher portion of the bitrate to color were considered. 
Instead, the $aqp$ value for the rate allocation strategy P2 is set to a value of 5 units higher than what is defined in the CTC, with \emph{occupancyPrecision} equal to 4 and $gqp$ as low as possible while keeping the bitrate equal or lower. 
For P3, the value of $aqp$ is kept the same but instead \emph{occupancyPrecision} is set to 2, and the value of $gqp$ is adjusted to keep bitrate equal or lower. 
This rule was applied for the entire dataset at all rates except for \emph{Boxer} and \emph{House\_without\_roof}, for which, at some rates, even the highest possible value for $gqp$ would result in higher bitrates for P3. 
For this reason, the value of $aqp$ was also adapted together with $gqp$ for some rate points, while for others the points were excluded from the analysis, resulting in incomplete curves for P3. 
The resulting parameter and metric values for P1, P2, and P3 are indicated in Figure \ref{fig:vpcc_isorate_curves}. 

\subsection{JPEG Pleno compression}

The VM version 3.0 was used to compress the test set with the JPEG Pleno codec. 
Unlike the MPEG compression standards, this encoding engine is not able to achieve its full range of bitrates only by varying two compression parameters. 
The parameter $\lambda$ allows the selection of a geometry compression model trained for a specific rate-distortion trade-off. 
Each model was trained with the loss function $L = \lambda R + D$, where $R$ is the estimated rate measured by the entropy of the latent features and $D$ is the distortion measured by the focal loss \cite{lin2017focal}. 
Therefore, choosing a lower $\lambda$ would increase both the bitrate and the quality of the point cloud, with a total of five values being available in the current version of the VM. 
However, even the highest $\lambda$ is not able to achieve the lowest desired bitrate range, especially for sparser point clouds. 
For that reason, downsampling can be applied prior to compression according to a sampling factor (SF), which drastically reduces the amount of points to be encoded. 
On the decoder side, learning-based super resolution can be used for upsampling, with the VM including super resolution modules trained for SF equal to 2 and 4. 
Moreover, the quantization step (q) used to produce integer coefficients in the latent space for entropy coding can also be adjusted, allowing for a finer control of the rate-distortion trade-off. 
In this study, only $\lambda$ and SF were varied, with the latter always being set to a power of 2, while q was kept to the default value of 1. 
Regarding the compression of color attributes, the parameter \emph{color\_rate\_index} (CRI) is used for rate control, which can be set with a value from 0 to 4 allowing the selection of one of the five available configurations for JPEG AI compression. 
\textcolor{black}{While the recent versions of JPEG AI allow for finer rate control thanks to the bitrate matcher, this functionality is not incorporated in the version of the JPEG Pleno VM used here.}

According to the CTTC document \cite{JPEG-Pleno-PC-CTTC}, the target rates for JPEG Pleno should approximately follow those of G-PCC. 
The same bitrate values obtained for G-PCC at R1 to R4 were therefore set as targets for JPEG Pleno. 
For each point cloud and target rate, the three parameters $\lambda$, SF, and CRI were adjusted in order to match the target rate as accurately as possible. 
Since a given set of parameter values can lead to broadly different bitrates depending on the characteristics of the point cloud, the selection of parameter values was separately conducted for each point cloud and rate. 
Given the limited number of choices for the compression parameters, in particular for color, the VM does not allow for fine granularity in the same way as the MPEG codecs. 
For that reason, detailed analysis with isorate curves could not be conducted. 
Instead, the proportion $p_g$ of the bitrate used for the representation of the geometry was used to define rate allocation strategies, where $p_g$ is equal to the size of the geometry bitstream divided by the size of the entire bitstream. 
The three different rate allocation strategies were defined as follows: P1 corresponded to $p_g < 0.4$, P2 corresponded to $0.4 < p_g < 0.6$, and P3 corresponded to $p_g > 0.6$. 
Even if these rules were followed for the majority of the selected configurations, there were cases for which there were not many available configurations leading to a bitrate reasonably close to the target, leading to an insufficient amount of available configurations matching these rules. 
For these cases, different values of $p_g$ were allowed, but the order between the three strategies was maintained, i.e. P3 assigned a higher weight to geometry and P1 assigned a higher weight to color.

\begin{table*}[]

\centering
\resizebox{\textwidth}{!}{
\begin{tabular}{c c c c c c c c c c c c c c}

\multirow{2}{*}{Point cloud} & \multirow{2}{*}{Rate} && \multicolumn{3}{c}{P1} && \multicolumn{3}{c}{P2} && \multicolumn{3}{c}{P3}  \\ 
\cmidrule{4-6} \cmidrule{8-10} \cmidrule{12-14}
& && $\lambda$ & SF & CRI && $\lambda$ & SF & CRI && $\lambda$ & SF & CRI  \\
\toprule
\multirow{4}{*}{\emph{Bouquet}} & R1 && 0.05 & 2 & 1 && 0.025 & 2 & 0 && 0.01 & 2 & 0 \\
& R2 && 0.05 & 1 & 1 && 0.025 & 1 & 0 && 0.01 & 1 & 0\\
& R3 && 0.005 & 1 & 3 && 0.005 & 1 & 2 && 0.0025 & 1 & 2 \\
& R4 && 0.005 & 1 & 4 && 0.0025 & 1 & 4 && 0.0025 & 1 & 3 \\
\midrule
\multirow{4}{*}{\emph{StMichael}} & R1 && 0.05 & 2 & 1 && 0.025 & 2 & 0 && 0.01 & 2 & 0 \\
& R2 && 0.05 & 1 & 2 && 0.025 & 1 & 1 && 0.01 & 1 & 0\\
& R3 && 0.01 & 1 & 3 && 0.005 & 1 & 2 && 0.0025 & 1 & 2\\
& R4 && 0.005 & 1 & 4 && 0.0025 & 1 & 4 && 0.0025 & 1 & 3\\
\midrule
\multirow{4}{*}{\emph{Soldier}} & R1 && 0.01 & 4 & 3 && 0.005 & 4 & 2 && 0.025 & 2 & 0 \\
& R2 && 0.05 & 1 & 2 && 0.05 & 1 & 1 && 0.025 & 1 & 0 \\
& R3 && 0.025 & 1 & 3 && 0.01 & 1 & 3 && 0.005 & 1 & 2 \\
& R4 && 0.005 & 1 & 4 && 0.0025 & 1 & 4 && 0.0025 & 1 & 3 \\
\midrule
\multirow{4}{*}{\emph{Thaidancer}} & R1 && 0.05 & 8 & 1 && 0.025 & 8 & 1 && 0.025 & 8 & 0 \\
& R2 && 0.05 & 4 & 1 && 0.025 & 4 & 1 && 0.05 & 4 & 0 \\
& R3 && 0.025 & 2 & 2 && 0.025 & 2 & 1 && 0.01 & 2 & 0 \\
& R4 && 0.025 & 1 & 3 && 0.01 & 1 & 2 && 0.005 & 1 & 1 \\
\midrule
\multirow{4}{*}{\emph{Boxer}} & R1 && 0.05 & 8 & 2 && 0.05 & 8 & 1 && 0.025 & 8 & 0 \\
& R2 && 0.05 & 4 & 3 && 0.05 & 4 & 2 && 0.025 & 4 & 1 \\
& R3 && 0.05 & 2 & 3 && 0.05 & 2 & 2 && 0.025 & 2 & 1 \\
& R4 && 0.05 & 1 & 3 && 0.005 & 2 & 4 && 0.0025 & 2 & 3 \\
\midrule
\multirow{4}{*}{\emph{House\_without\_roof}} & R1 && 0.05 & 8 & 2 && 0.025 & 8 & 1 && 0.01 & 8 & 0 \\
& R2 && 0.025 & 4 & 2 && 0.025 & 4 & 1 && 0.01 & 4 & 1 \\
& R3 && 0.025 & 2 & 3 && 0.01 & 2 & 2 && 0.005 & 2 & 1 \\
& R4 && 0.01 & 1 & 3 && 0.005 & 1 & 3 && 0.005 & 1 & 2 \\
\bottomrule
\end{tabular}
}
\caption{Configuration parameters used for JPEG Pleno compression}
\label{tab:jpeg_params}

\end{table*}

\begin{figure*}
    \centering
    \begin{tabular}{ccccc} 
    Point cloud & R1 & R2 & R3 & R4 \\ \hline
    
    \emph{Bouquet}&
    \begin{minipage}[b]{0.11\linewidth}
    \includegraphics[width=\linewidth]{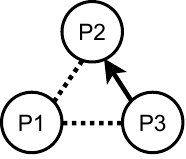}
    \end{minipage} &
    \begin{minipage}[b]{0.11\linewidth}
    \includegraphics[width=\linewidth]{Figures--JPEG_profiles--P3P2.pdf}
    \end{minipage} &
    \begin{minipage}[b]{0.11\linewidth}
    \includegraphics[width=\linewidth]{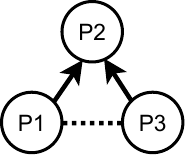}
    \end{minipage} &
    \begin{minipage}[b]{0.11\linewidth}
    \includegraphics[width=\linewidth]{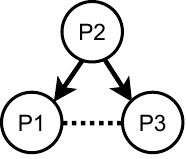}
    \end{minipage} \\
    
    \emph{StMichael}&
    \begin{minipage}[b]{0.11\linewidth}
    \includegraphics[width=\linewidth]{Figures--JPEG_profiles--P3P2.pdf}
    \end{minipage} &
    \begin{minipage}[b]{0.11\linewidth}
    \includegraphics[width=\linewidth]{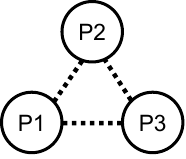}
    \end{minipage} &
    \begin{minipage}[b]{0.11\linewidth}
    \includegraphics[width=\linewidth]{Figures--JPEG_profiles--P3P2.pdf}
    \end{minipage} &
    \begin{minipage}[b]{0.11\linewidth}
    \includegraphics[width=\linewidth]{Figures--JPEG_profiles--P2P1-P2P3.pdf}
    \end{minipage} \\
    
    \emph{Soldier}&
    \begin{minipage}[b]{0.11\linewidth}
    \includegraphics[width=\linewidth]{Figures--JPEG_profiles--Equal.pdf}
    \end{minipage} &
    \begin{minipage}[b]{0.11\linewidth}
    \includegraphics[width=\linewidth]{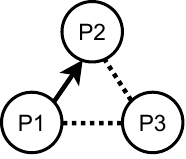}
    \end{minipage} &
    \begin{minipage}[b]{0.11\linewidth}
    \includegraphics[width=\linewidth]{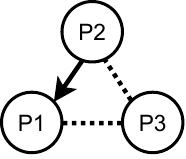}
    \end{minipage} &
    \begin{minipage}[b]{0.11\linewidth}
    \includegraphics[width=\linewidth]{Figures--JPEG_profiles--P2P1-P2P3.pdf}
    \end{minipage} \\
    
    \emph{Thaidancer}&
    \begin{minipage}[b]{0.11\linewidth}
    \includegraphics[width=\linewidth]{Figures--JPEG_profiles--P2P1-P2P3.pdf}
    \end{minipage} &
    \begin{minipage}[b]{0.11\linewidth}
    \includegraphics[width=\linewidth]{Figures--JPEG_profiles--P2P1.pdf}
    \end{minipage} &
    \begin{minipage}[b]{0.11\linewidth}
    \includegraphics[width=\linewidth]{Figures--JPEG_profiles--P1P2.pdf}
    \end{minipage} &
    \begin{minipage}[b]{0.11\linewidth}
    \includegraphics[width=\linewidth]{Figures--JPEG_profiles--Equal.pdf}
    \end{minipage} \\

    \emph{Boxer}&
    \begin{minipage}[b]{0.11\linewidth}
    \includegraphics[width=\linewidth]{Figures--JPEG_profiles--P1P2.pdf}
    \end{minipage} &
    \begin{minipage}[b]{0.11\linewidth}
    \includegraphics[width=\linewidth]{Figures--JPEG_profiles--P2P1.pdf}
    \end{minipage} &
    \begin{minipage}[b]{0.11\linewidth}
    \includegraphics[width=\linewidth]{Figures--JPEG_profiles--P1P2.pdf}
    \end{minipage} &
    \begin{minipage}[b]{0.11\linewidth}
    \includegraphics[width=\linewidth]{Figures--JPEG_profiles--Equal.pdf}
    \end{minipage} \\

    \emph{House\_without\_roof}&
    \begin{minipage}[b]{0.11\linewidth}
    \includegraphics[width=\linewidth]{Figures--JPEG_profiles--Equal.pdf}
    \end{minipage} &
    \begin{minipage}[b]{0.11\linewidth}
    \includegraphics[width=\linewidth]{Figures--JPEG_profiles--P1P2-P3P2.pdf}
    \end{minipage} &
    \begin{minipage}[b]{0.11\linewidth}
    \includegraphics[width=\linewidth]{Figures--JPEG_profiles--Equal.pdf}
    \end{minipage} &
    \begin{minipage}[b]{0.11\linewidth}
    \includegraphics[width=\linewidth]{Figures--JPEG_profiles--P2P1-P2P3.pdf}
    \end{minipage} \\

    \end{tabular}
    
    \caption{\textcolor{black}{Relationship between rate allocation strategies for each point cloud and rate point defined for JPEG Pleno compression. When they appear, arrows denote that the strategy corresponding to their source has \textit{strictly better} quality than the strategy at the end of the error. When the dotted line is used, there is a real \textit{trade-off} between geometry and color quality and comparing the two strategies.}}
    \label{fig:allocation_relationship}
\end{figure*}

Table \ref{tab:jpeg_params} includes the configuration parameters employed for all compressed point clouds. 
The scarcity of available configurations also meant that the bitrate variation between different configurations at the same rate point was higher than what could be achieved with the MPEG standards. 
Moreover, in many cases, two configurations for the same target rate using different rate allocation strategies would have the same geometry quality but different color quality, or vice-versa. 
For instance, the compression of \emph{Bouquet} at R1 and P2 used $\lambda=0.025$, SF$=2$, and CRI$=0$ resulting in a bitrate of 0.08, while for the same target rate at P3, only the parameter $\lambda=0.01$ was changed, with a bitrate of 0.12. 
In this case, the geometry quality for P3 is higher than for P2, but the parameter controlling the color quality is the same. 
Therefore, the subjective quality of P3 could not be lower than the quality of P2 in reasonable circumstances, and for that reason, any comparison between them has to take into account their difference in bitrate as well. 
\textcolor{black}{In this example, P3 is defined as being \emph{strictly better} than P2. 
In the remaining cases, there is a \emph{trade-off} between color and geometry quality, such as when that same point cloud compressed at P2 is compared to P1. 
While the geometry of the latter was compressed at $\lambda=0.05$, its color was compressed with a higher quality parameter of CRI$=1$. 
For each point cloud and rate, the relationship between different rate allocation strategies is illustrated in Figure \ref{fig:allocation_relationship}, where a dotted connection indicates there is a \emph{trade-off} between geometry and color and solid arrows indicate that the source of the arrow is \emph{strictly better} when compared to the other.}



\textcolor{black}{Since the rate allocation strategies P1, P2, and P3 have different meanings depending on the codec for which they were defined, a summary of their descriptions can be found in Table \ref{tab:all_strategies} as a quick reference. }

\begin{table*}

\centering

\resizebox{\textwidth}{!}{
\begin{tabular}{c c c c}

Codec &  P1 & P2 & P3 \\ 
\toprule
G-PCC & CTC & Higher importance to color* & Even higher importance to color \\
V-PCC & CTC &  Higher importance to geometry & Higher importance to occupancy \\
JPEG Pleno & $p_g < 0.4$ &  $0.4 < p_g < 0.6$ & $p_g > 0.6$

\end{tabular}
}
\caption{\textcolor{black}{Short description of rate allocation strategies for each codec. For G-PCC and V-PCC, P2 and P3 are defined using P1, i.e. the CTC configurations, as the baseline. For JPEG Pleno, $p_g$ refers to the percentage of the bitstream allowed for the geometry.
*For the configuration P2 of G-PCC, a larger portion of the bitstream is assigned to the color attributes when compared to P1 for all rates except for R4.}}
\label{tab:all_strategies}

\end{table*}

\section{Objective evaluation}

\begin{figure}
    \centering
    \subfloat[\emph{Bouquet}]{
    \begin{minipage}[b]{0.45\linewidth}
    \centering
    \includegraphics[width=\linewidth]{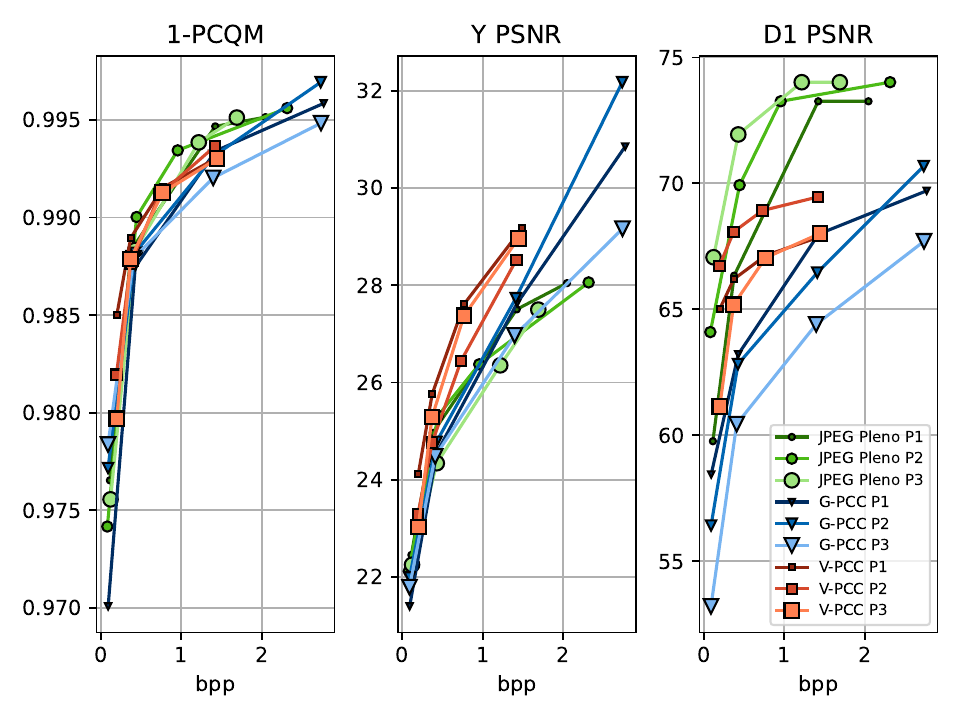}
    \end{minipage}}
    \subfloat[\emph{StMichael}]{
    \begin{minipage}[b]{0.45\linewidth}
    \centering
    \includegraphics[width=\linewidth]{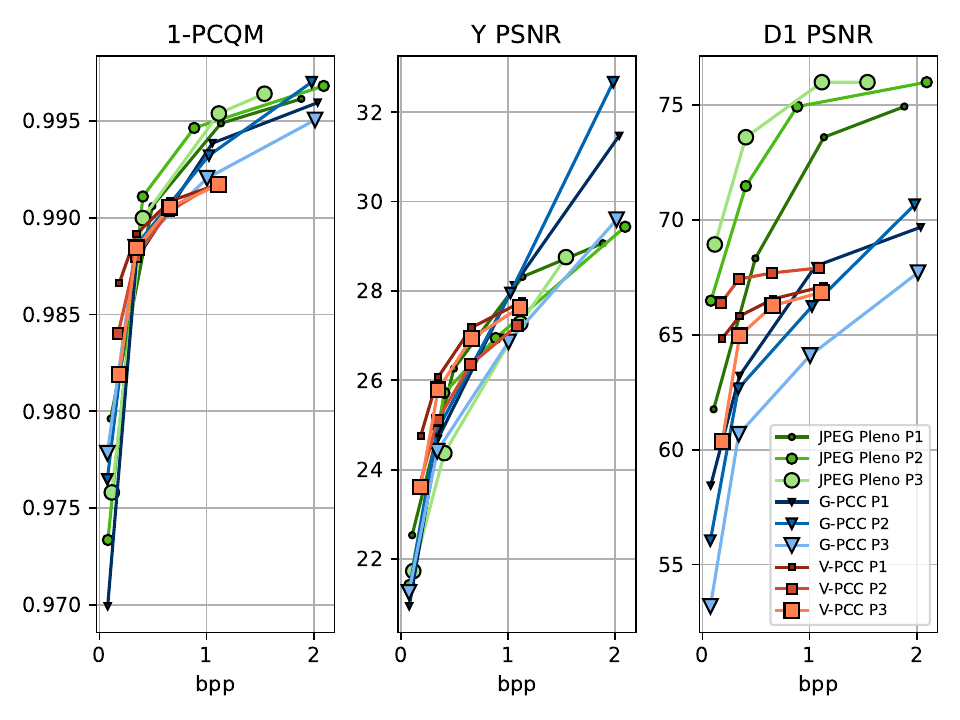}
    \end{minipage}}
    
    \subfloat[\emph{Soldier}]{
    \begin{minipage}[b]{0.45\linewidth}
    \centering
    \includegraphics[width=\linewidth]{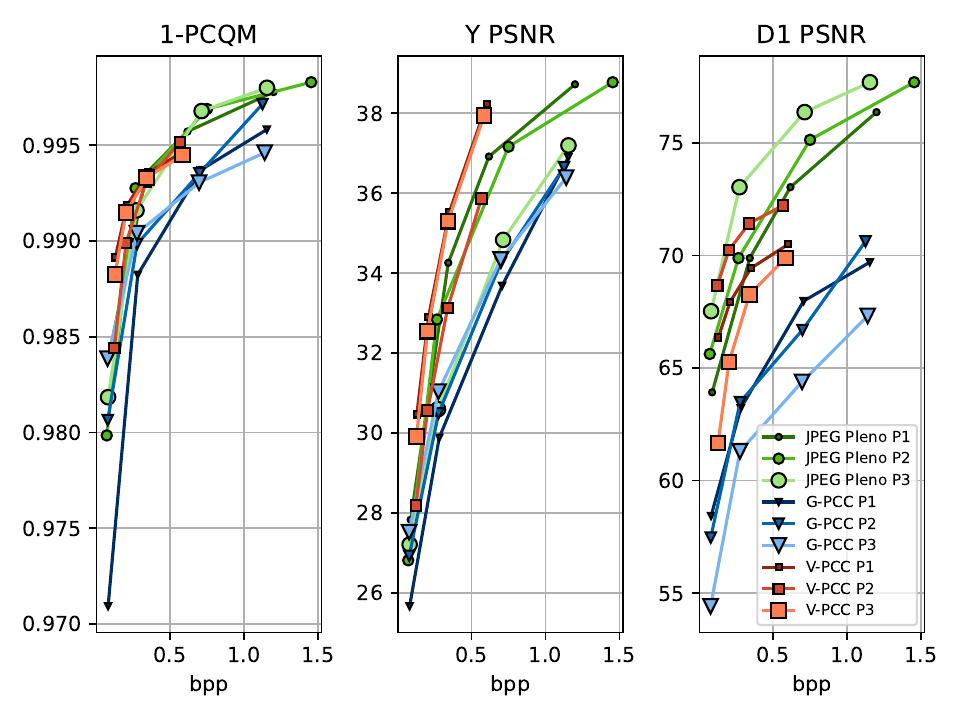}
    \end{minipage}}
    \subfloat[\emph{Thaidancer}]{
    \begin{minipage}[b]{0.45\linewidth}
    \centering
    \includegraphics[width=\linewidth]{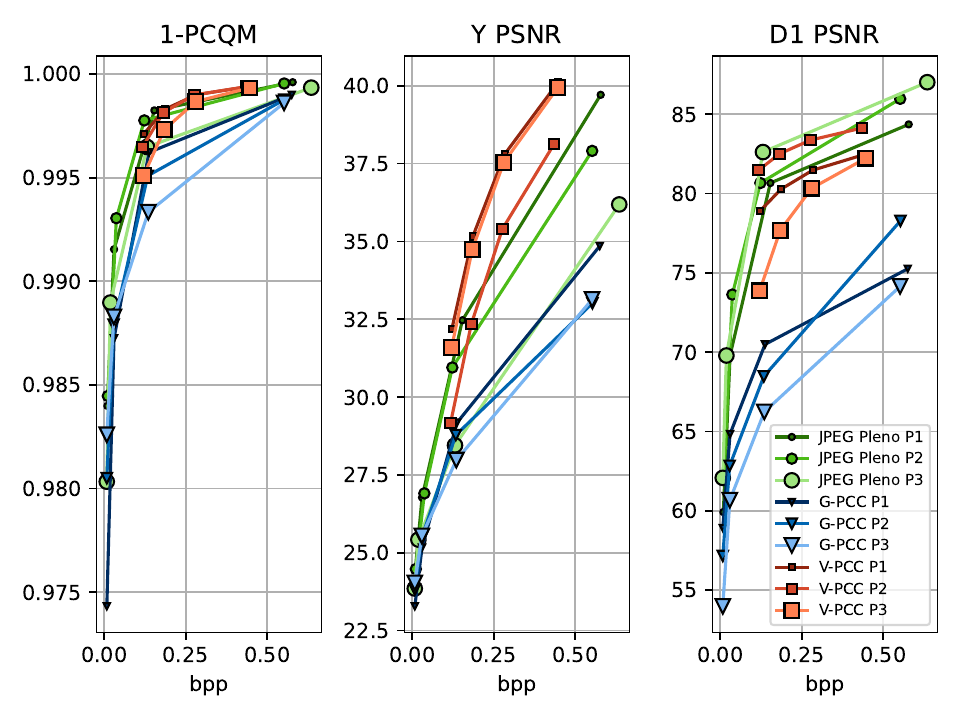}
    \end{minipage}}

    \subfloat[\emph{Boxer}]{
    \begin{minipage}[b]{0.45\linewidth}
    \centering
    \includegraphics[width=\linewidth]{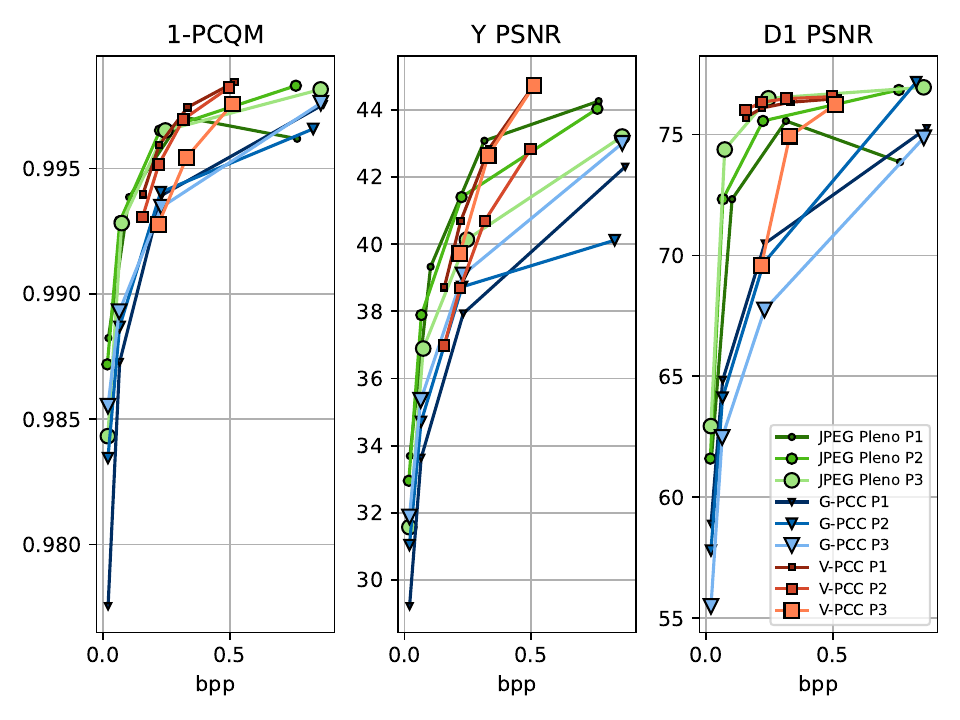}
    \end{minipage}}
    \subfloat[\emph{House\_without\_roof}]{
    \begin{minipage}[b]{0.45\linewidth}
    \centering
    \includegraphics[width=\linewidth]{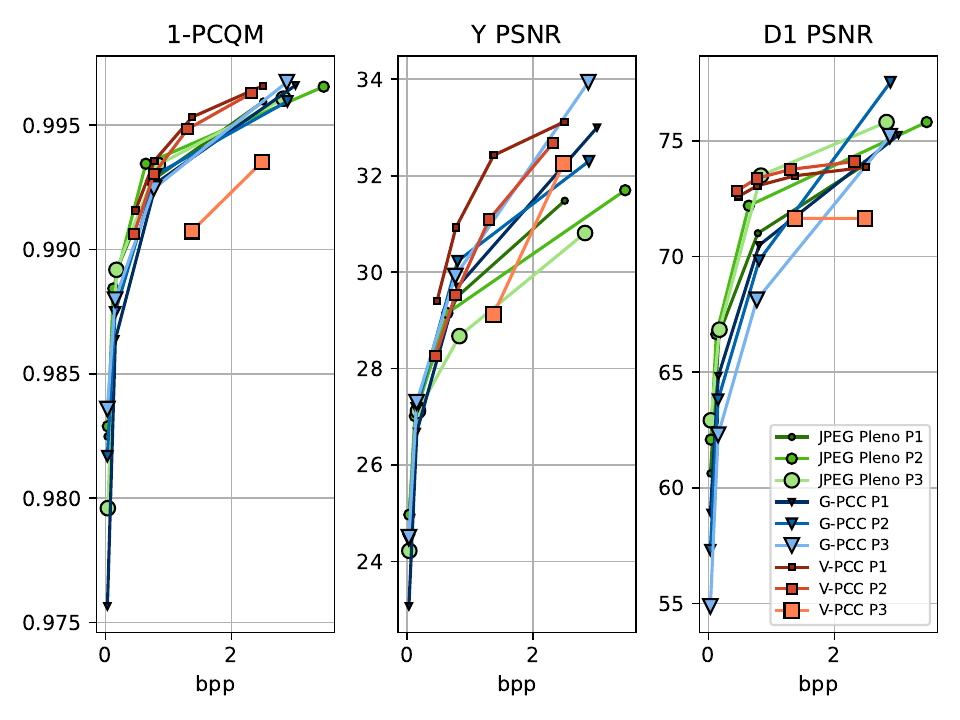}
    \end{minipage}}
    
    \caption{Rate-distortion curves for 1-PCQM, Y PSNR and D1 PSNR for each selected compressed point cloud for the subjective experiment.}
    \label{fig:selected_metrics}
\end{figure}

The distorted point clouds selected for the subjective experiment are evaluated using three objective quality metrics: D1 PSNR for the evaluation of geometry distortion, Y PSNR for the evaluation of color distortion, and PCQM for a joint evaluation of both kinds of distortion. 
The plots of these metrics against the bitrates for all point clouds and codecs can be observed in Figure \ref{fig:selected_metrics}. 

%
According to PCQM, the performance of G-PCC is slightly worse than the other codecs for \emph{Soldier}, \emph{Boxer}, and \emph{Thaidancer}. 
V-PCC, on the other hand, is able to achieve higher scores than its counterparts for the sparser models \emph{House\_without\_roof} and \emph{Boxer} at some bitrates. 
The PCQM values of both V-PCC and JPEG Pleno are similar for \emph{Soldier} and \emph{Thaidancer}, but JPEG Pleno achieves a slight advantage for \emph{Bouquet} and a higher advantage for \emph{StMichael}, in which case V-PCC stagnates at a lower quality level. 
The unusual behavior of V-PCC for \emph{StMichael} is evaluated through visual examples in the following sections. 
Regarding differences between rate allocation strategies, P1 tends to have lower performance at R1 for G-PCC, but there is no clear tendency at higher bitrates. 
For V-PCC, P1 usually achieves better PCQM values at low bitrates, with the gap being closed at high bitrates. 
The sparser point clouds \emph{Boxer} and \emph{House\_without\_roof} are the exception, for which P3 is consistently worse, probably because a larger increase in $gqp$ was needed to achieve similar bitrates as the CTC with \emph{occupancyPrecision} set to 2, as it can be observed in the isorate plots of Figure \ref{fig:vpcc_isorate_curves}. 
On the other hand, there is no rate allocation strategy that consistently outperform the others for JPEG Pleno, with P3 having worse metric values at R1 for some point clouds such as \emph{Thaidancer}, \emph{Boxer} and \emph{House\_without\_roof}, but not for the remaining ones. 

The Y PSNR scores present a different behavior, usually attributing relatively lower scores for JPEG Pleno when compared to the MPEG standards. 
For instance, G-PCC is able to achieve a large advantage for \emph{Bouquet} and \emph{StMichael} at the highest rate. 
Still, V-PCC is the best-performing codec across the entire range overall, especially for P1. 
The exceptions are \emph{Boxer}, for which JPEG Pleno achieves better metric values for a part of the range, and \emph{StMichael}, where V-PCC saturates at metric values lower than the other codecs. 
The difference between rate allocation strategies can be better observed for Y PSNR than for PCQM, with P1 achieving the highest color quality for JPEG Pleno since it assigns a larger portion of the bitstream to the representation of color attributes. 
For V-PCC, P1 usually achieves the best performance, while for G-PCC it is harder to discern any consistent pattern. 
The latter observation can also be visualized in the differences between isorate plots across point clouds as displayed in Figure \ref{fig:gpcc_isorate_curves}. 

When considering geometry-only distortion, D1 PSNR indicates that JPEG Pleno presents the best performance for the solid point clouds, while for the dense models, it is comparable with V-PCC. 
On the other hand, G-PCC trails the other codecs at low and mid-range bitrates, achieving however competitive performance at the highest rates for \emph{Boxer} and \emph{House\_without\_roof}. 
For JPEG Pleno, P3 delivers the best geometry quality, while for V-PCC, P2 outperforms the remaining rate allocation strategies due to its lower $gqp$ values. 
For G-PCC, P1 presents the best performance except for the last rate, which is expected given that P2 and P3 assign a relatively higher importance to color. 

Overall, the employed metrics usually provide different rankings between codecs, with Y PSNR indicating an advantage to V-PCC and D1 PSNR putting JPEG Pleno on the top. 
As for PCQM, the two codecs achieve similar performance, with JPEG Pleno having an advantage for \emph{Bouquet} and \emph{StMichael} and V-PCC with the upper hand for the sparser point clouds \emph{Boxer} and \emph{House\_without\_roof}.  
The latter metric has demonstrated a better correlation with subjective perception in previous studies and is therefore expected to provide a better description of the perceptual quality of the evaluated codecs. 
This assumption is evaluated in Section \ref{sec:results} through comparison against the scores obtained in the subjective experiment described in this paper.

\section{Subjective experiments}

The visual quality of the point clouds compressed with the different configurations defined in Section~\ref{sec:dataset_construction} was assessed employing two subjective visual quality assessment experiments, namely using the DSIS and PWC protocols in a controlled lab environment.

In the DSIS methodology, test subjects visualize pairs of point clouds displayed side-by-side. 
One model always consists of the original point cloud, while the other is the distorted one. 
The position of each point cloud is randomly selected and disclosed to the test subjects. 
The test subjects are asked to assess the impairment between the original and the distorted model on a 5-level discrete quality scale with values ``5 - Imperceptible'', ``4 - Perceptible, but not annoying'', ``3 - Slightly annoying'', ``2 - Annoying" and ``1 - Very annoying''. 
A hidden reference was added to the experiment for sanity check. 
In the context of this paper, this DSIS experiment was conducted following a \textit{passive} inspection modality, where the test subjects observed the rotating point clouds without the possibility of interacting with the models.

In the PWC experiment, the pairs of point clouds are displayed side-by-side, similar to DSIS. 
However, instead of a grading scale, the test subjects are requested to select the model with the highest visual quality among the two given options, both of which might contain distortions. 
To release the \textit{Mental Demand} of the test subjects, the experiment followed a relaxed forced choice format by introducing the \textit{``Not Sure"} option~\cite{jenadeleh2023relaxed}. 
The PWC experiment was conducted following an \textit{interactive} inspection modality, where the test subjects were asked to freely interact with the point clouds by exploring different views before submitting their preference. 
Due to the large number of distorted point clouds contained in the dataset, comparing each model against all the others would cause the experiment to last for an impractical time. 
For that reason, only point clouds compressed with the same codec and the same rate were compared to each other. 
For instance, the subjects would visualize side-by-side the compressed \emph{Bouquet} point cloud with rate R1 and rate allocation strategies P1 and P2, but they would not compare \emph{Bouquet} compressed with rates R1 and R2. 
For this reason, the PWC experiment only allows to draw conclusions about the relative difference in quality between stimuli within the same rate and codec, without any comparison being made across rate or codecs. 
Moreover, only four point clouds were used for this experiment, namely \textit{Bouquet}, \textit{StMichael}, \textit{Soldier}, and \textit{Thaidancer}.

\subsection{Visualization Framework}

\textcolor{black}{The visualization framework produced in a previous study~\cite{lazzarotto2022impact} was used as the baseline} for the experiments. 
Notably, the framework was implemented in Unity where the \textit{Pcx package}~\footnote{https://github.com/keijiro/Pcx} was adopted for rendering the point clouds. Each point was rendered as a disk with a variable size, manually determined for each content and degradation level. Identical ground plane tonality and camera position as \textcolor{black}{the baseline framework}~\cite{lazzarotto2022impact} was adopted in this experiment, with a scaling factor depending only on the voxelization bit depth being applied to fit the point clouds into the screen. 
During the experiment, the interface presented to the test subjects all the necessary instructions in written form before proceeding with a mandatory training session.

As the DSIS experiment was conducted following a \textit{passive} inspection modality, each model was rotated 360\textdegree around their vertical axis for 12 seconds. 
The model would then stand static for one second before automatically moving to the voting interface. 
The subjects did not have the choice to replay the video, therefore ensuring that all the test subjects would have the same viewing time as recommended in \textcolor{black}{ITU-R Recommendations P.919} \cite{p919}. 
On the other hand, as the PWC experiment was conducted following a \textit{interactive} inspection modality, the framework was adapted to allow test subjects to actively interact with the models. 
In particular, they could rotate the models with the mouse to allow for inspection from any angle on the upper hemisphere around the point cloud. 
Inspection from below the model was not allowed, avoiding the visualization of acquisition artifacts present in some models. 
In order to limit the duration of the experiment, a maximum inspection time of 12 seconds was imposed, after which the subjects would be automatically directed to the voting interface. 
For both protocols, the time for the subjects to decide a rate was not restricted.

\subsection{Experiment setup}
The subjective experiments were conducted in a lab environment with controlled conditions. 
A DELL UltraSharp U3219Q monitor with 31.5 inches of diagonal size and a native resolution of 3840 x 2160 pixels was adopted for both experiments. 
The monitor was calibrated with an X-Rite i1Display Pro calibration device following the guidelines provided in ITU-R Recommendations BT.500~\cite{bt.500}, i.e. by setting a D65 white point and 120 $cd/m^2$ maximum brightness. The room lighting condition was kept low, at approximately 15 lux. The viewing distance was set proportionally to the picture height to a value of 3.2H \cite{bt.500}, corresponding to 48 cm. Prior to the experiments, the test subjects signed a mandatory consent form, and their vision was tested through a Snellen visual acuity test and Ishihara color vision test. 

A total of 20 naive test subjects participated in the DSIS experiment, where 15 test subjects identified as males and 5 as females. All the subjects demonstrated normal or corrected-to-normal visual capabilities. The average age was 23.5 years and the median age was 22.5 years, with a minimum age of 18 years and a maximum age of 36 years. To avoid fatiguing the test subjects, the experiment was organized over two consecutive days of approximately half an hour each. 
Moreover, a total of 15 test subjects participated in the PWC experiment, where 8 test subjects identified as males and 7 identified as females. All the test subjects demonstrated normal or corrected-to-normal visual capabilities. The average age was 22.4 years and the median age was 22 years, with a minimum age of 19 years and a maximum age of 27 years.

\section{Data processing}
\label{sec:data_processing}

The subjective visual quality scores collected with the DSIS protocol were first analyzed for outliers using the methodology proposed in ITU-R Recommendations BT.500~\cite{bt.500}. 
The analysis did not reveal any outliers, and therefore the scores of all 20 test subjects were used for the analysis. 
The raw scores were processed by computing the Mean Opinion Score (MOS) and 95\% confidence intervals (CI) according to a Student's t-distribution. 

For the PWC experiment, the preference probability for each stimulus in the stimuli pairs was computed by dividing the number of received votes by the total number of test subjects. 
Likewise, the proportion of \textit{``Not Sure"} votes was also computed. 
The \textit{Thurstone} Case V model was then used to compute scale values via maximum-likelihood estimation using the same procedure as \textcolor{black}{in a previous study} \cite{testolina2023jpeg}. 
For this purpose, a value of 1 was added to the score of a stimulus each time that it was preferred over its pair. 
In the case of \textit{``Not Sure"} answers, a value of 0.5 was added to the score of both stimuli. 
To avoid the \textit{zero-frequency problem} for stimuli that were never selected, all scores were initialized to a value of 0.1, corresponding to a \textit{``Not Sure"} answer weighted by 0.2. 
The statistical model produces results reported in terms of Just-Objectionable-Differences (JOD), where a difference of 1 JOD between two stimuli indicates that the stimulus with the highest score would be preferred over the other 75\% of the time~\cite{perez2019pairwise}. 

The JOD scores were shifted so that the rate allocation strategy P1 was set to 0 JOD for both G-PCC and V-PCC. 
Therefore, a positive score for P2 and P3 means that the point cloud compressed with an alternative rate allocation strategy is preferred over the CTC configurations, and vice-versa. 
For JPEG Pleno, the scores were shifted to set P2 to 0, with JOD values of P1 and P3 reflecting whether allocating a higher proportion of the bitstream to geometry or color can be favorable. 
Since the comparisons were conducted only for the same rate and the same codec, the obtained JOD values can only be compared within those conditions. 
In other words, it is not possible to obtain the difference in JOD between one point cloud compressed at a given rate and another point cloud compressed at another rate. 
The same statement holds for different codecs as well. 
Bootstrapping was conducted using a sample size of 1000 to obtain 95\% confidence intervals for each stimulus. 
Since the scores were shifted at each bootstrap iteration, the stimuli set to 0 would always have the same JOD value and therefore their CI is always equal to 0.

\section{Results and discussion}
\label{sec:results}

\subsection{DSIS Experiment}
\label{sec:dsis}

\begin{figure}
    \centering
    \begin{minipage}[b]{\linewidth}
    \centering
    \includegraphics[width=0.48\linewidth]{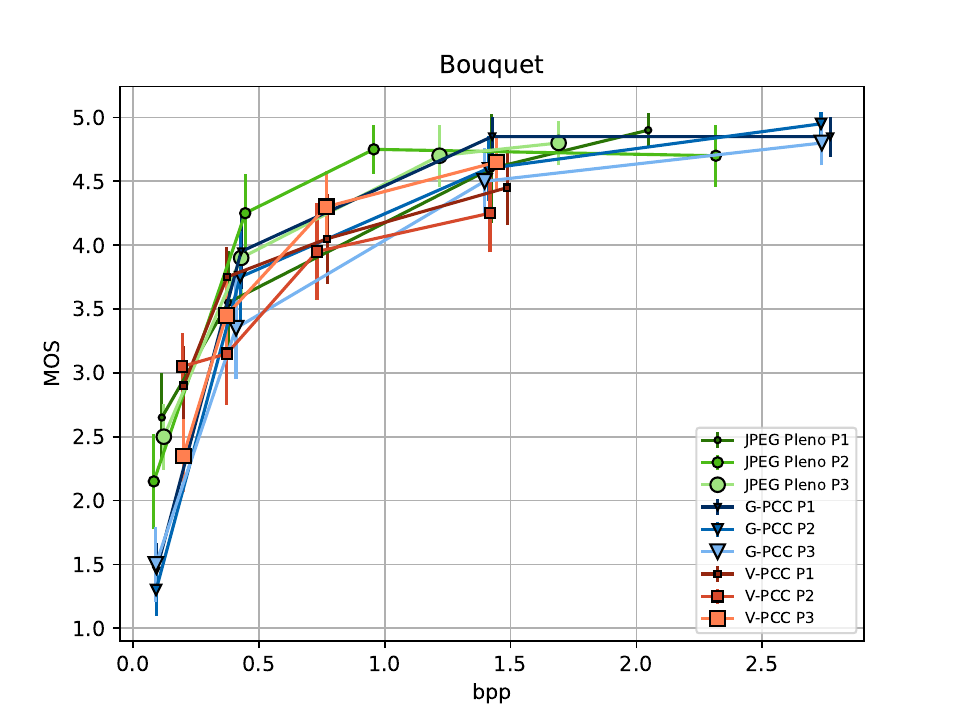}
    \includegraphics[width=0.48\linewidth]{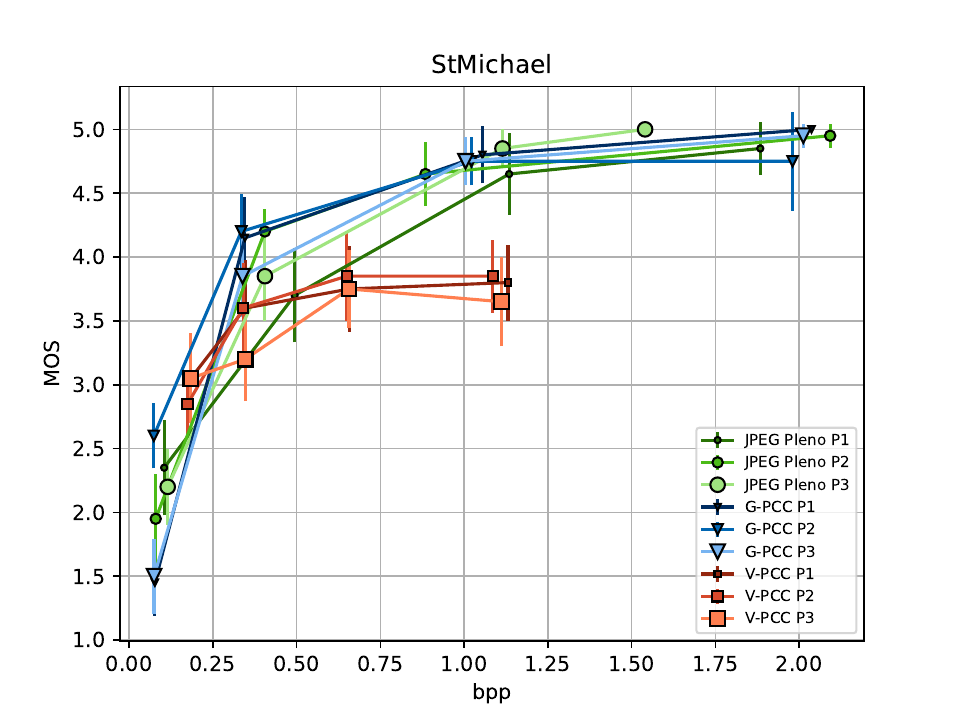}
    \end{minipage}
    
    \begin{minipage}[b]{\linewidth}
    \centering
    \includegraphics[width=0.48\linewidth]{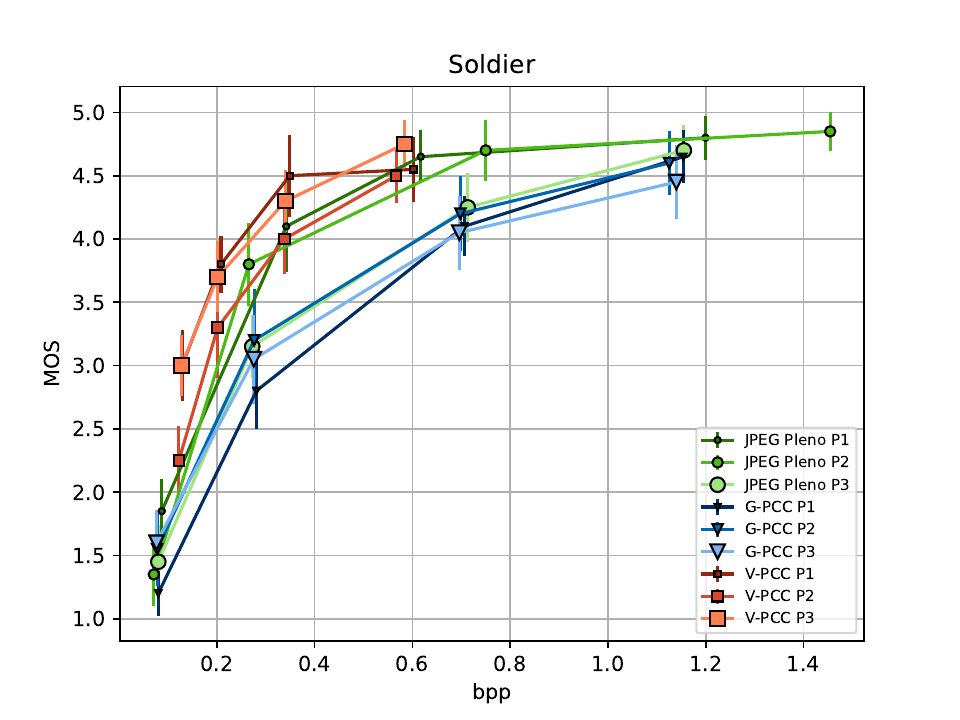}
    \includegraphics[width=0.48\linewidth]{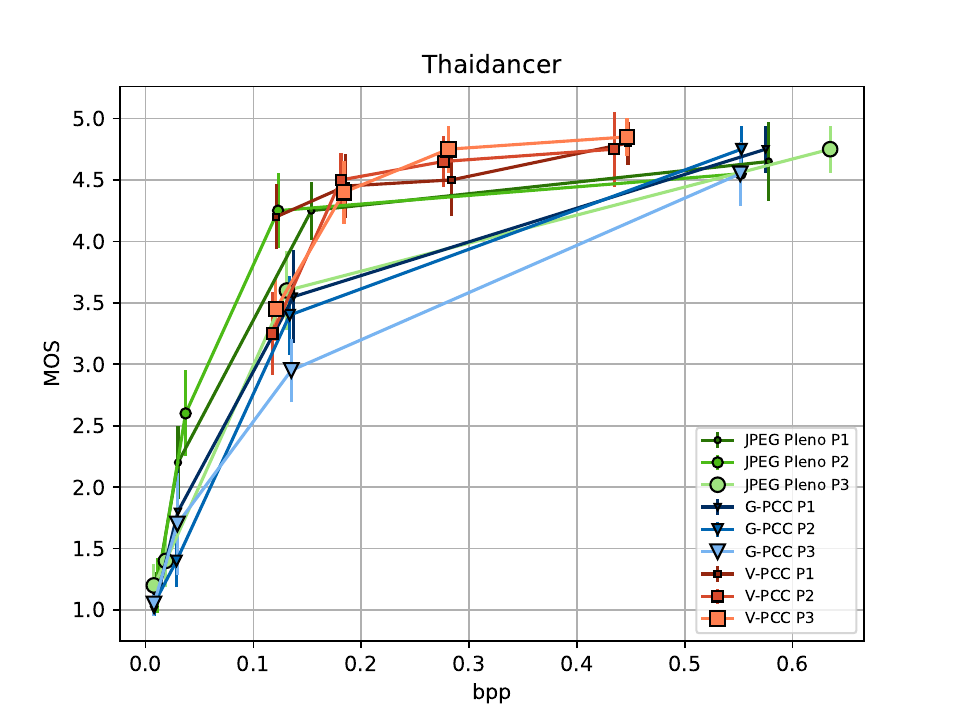}
    \end{minipage}

    \begin{minipage}[b]{\linewidth}
    \centering
    \includegraphics[width=0.48\linewidth]{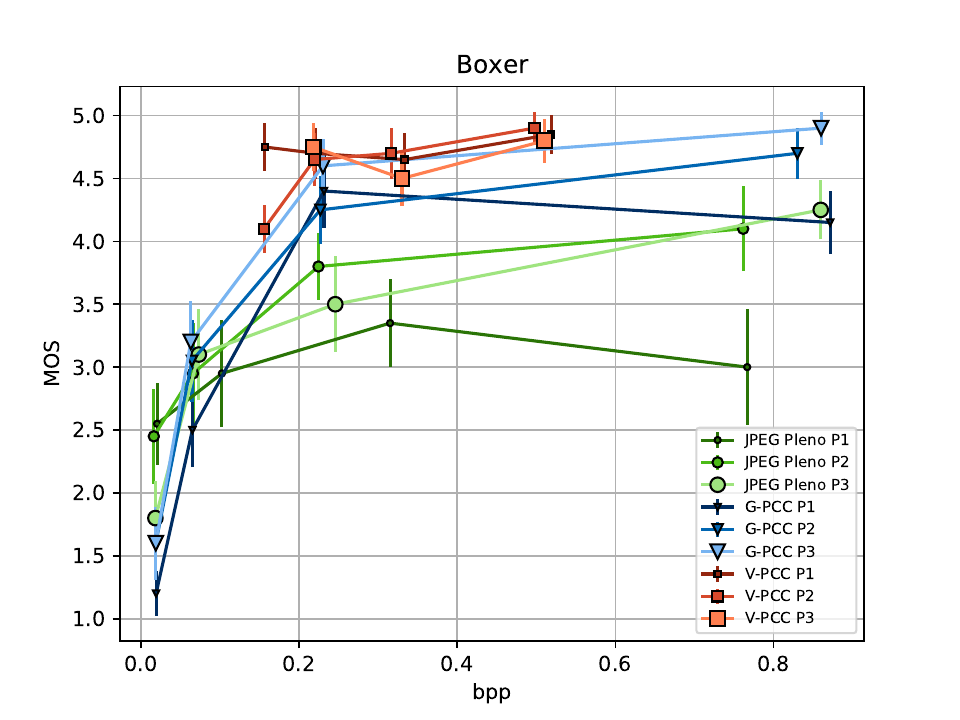}
    \includegraphics[width=0.48\linewidth]{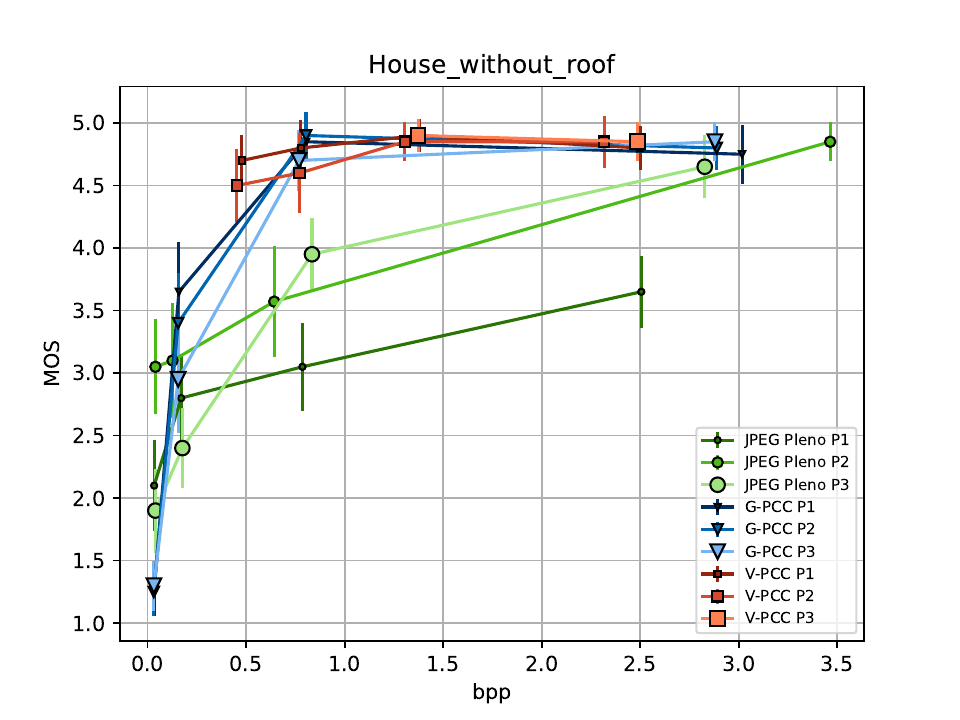}
    \end{minipage}
    
    \caption{MOS scores against bitrate for each codec, grouped per content.}
    \label{fig:dsis_results}
\end{figure}

Figure~\ref{fig:dsis_results} shows the obtained MOS and CI values obtained according to Section \ref{sec:data_processing} against the bitrate for each content. 
Several conclusions regarding the performance of the three evaluated codecs can be derived from these plots:

\begin{itemize}
    \item \textbf{Comparison between V-PCC and G-PCC:} Previous studies suggested that V-PCC has better rate-distortion performance than G-PCC in most scenarios. 
    The results of this study corroborate such conclusions for most cases but also indicate that features of the point clouds have to be taken into account when running this comparison:
    \begin{itemize}
        \item For \emph{Soldier} and \emph{Thaidancer}, V-PCC achieves MOS values superior to G-PCC at comparable bitrates. These point clouds have points uniformly sampled over a surface with high point density, indicating that these features favor V-PCC in terms of rate-distortion performance. 
        
        \item For \emph{StMichael} V-PCC saturates at MOS values lower than 4, being effectively outperformed by G-PCC. While these results may be surprising, the reason for this anomaly can easily be observed through subjective inspection in Figure \ref{fig:stMichael_vpcc}: a section of the decoded point cloud is distinctly degraded even at the highest bitrate, causing visual distortions that likely prevent most test subjects from submitting higher ratings. 
        As mentioned in Section \ref{sec:vpcc_compression}, the sequence-specific configuration parameters were not fine-tuned for each point cloud and therefore they may not be optimal for \emph{StMichael}, \textcolor{black}{resulting in poorer performance due to the set of parameters adopted in this study. 
        It is likely that another choice of parameters exists that would considerably increase the performance of the V-PCC codec for this point cloud.} 
        Regarding why the distortion is found precisely at the corner of the roof of the depicted model, it is observed in Figure \ref{fig:stMichael_vpcc_cut} that there is another layer of points behind the exterior of the roof only at that specific region. Since V-PCC is based on mapping 3D points into 2D maps through projection, the position of such interior points could be projected to the same region in the 2D maps as the exterior of the roof, somehow affecting the quality of the decoded point cloud regardless of the quality parameter used on the compression of the 2D maps themselves. 

        \begin{figure}
            \centering
            \subfloat[]{
            \begin{minipage}[b]{0.48\linewidth}
                \centering
                \includegraphics[width=0.48\linewidth]{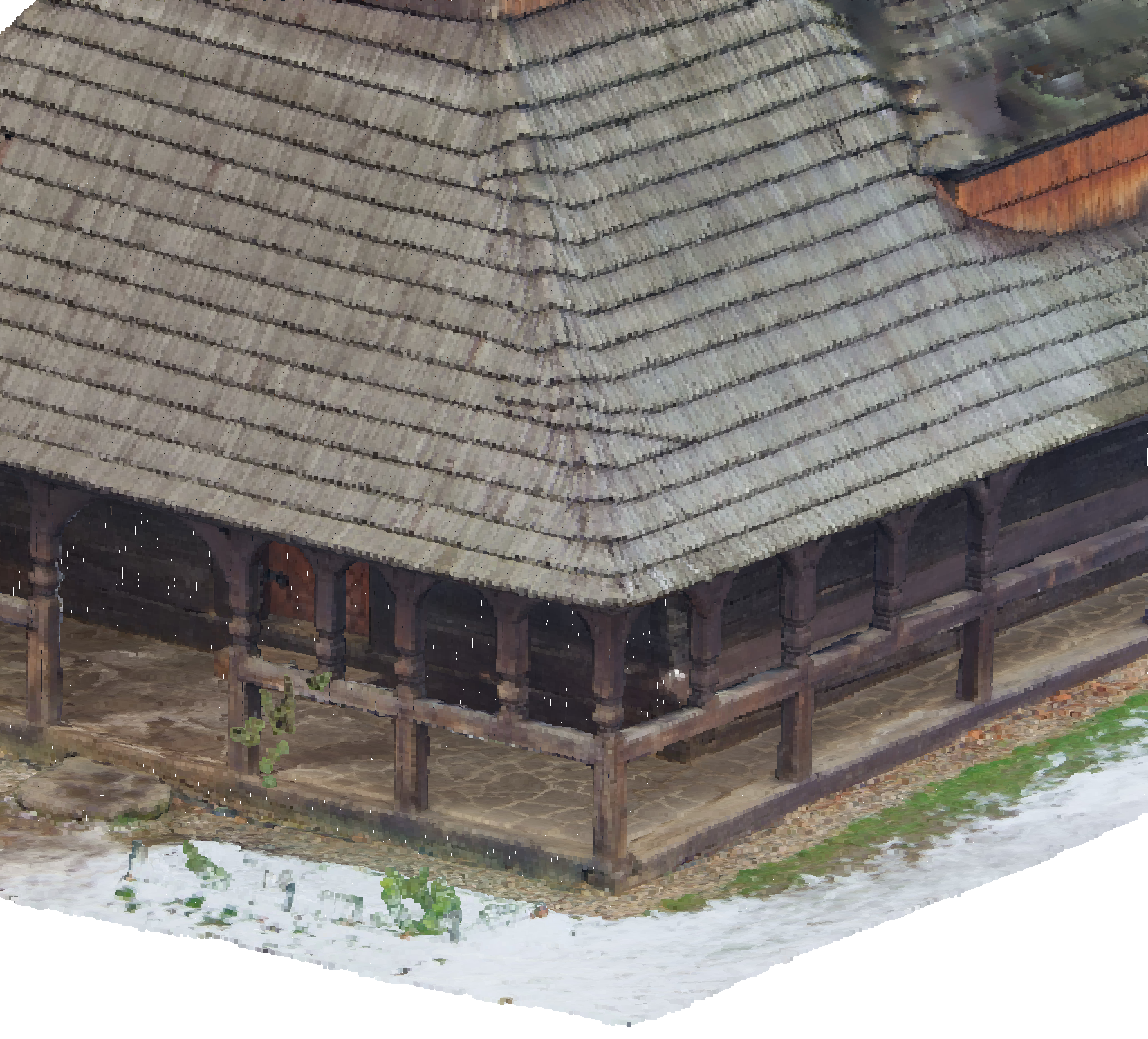}
                \includegraphics[width=0.48\linewidth]{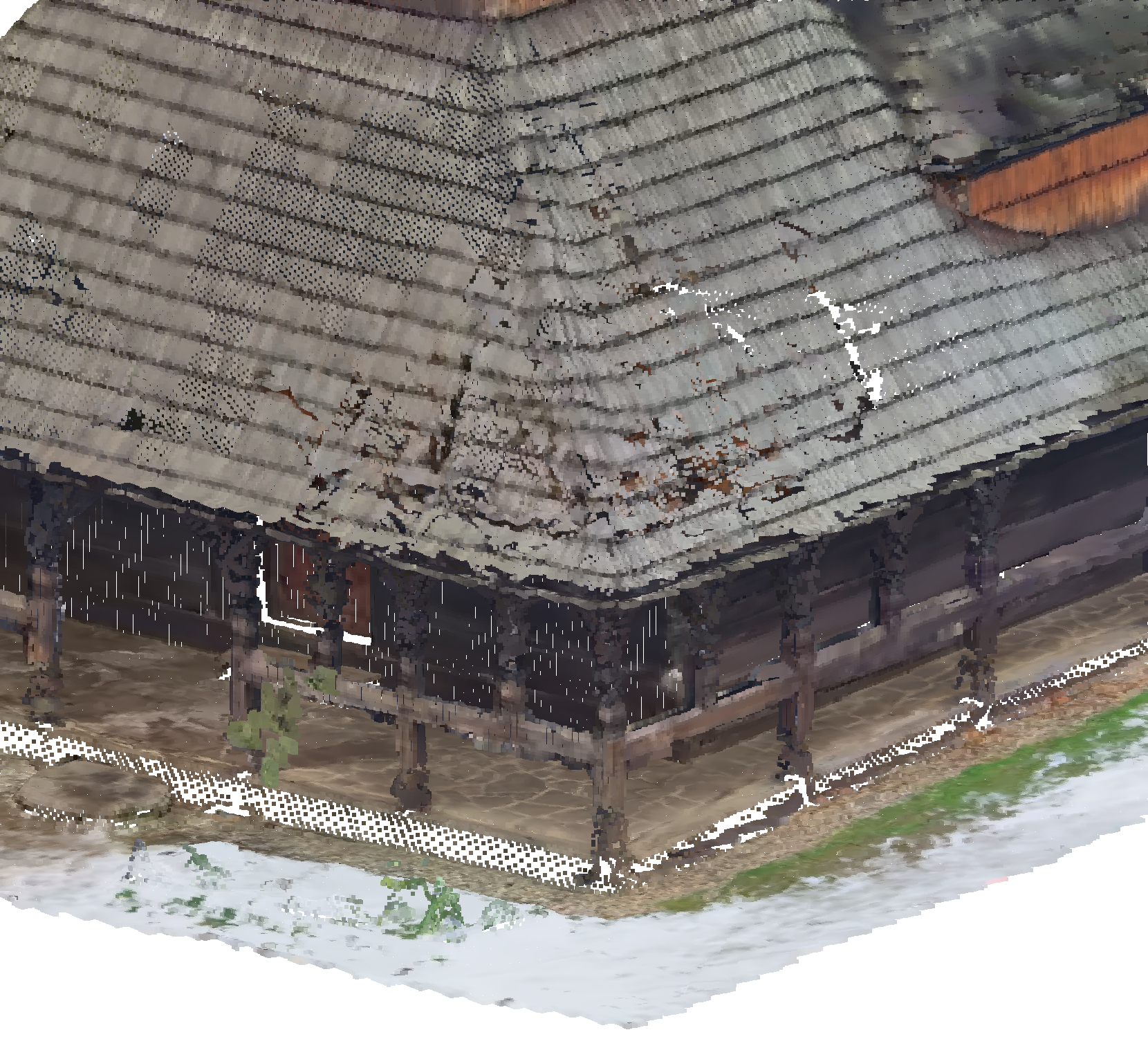}
            \end{minipage}}
            \subfloat[]{
            \begin{minipage}[b]{0.48\linewidth}
                \centering
                \includegraphics[width=0.48\linewidth]{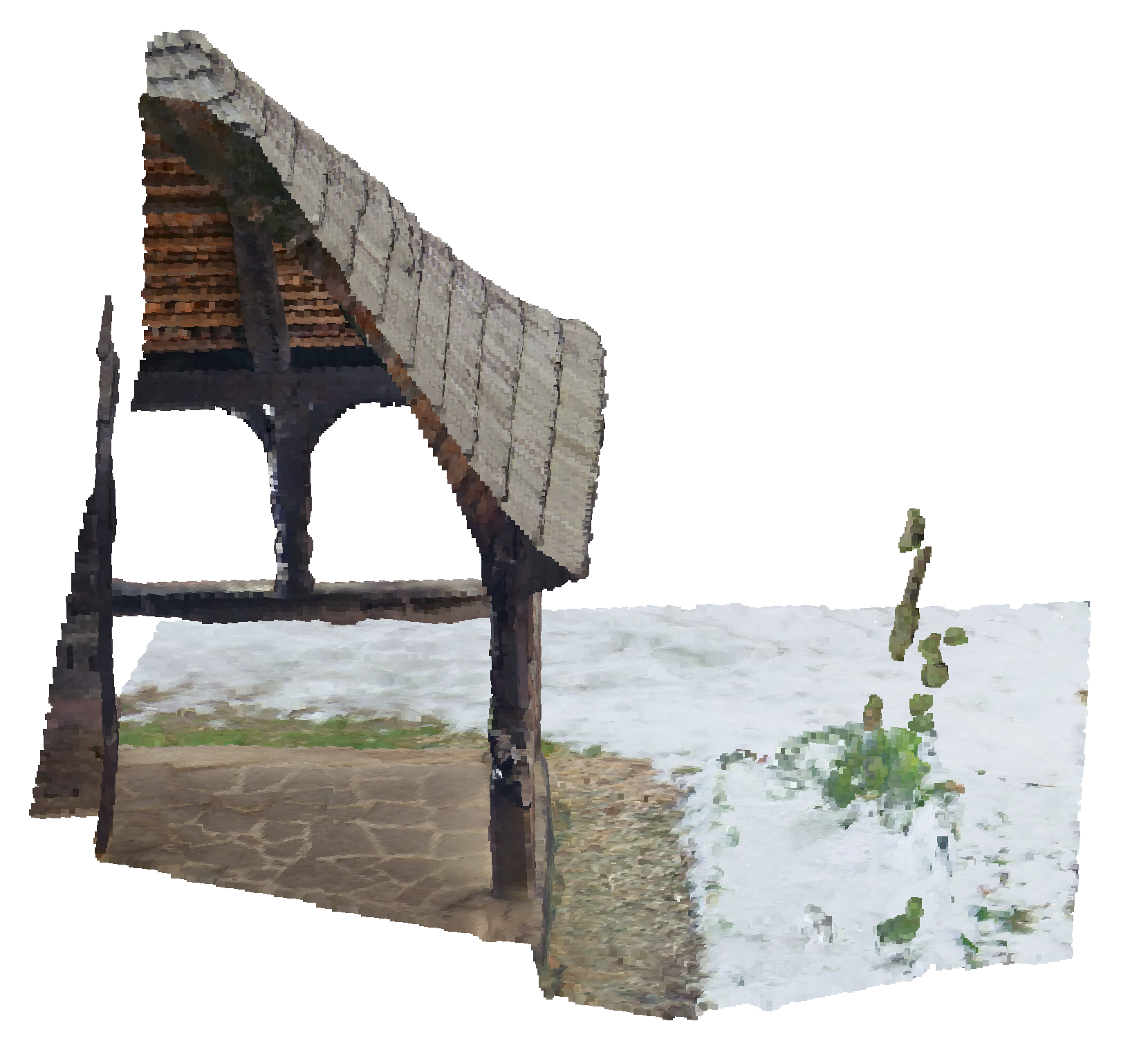}
                \label{fig:stMichael_vpcc_cut}
            \end{minipage}}
            
            \caption{(a) Region of \emph{StMichael} in original point cloud and in V-PCC decoded for P1 and rate R4. (b) Cut of the corresponding region in the original point cloud displaying the double layer of points behind the outer visible surface.}
            \label{fig:stMichael_vpcc}
        \end{figure}
        
        \item For \emph{Bouquet}, V-PCC and G-PCC achieve similar MOS values at the same bitrate range. In this case, it is again observed that V-PCC could have been affected by two surfaces that are close together similar to \emph{StMichael}. Moreover, even if this point cloud density is considered solid, it contains regions with irregular geometry that could challenge the projection mechanism of V-PCC. 

        \item For \emph{Boxer} and \emph{House\_without\_roof}, which are sparser than the remaining ones, V-PCC achieves very high quality for the entire evaluated range. This is not the case for G-PCC, which has lower MOS values for the lower rates. Interestingly, the point clouds decoded with G-PCC at higher rates have a number of points closer to the original than V-PCC. For instance, \emph{Boxer} decoded at R4 using the rate allocation strategy P1 has 12.5\% less points than the original. On the other hand, the same point cloud decoded with V-PCC at R4 has more than 240\% more points than the original for all rate allocation strategies. It is therefore observed that even if V-PCC increases drastically the number of points in sparser point clouds, this effect does not appear to have a heavy negative impact on the quality. 
    
    \end{itemize}
    Overall, this study corroborates, for most of the dataset, the conclusions reached in previous studies that V-PCC achieves higher performance than G-PCC. However, similar results were not observed for \emph{StMichael}, \textcolor{black}{which may be credited to the adopted codec configuration combined with the underlying characteristics of the point cloud.}

    \item \textbf{Comparison between JPEG Pleno and the MPEG standards:} Similarly to the previous comparison, the performance of the JPEG Pleno codec is also observed to be highly dependent on the content: 

    \begin{itemize}
        \item For \emph{Soldier} and \emph{Thaidancer}, JPEG Pleno achieves on average better performance than G-PCC, except for P3, for which the performance is lower. When compared to V-PCC, JPEG Pleno needs slightly higher bitrates to achieve the same MOS scores, and can therefore be ranked between the two MPEG standards for these point clouds.  
        \item For \emph{Bouquet}, JPEG Pleno achieves very similar performance to the MPEG standards, with a small advantage over them at the mid-range bitrates. 
        \item For \emph{StMichael}, for which the observed performance of V-PCC is lower, JPEG Pleno achieves similar performance to G-PCC, being slightly worse at mid-range bitrates, and higher performance than V-PCC. 
        \item For \emph{Boxer} and \emph{House\_without\_roof}, JPEG Pleno has consistently lower performance than the MPEG standards, especially at higher bitrates. For R4, the JPEG Pleno decoded point clouds have a number of points much higher than the original point cloud, with a difference of approximately 85\% for P2 and P3, and of more than 580\% for P1. The discrepancy between the number of points across different rate allocation strategies is due to the value used for SF, as it can be observed in Table \ref{tab:jpeg_params}. However, unlike V-PCC, the additional points are not evenly distributed across the underlying surface, with various holes being observed in the decoded point cloud, as depicted in Figure \ref{fig:boxer_jpeg}.         
    \end{itemize}

     \begin{figure}
            \centering
            \subfloat[]{
            \begin{minipage}[b]{0.24\linewidth}
            \includegraphics[width=\linewidth]{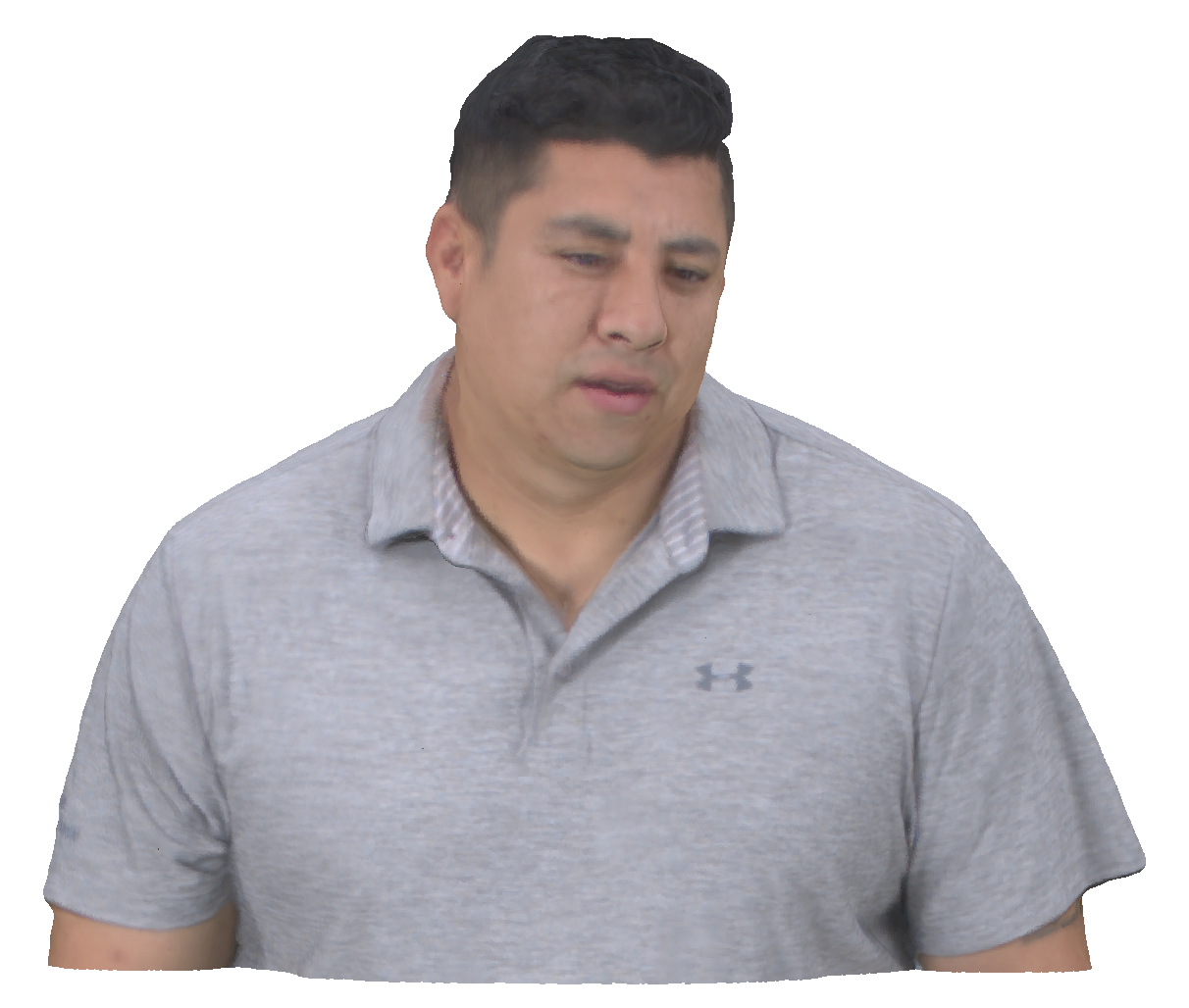}
            \end{minipage}
            }
            \subfloat[]{
            \begin{minipage}[b]{0.24\linewidth}
            \includegraphics[width=\linewidth]{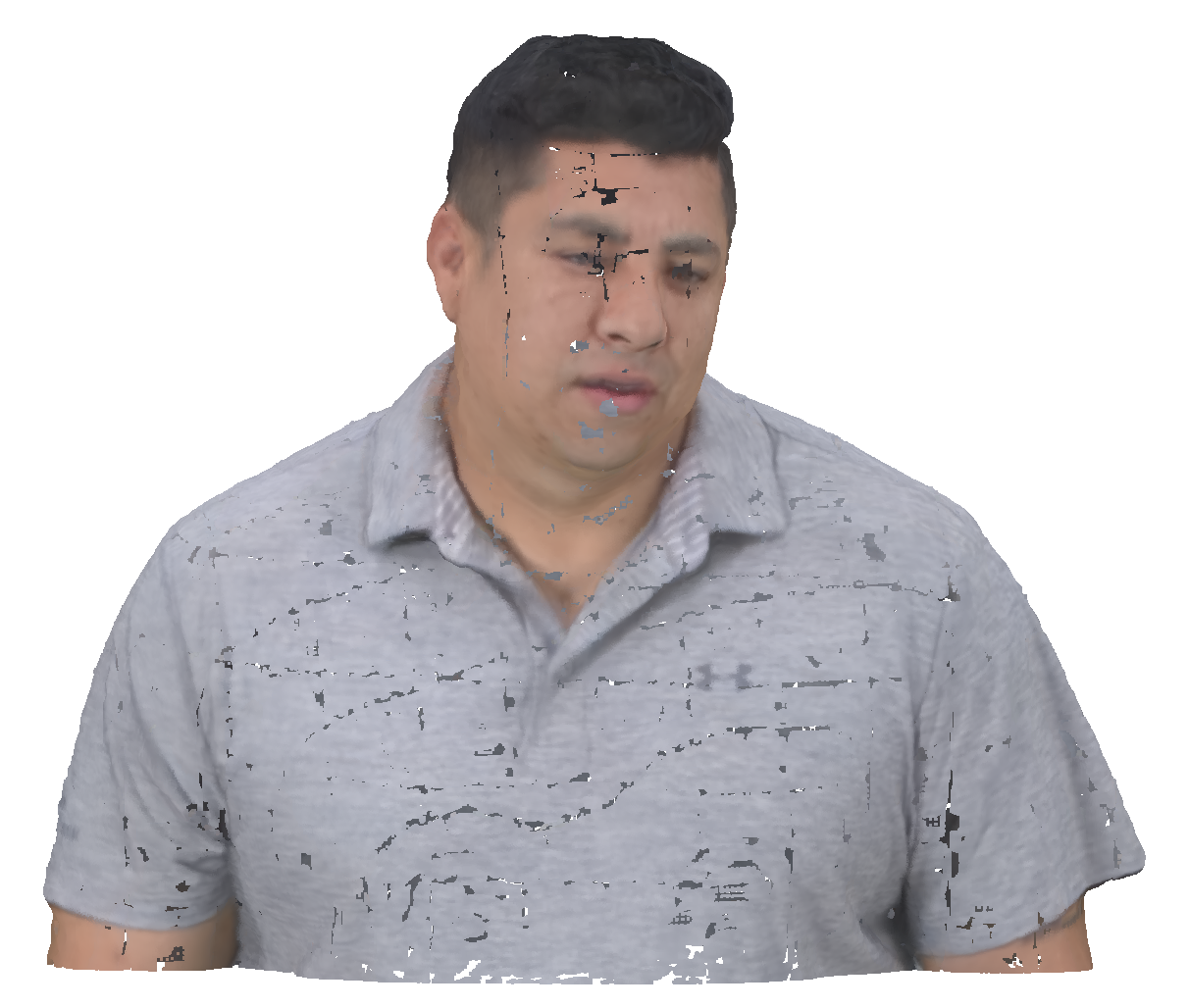}
            \end{minipage}
            }
            \subfloat[]{
            \begin{minipage}[b]{0.24\linewidth}
            \includegraphics[width=\linewidth]{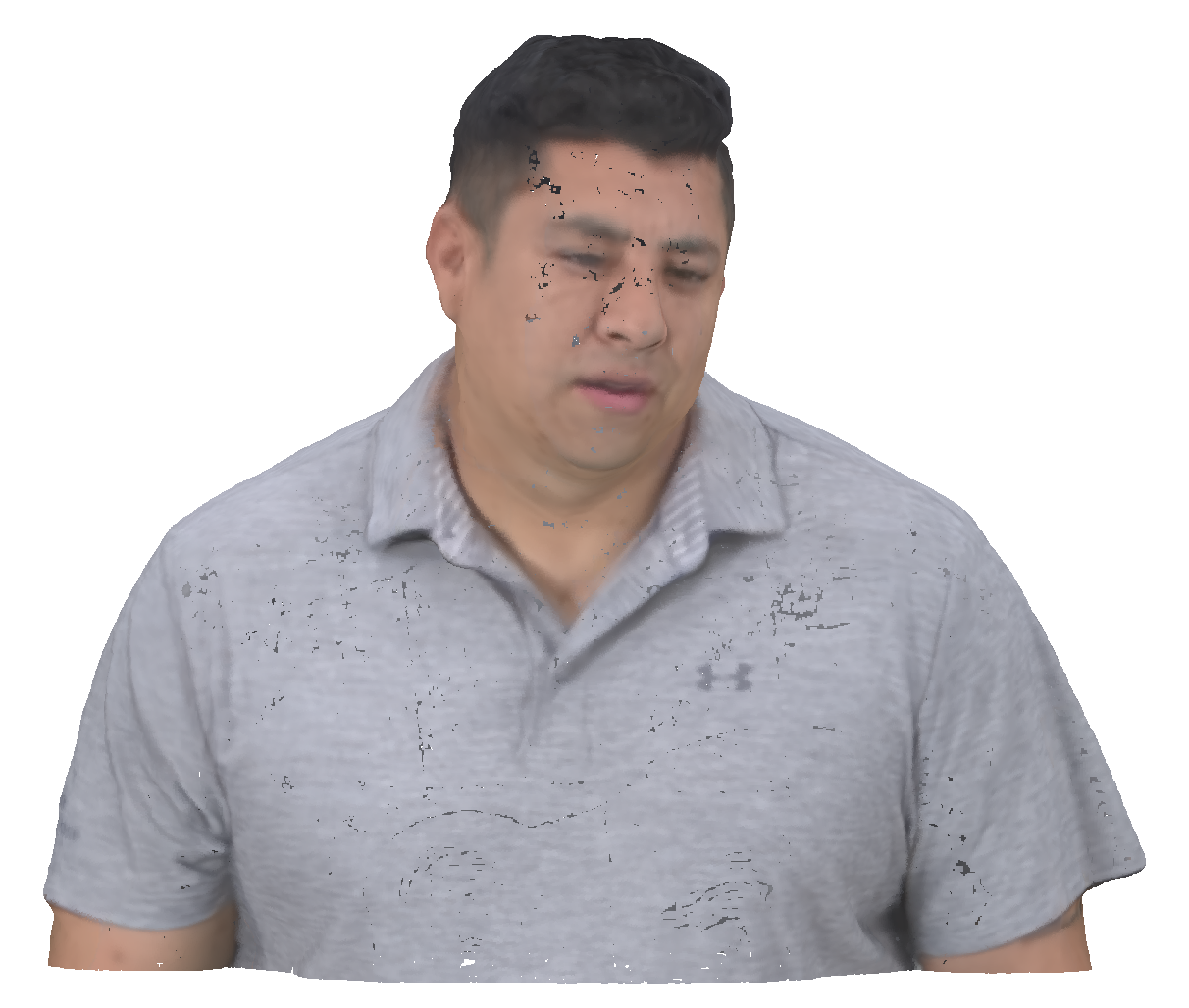}
            \end{minipage}
            }
            \subfloat[]{
            \begin{minipage}[b]{0.24\linewidth}
            \includegraphics[width=\linewidth]{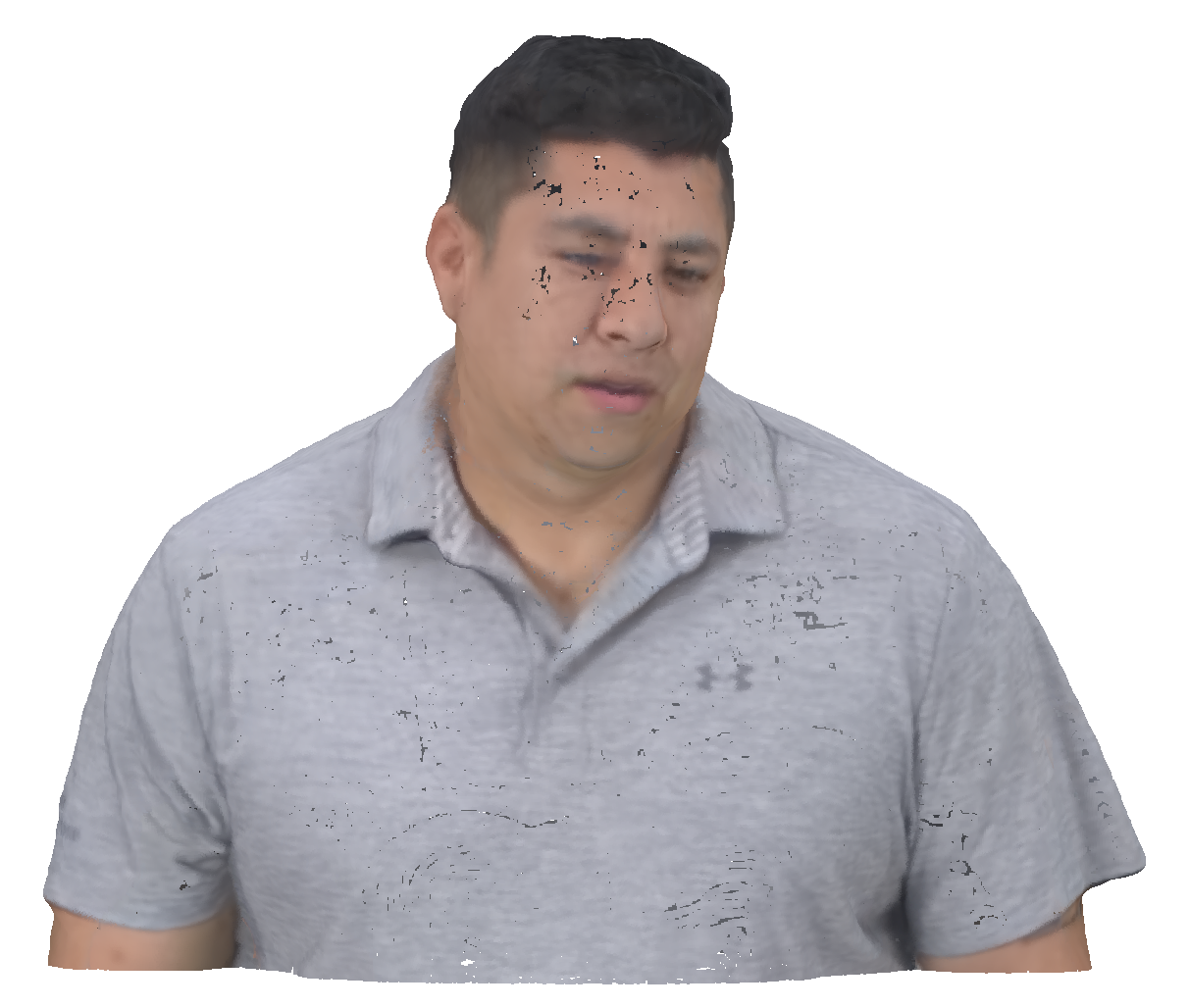}
            \end{minipage}
            }

            \caption{(a) Original \emph{Boxer} point cloud. (b, c, d) \emph{Boxer} decoded with JPEG Pleno at R4. The three images depict three rate allocations used during compression, namely P1, P2, and P3.}
            \label{fig:boxer_jpeg}
        \end{figure}

    In essence, while JPEG Pleno achieves comparable performance to the MPEG standards for the solid point clouds, the reduced point density plays a key role in reducing its performance, especially at higher bitrates for \emph{Boxer} and \emph{House\_without\_roof}, with the decoded point cloud geometry not being able to represent the original in a subjectively satisfying way. 

    \item \textbf{Comparison between different rate allocation strategies:} The difference in MOS scores between P1, P2, and P3 is small in most cases, with no rate allocation strategy consistently outperforming or underperforming the others for any codec. There are however differences in specific cases, for instance with P2 having lower overall performance than P1 and P3 for V-PCC in \emph{Soldier}, or P1 slightly trailing P2 and P3 for G-PCC in \emph{Boxer}. The larger differences are observed for JPEG Pleno, with P3 having lower performance for \emph{Soldier} and \emph{Thaidancer} and P1 underperforming for the sparser point clouds \emph{Boxer} and \emph{House\_without\_roof}. These results could suggest that a more balanced allocation may be best suited to consistently achieve better results, especially for the sparser point clouds. Since the codec struggles to faithfully reproduce sparse geometry, allocating a smaller portion of the bitrate to its representation may not be beneficial.

    \textcolor{black}{
    Although further conclusions could be derived by observing only the MOS values, the confidence intervals from many stimuli overlap.
    It is therefore not possible to affirm with a high degree of confidence the relationship between rate allocation strategies for the majority of cases. 
    Instead, a more detailed analysis is conducted using the results from the PWC in Section \ref{sec:pwc} since subjects were able to directly compare the quality of these stimuli with each other.
    Additionally, a comparison between rate allocation strategies is also described in Section \ref{sec:joint_analysis} using the results obtained in both experiments through an evaluation of the statistical significance.}

    \item \textbf{Comparison between the MOS scores and objective metrics:} When comparing the MOS scores from Figure \ref{fig:dsis_results} to the objective metrics from Figure \ref{fig:selected_metrics}, it is apparent that none of the employed metrics allow to correctly estimate subjective quality at all situations. 
    For the four solid point clouds, PCQM appears to be the best predictor, correctly ranking the performance of the codecs in most cases, but slightly overestimating the quality of \emph{StMichael} compressed with JPEG Pleno when compared to G-PCC. 
    The geometry-only metric D1 PSNR overestimated the performance JPEG Pleno even more for these models, predicting a large gap against the other codecs that is not observed in the subjective scores. 
    On the other hand, Y PSNR tends to underestimate JPEG Pleno for \emph{Bouquet} and \emph{StMichael}. 
    While better insights regarding the performance of objective quality metrics could be derived through the analysis of the correlation between subjective and objective scores, this detailed study is deferred to future works.

\end{itemize}

\subsection{PWC Experiment}
\label{sec:pwc}

\begin{figure}
    \centering
    \begin{minipage}[b]{\linewidth}
    \centering
    \includegraphics[width=0.32\linewidth]{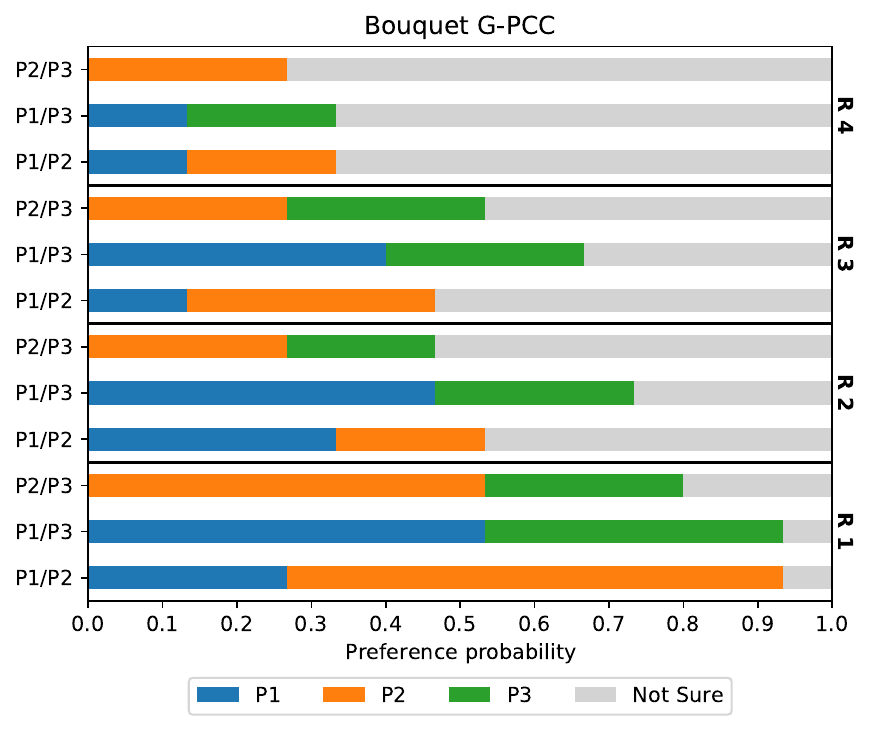}
    \includegraphics[width=0.32\linewidth]{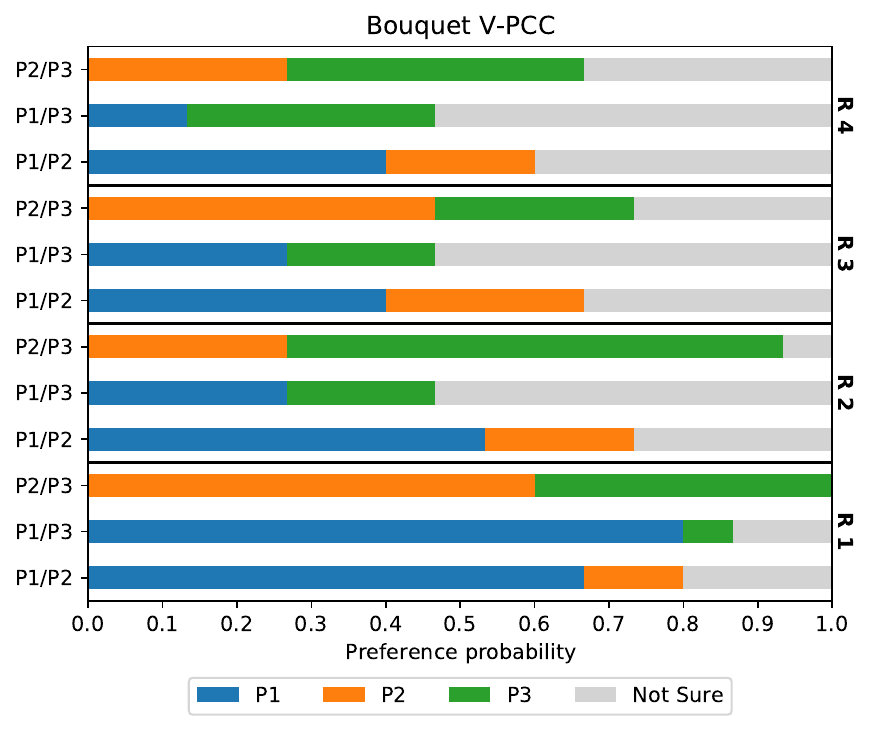}
    \includegraphics[width=0.32\linewidth]{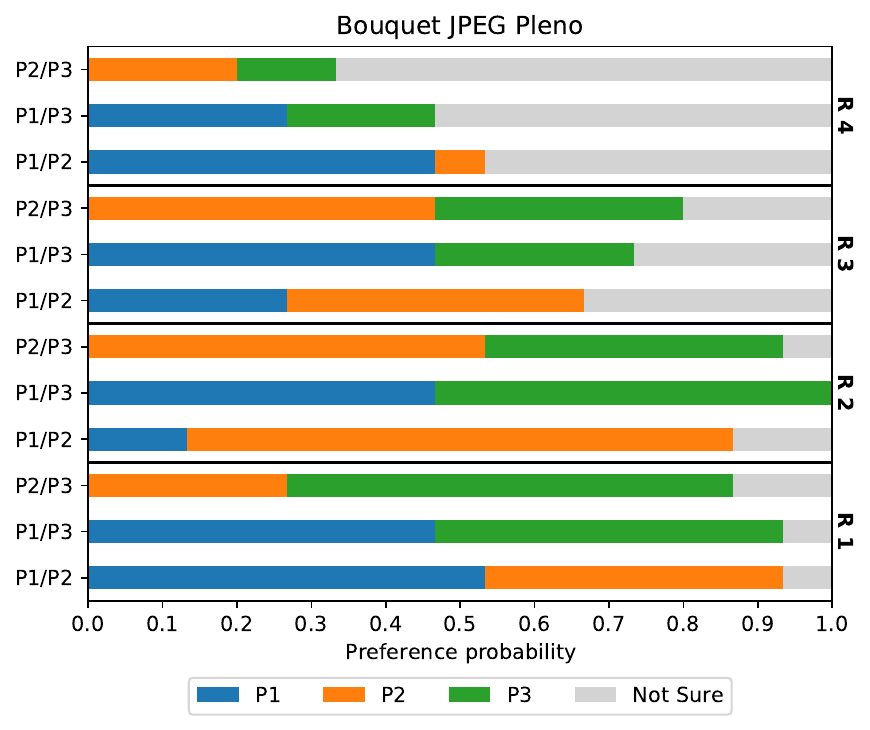}
    \end{minipage}

    \begin{minipage}[b]{\linewidth}
    \centering
    \includegraphics[width=0.32\linewidth]{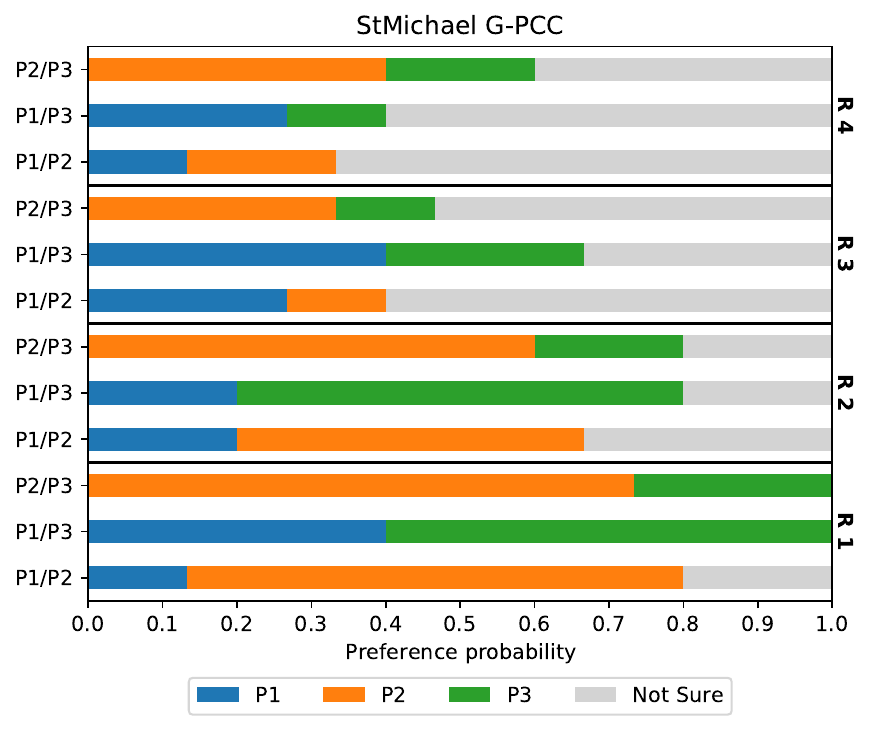}
    \includegraphics[width=0.32\linewidth]{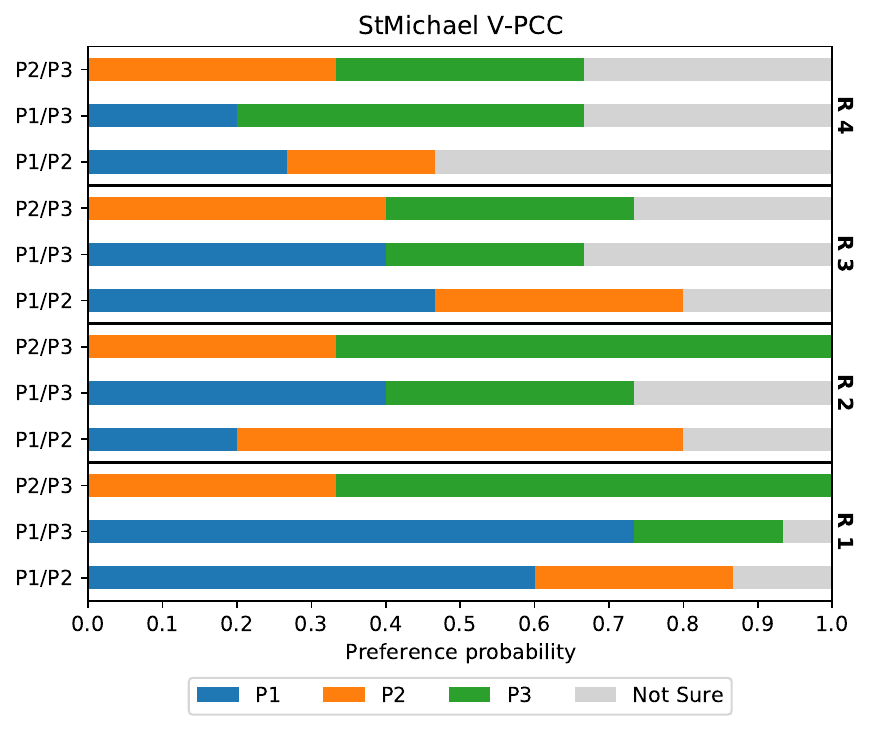}
    \includegraphics[width=0.32\linewidth]{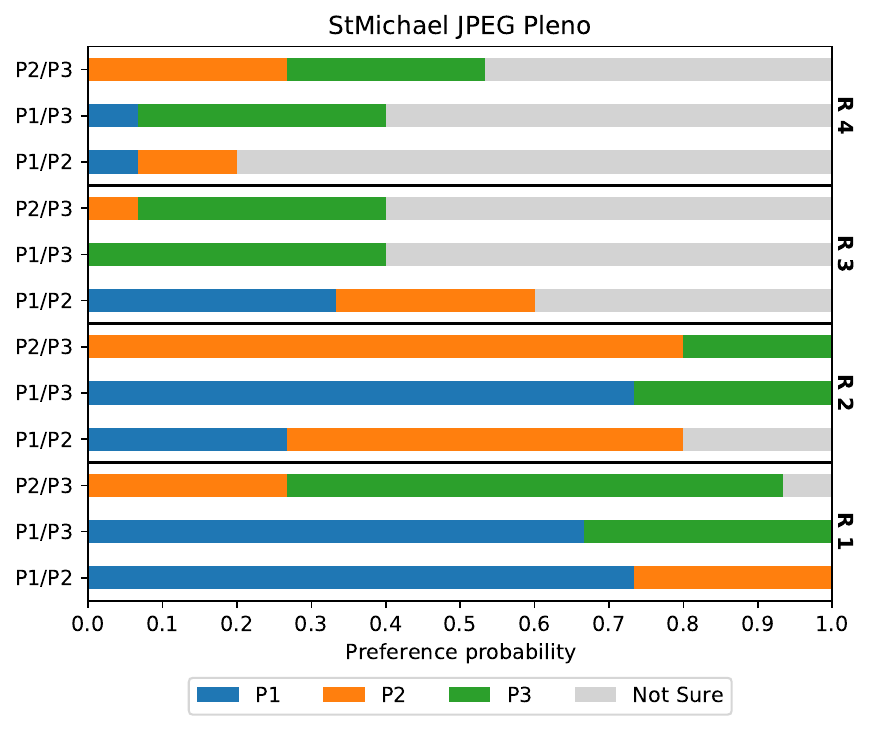}
    \end{minipage}
    
    \begin{minipage}[b]{\linewidth}
    \centering
    \includegraphics[width=0.32\linewidth]{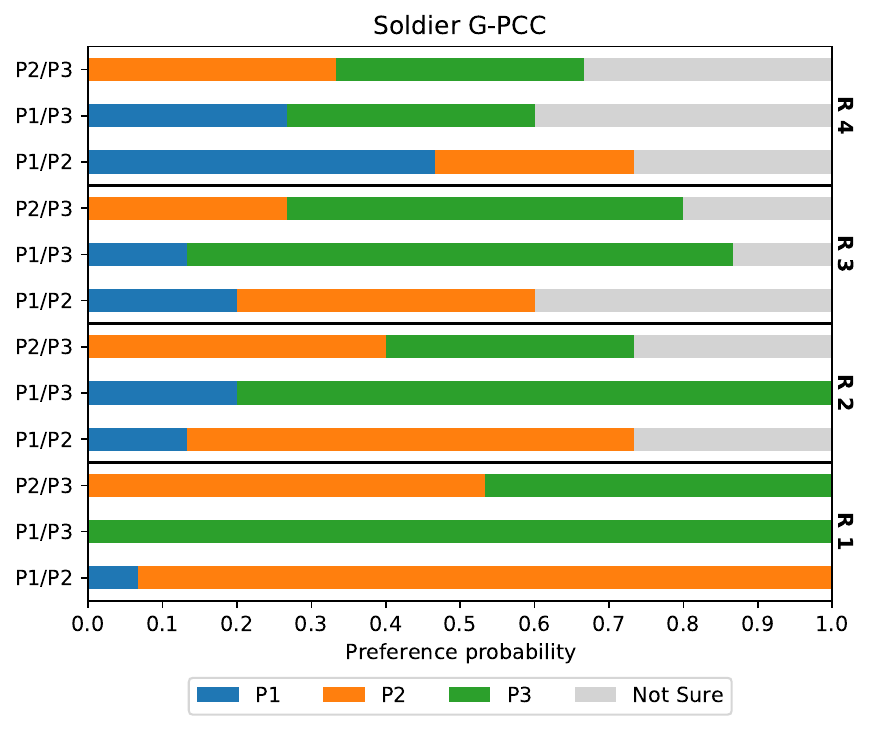}
    \includegraphics[width=0.32\linewidth]{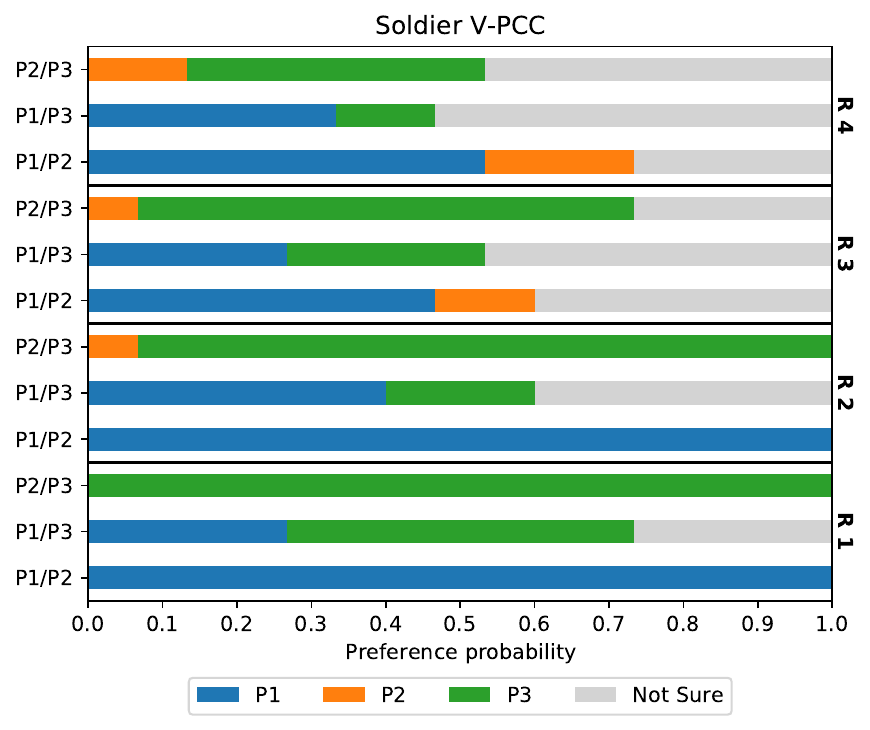}
    \includegraphics[width=0.32\linewidth]{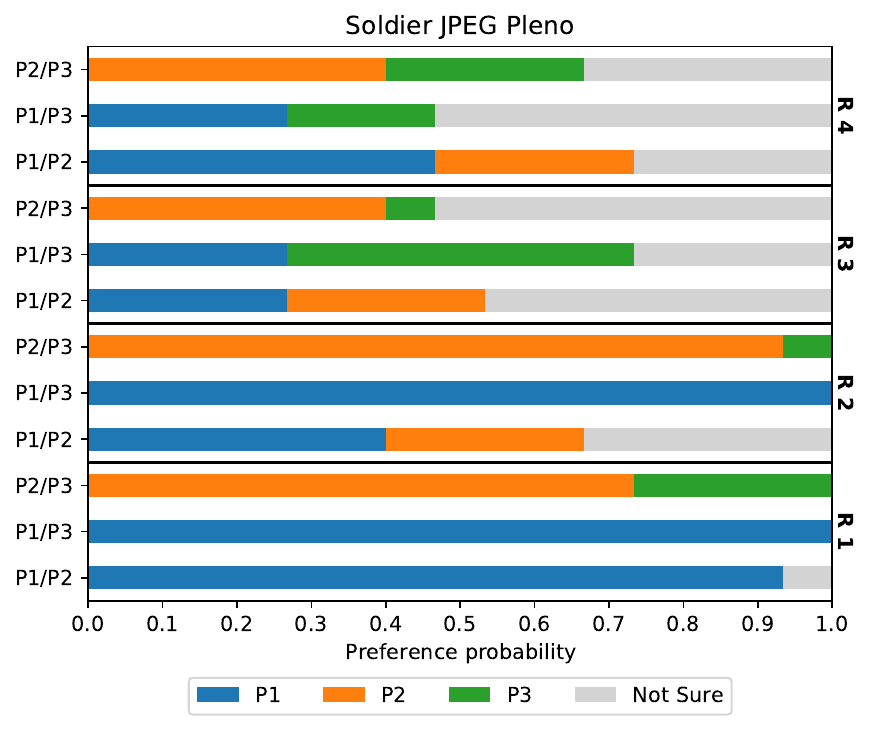}
    \end{minipage}

    \begin{minipage}[b]{\linewidth}
    \centering
    \includegraphics[width=0.32\linewidth]{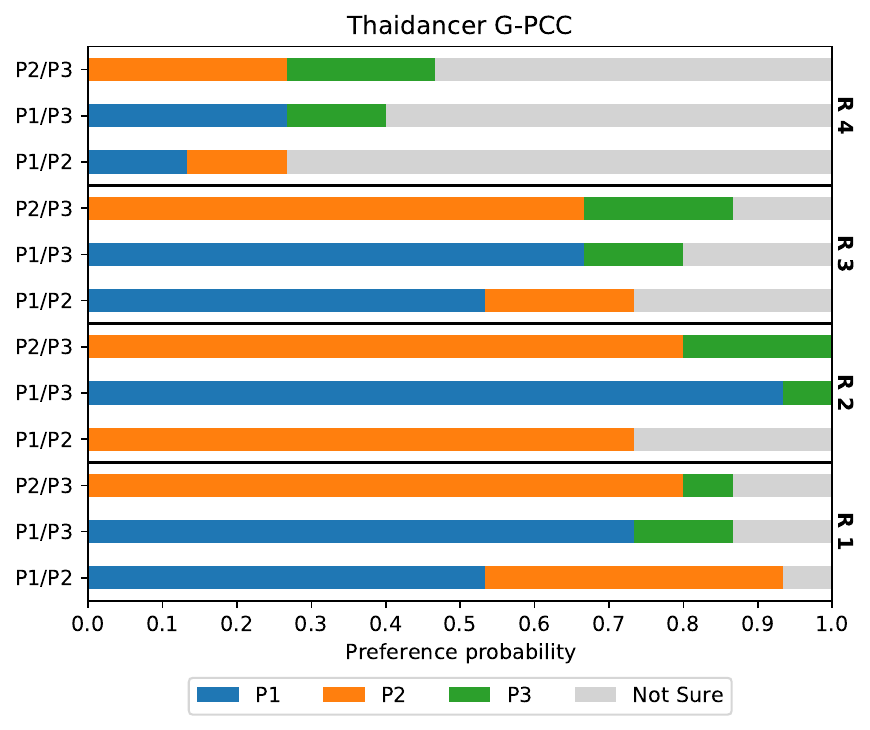}
    \includegraphics[width=0.32\linewidth]{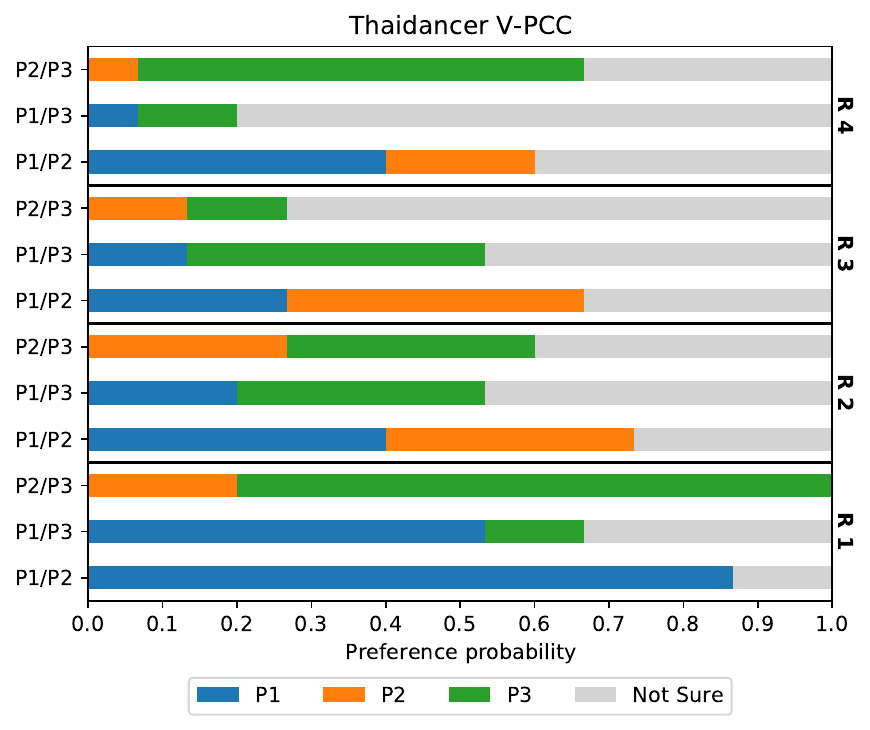}
    \includegraphics[width=0.32\linewidth]{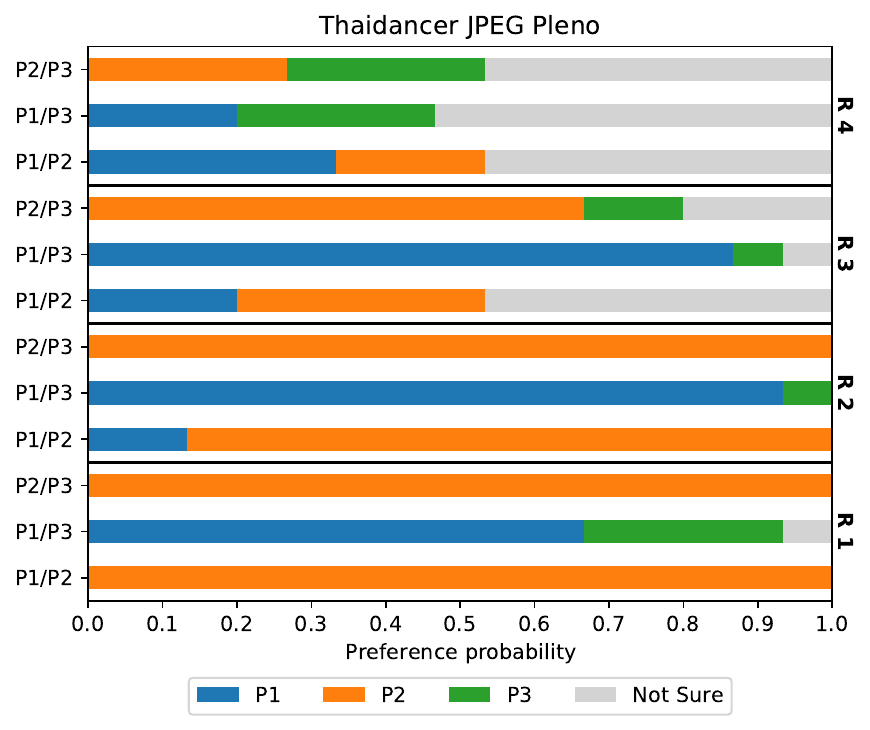}
    \end{minipage}
    
    \caption{Preference probability for each pair of configurations examined during the pariwise comparison experiment.}
    \label{fig:pwc_results}
\end{figure}

\begin{figure}
    \centering
    \subfloat[]{
    \begin{minipage}[b]{0.49\linewidth}
        \centering
        \includegraphics[width=0.32\linewidth]{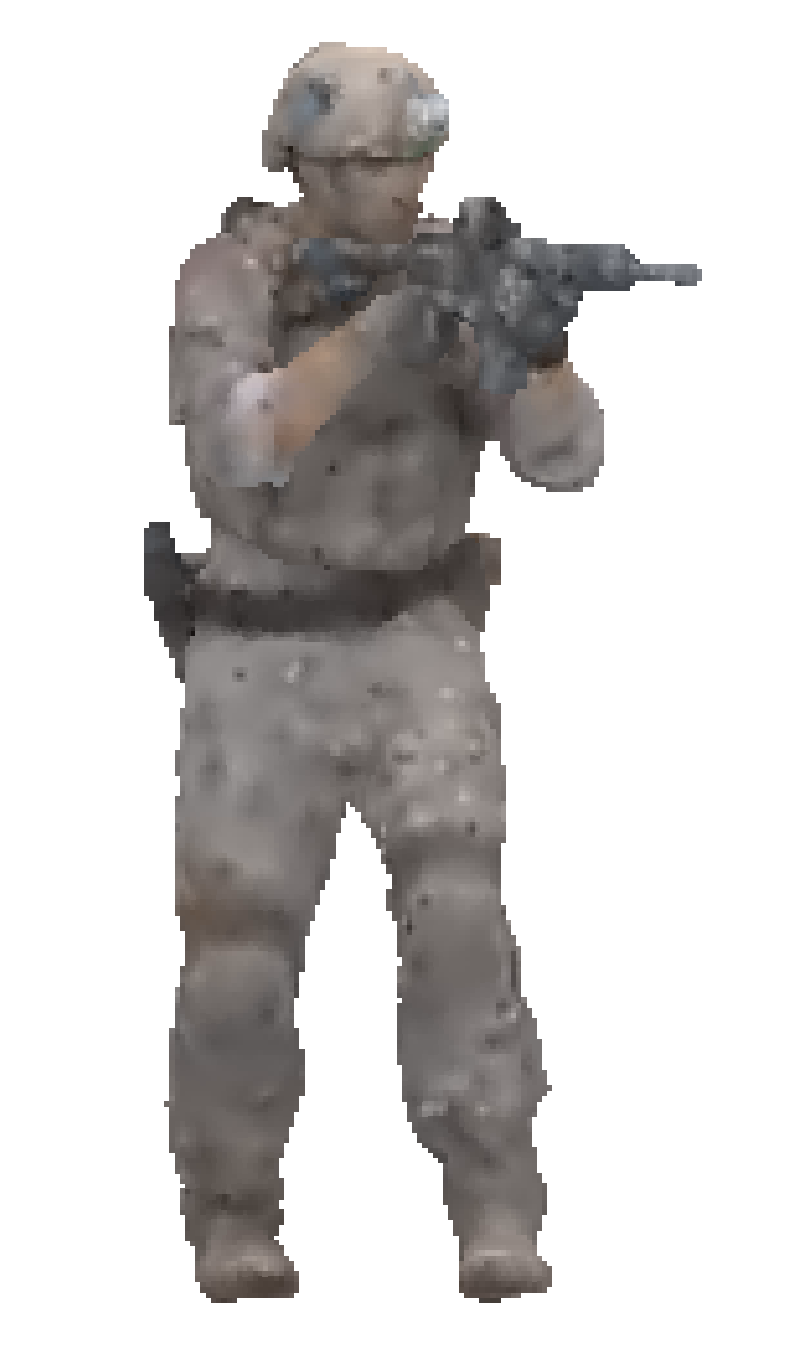}
        \includegraphics[width=0.32\linewidth]{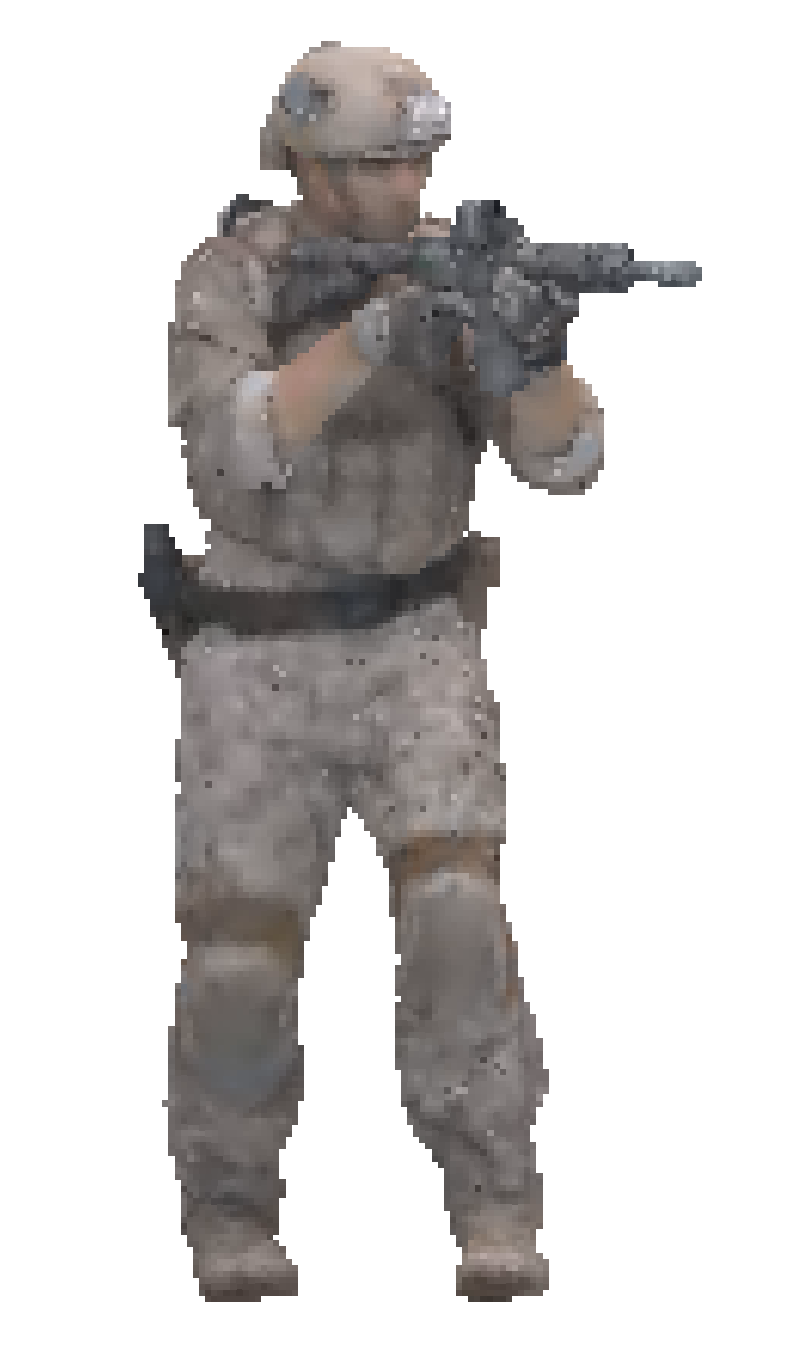}
        \includegraphics[width=0.32\linewidth]{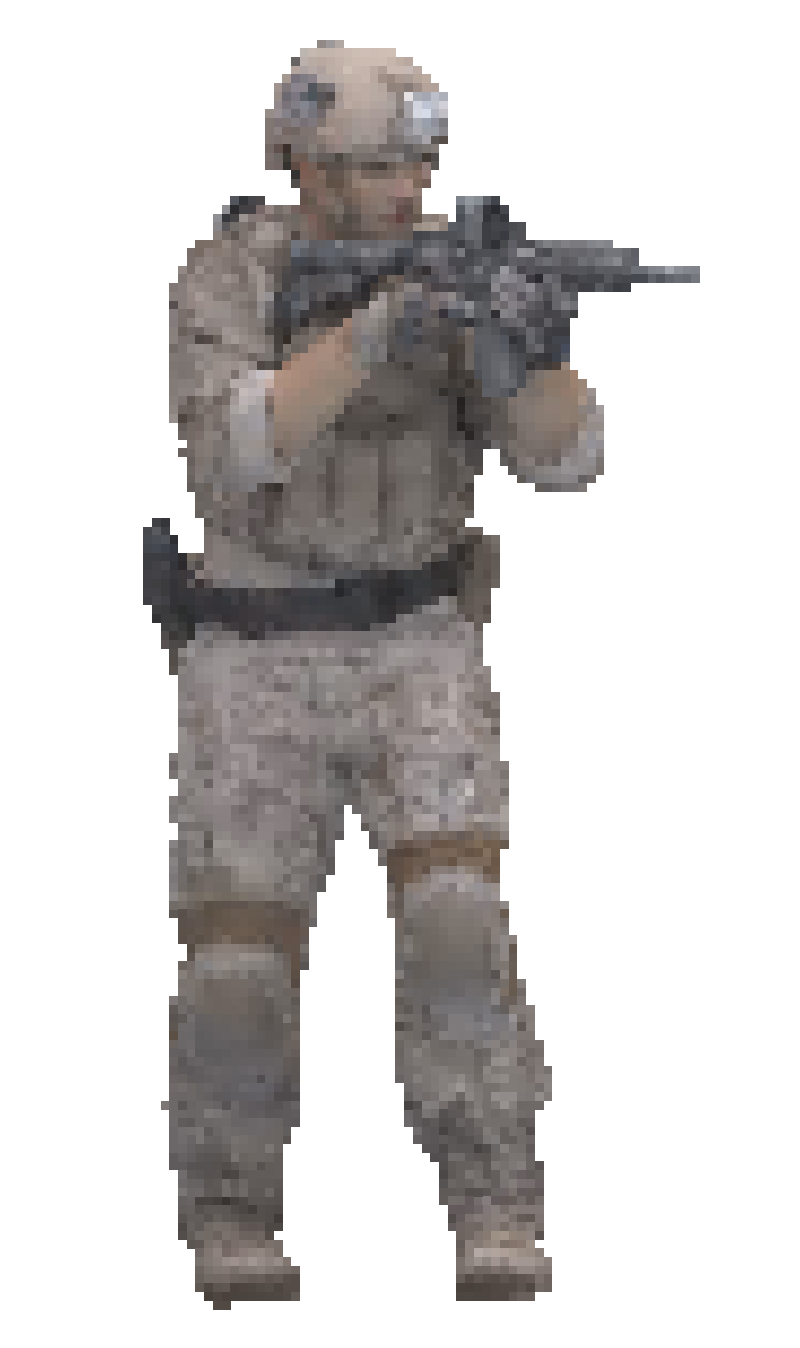}
    \end{minipage}}
    \subfloat[]{
     \begin{minipage}[b]{0.49\linewidth}
        \centering
        \includegraphics[width=0.32\linewidth]{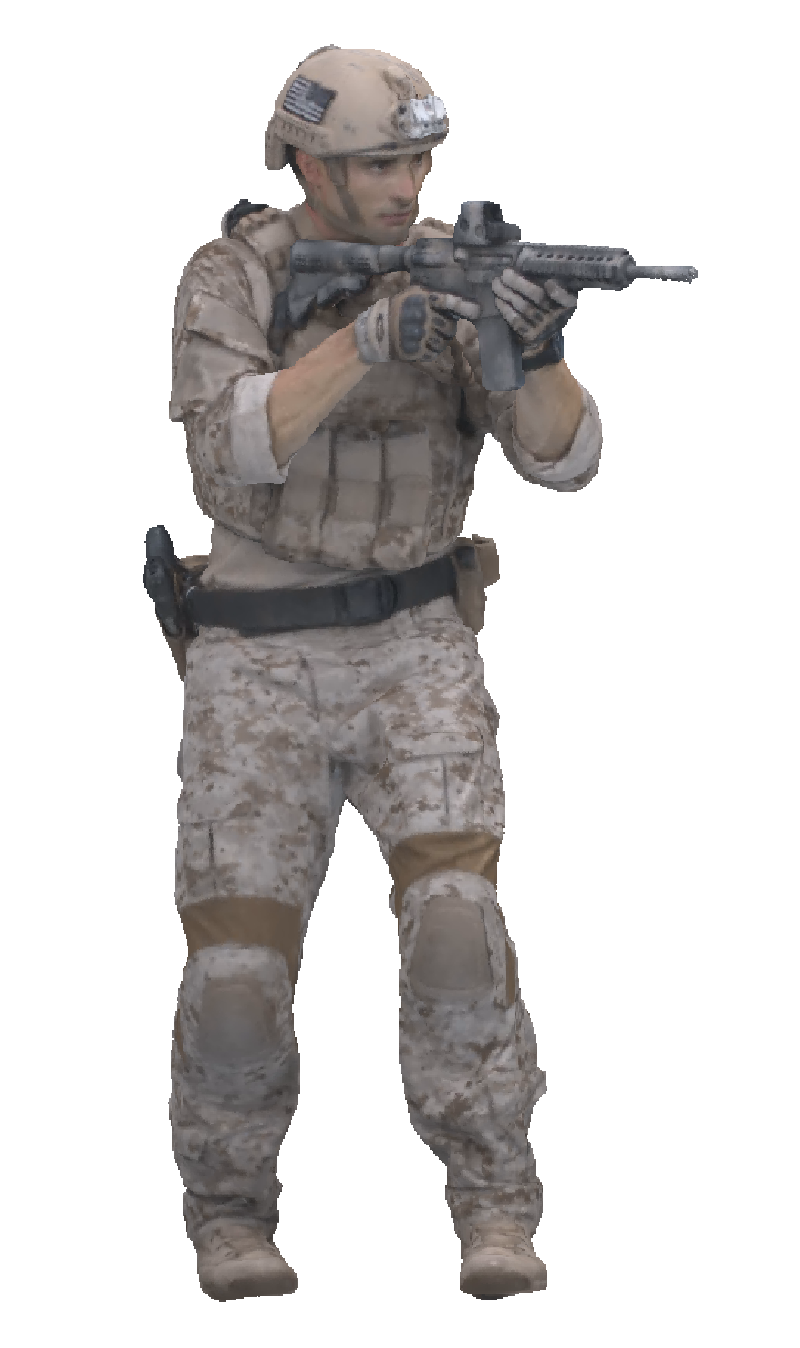}
        \includegraphics[width=0.32\linewidth]{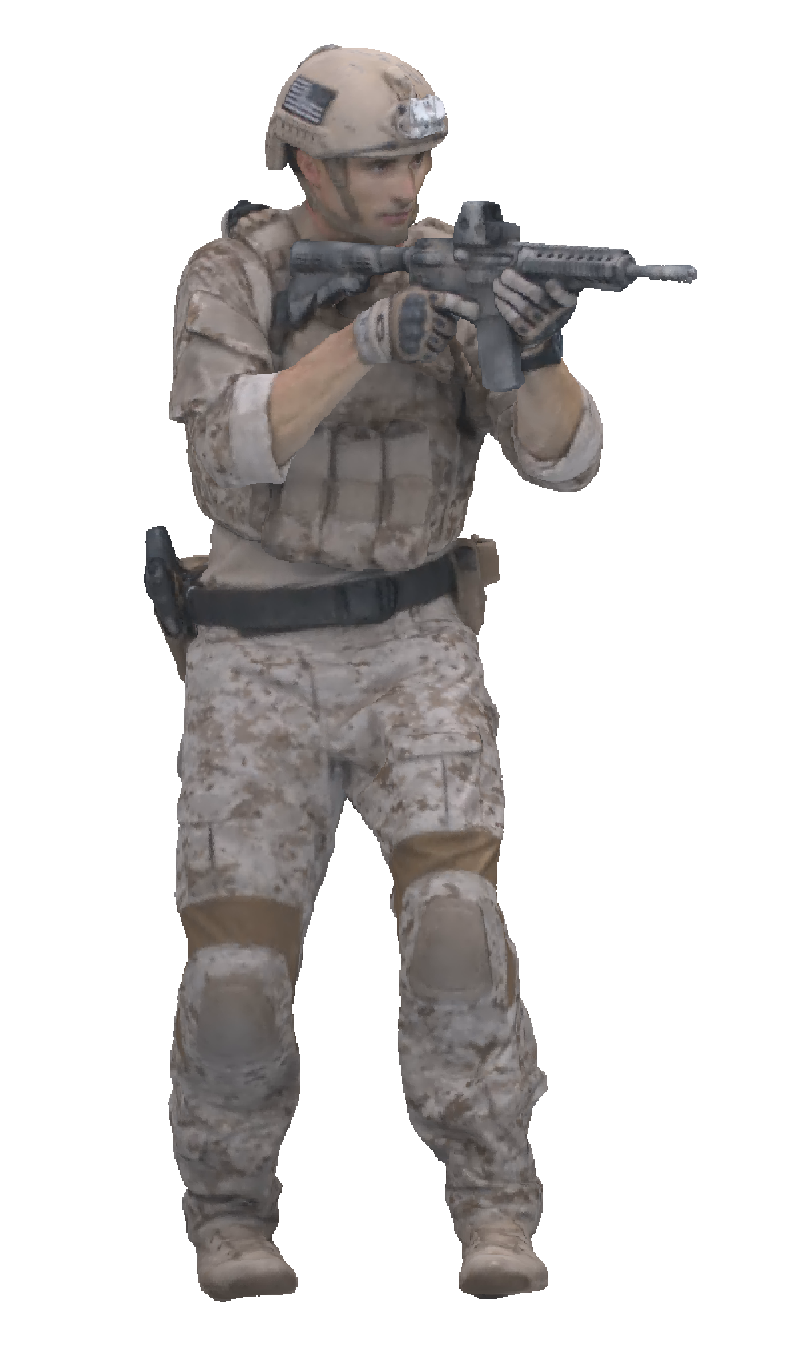}
        \includegraphics[width=0.32\linewidth]{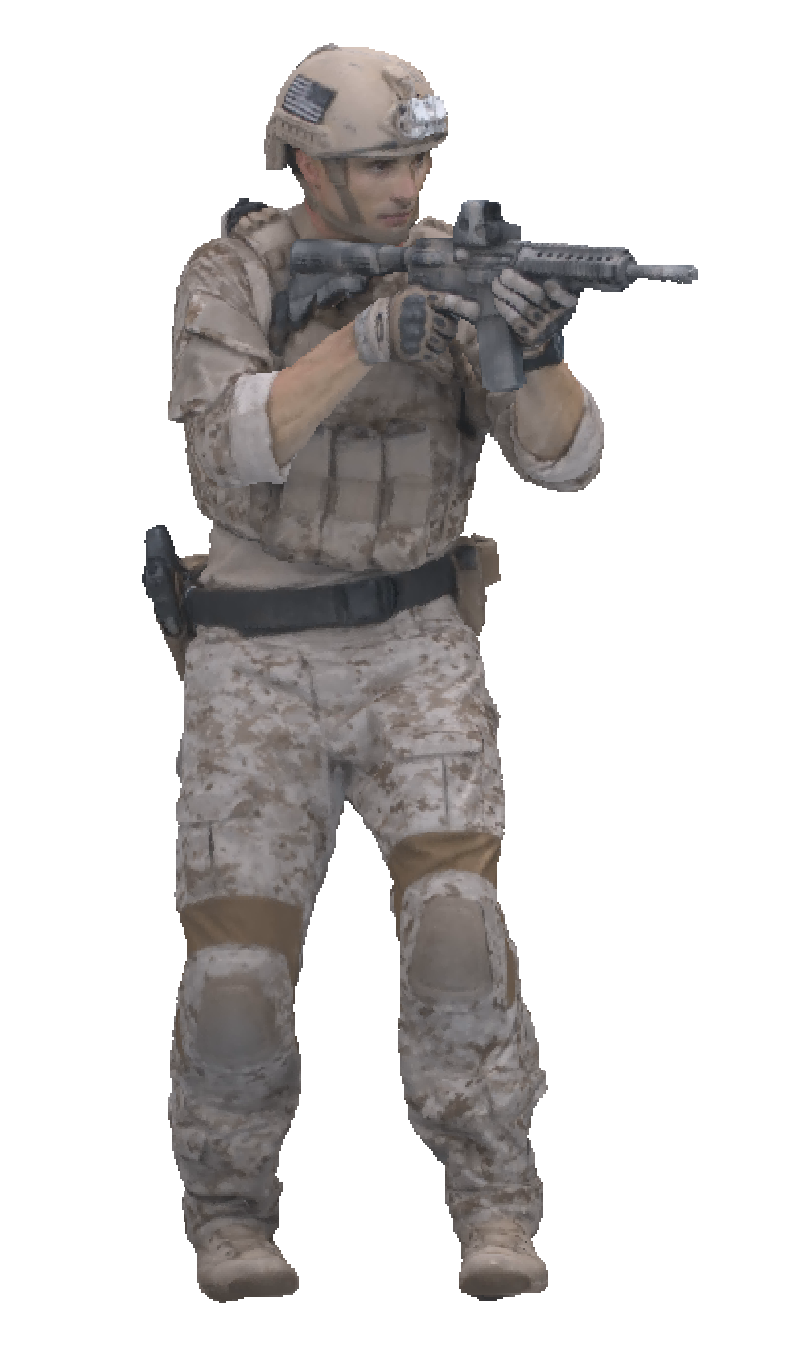}
    \end{minipage}}

    \caption{\emph{Soldier} compressed with G-PCC. (a) Compressed at rate R1 using rate allocation strategies P1, P2 and P3. (b) Compressed at rate R4 using rate allocation strategies P1, P2 and P3. }
    \label{fig:soldier_gpcc}
\end{figure}

For each compared pair of stimuli, the proportion of preference for each stimulus, as well as the proportion of \textit{``Not Sure"} answers, are displayed in Figure \ref{fig:pwc_results}. 
It is noticeable that the amount of \textit{``Not Sure"} votes is higher for the highest rates, with its proportion going from 7\% at R1 to 18\% at R2, 36\% at R3, and finally to 50\% at R4. 
This behavior is consistent across all codecs, suggesting that it does not depend on the nature of the artifacts but rather that stronger artifacts lead to a lower percentage of \textit{``Not Sure"} votes. 
These results indicate that many subjects may not be able to discern any artifacts at high bitrates, also suggested by the high MOS scores that these stimuli received in the DSIS experiment. 
A close inspection of the point clouds depicted in Figure \ref{fig:soldier_gpcc} also enforces this conclusion. 
For the lowest rate, the differences between all three compressed point clouds can be easily observed, with P1 having the highest geometric resolution and P3 having a more accurate color representation. 
On the other hand, the three point clouds compressed at the highest rate are nearly indistinguishable at the presented scale, since their compression distortion is very low and they resemble the original model very closely. 
Different boosting strategies such as allowing the subjects to zoom in on the point clouds during inspection may allow for more accurate visualization of artifacts and therefore reduce the amount of \textit{``Not Sure"} votes in future experiments.

The plots from Figure \ref{fig:pwc_results} do not allow the easy distinction of which rate allocation strategy has the best performance for each codec. 
For this reason, the average proportion of preference votes given to P1, P2, and P3, as well as to the \textit{``Not Sure"} option, was computed by pooling across different point clouds and rates. 
These values are displayed in Table~\ref{tab:preference_per_codec} and illustrate which rate allocation strategy was preferred for each codec. 
This analysis reveals that, for G-PCC, the subjects are more inclined to choose P2 over P1 and P3. 
Given that P1 employs the parameters suggested by the CTC, this study suggests that there are configurations that may result in better subjective performance than the CTC at equal or lower bitrates. 
Both P2 and P3 assign higher importance to color than P1 for most rates, with P2 being closer to the CTC and P3 having lower $qp$ values. 
It is important to note that neither of the alternative rate allocation strategies was the result of a systematic optimization of the performance according to objective quality measures. 
P2 and P3 were defined as means to explore whether different configurations could provide better subjective quality without the goal of maximizing it. 
There are therefore potentially other configurations that could provide even better results. 
As the isorate plots from Figure \ref{fig:gpcc_isorate_curves} suggest, these configurations would probably have to be defined separately for each point cloud to optimize rate-distortion performance.

For V-PCC, the results provided by the CTC seem to beat the alternative configurations, especially P2. 
As detailed in Section \ref{sec:vpcc_compression}, there were challenges associated with creating an alternative configuration giving a higher weight to color, i.e. with lower $aqp$, and for that reason, P2 was defined in the opposite direction and increased the value of $aqp$ in relation to P1. 
This modification is however observed to have a negative impact on subjective quality. 
On the other hand, adapting the \emph{occupancyPrecision} has little impact on the subjective quality, reinforcing that it is difficult to find better configurations than the CTC at similar bitrates. 
Finally, for JPEG Pleno, subjects are more inclined to choose P1 or P2 over P3, suggesting that, to obtain higher visual quality, it is preferable to allocate a similar or a higher bitrate to the color when compared to the geometry.

\begin{table*}
    \centering
    \begin{tabular}{ccccc}
          &  P1 & P2 & P3 & Not sure\\
          \hline
         G-PCC & 0.20 & 0.29 & 0.21 & 0.30\\
         V-PCC & 0.29 & 0.15 & 0.27 & 0.29\\
         JPEG Pleno & 0.29 & 0.30 & 0.17 & 0.24\\
    \end{tabular}
    \caption{Preference probability for the different configurations, grouped per codec.}
    \label{tab:preference_per_codec}
\end{table*}

\begin{figure}
    \centering
    \begin{minipage}[b]{\linewidth}
    \centering
    \includegraphics[width=0.29\linewidth]{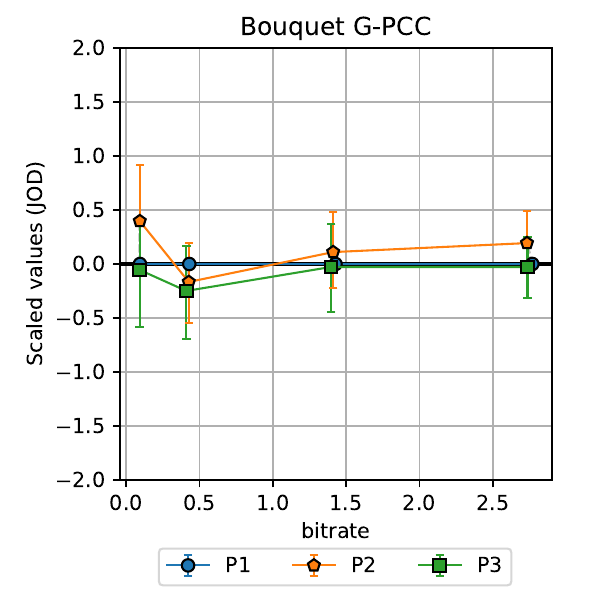}
    \includegraphics[width=0.29\linewidth]{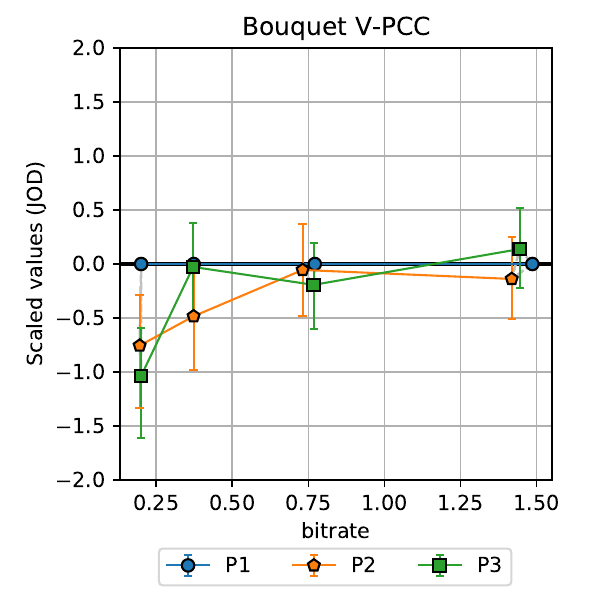}
    \includegraphics[width=0.29\linewidth]{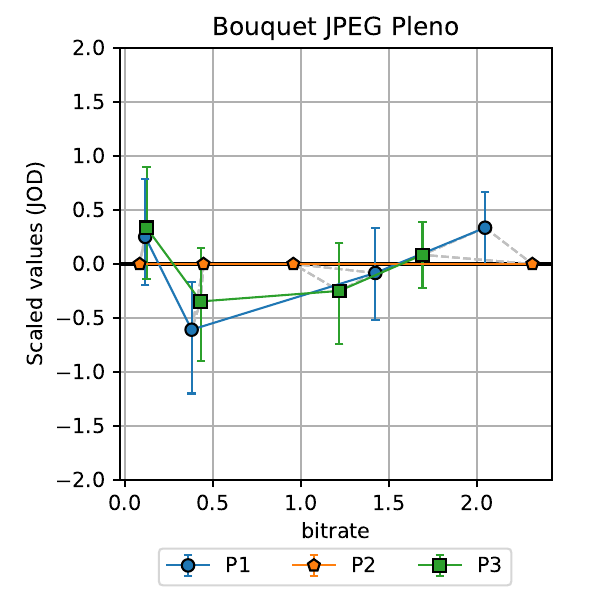}
    \end{minipage}

    \begin{minipage}[b]{\linewidth}
    \centering
    \includegraphics[width=0.29\linewidth]{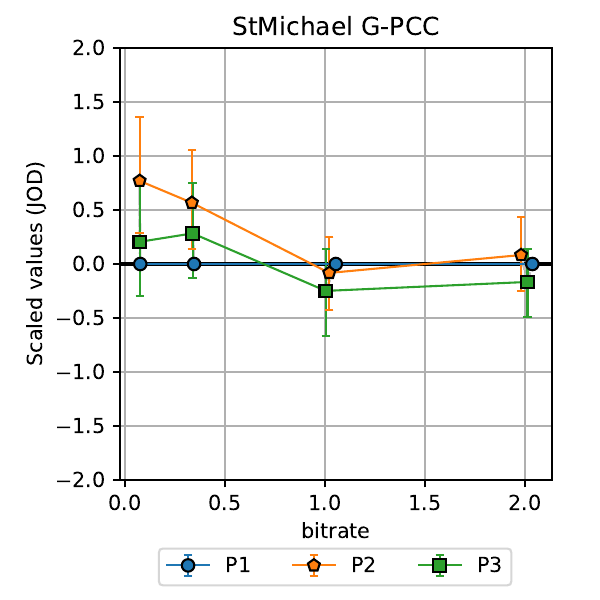}
    \includegraphics[width=0.29\linewidth]{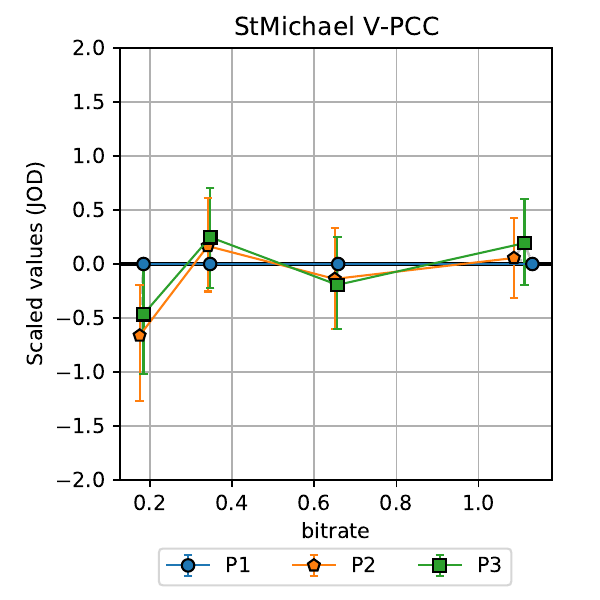}
    \includegraphics[width=0.29\linewidth]{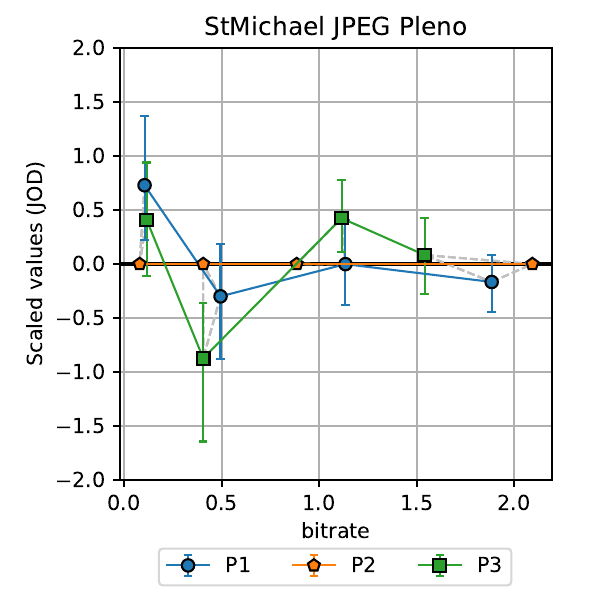}
    \end{minipage}
    
    \begin{minipage}[b]{\linewidth}
    \centering
    \includegraphics[width=0.29\linewidth]{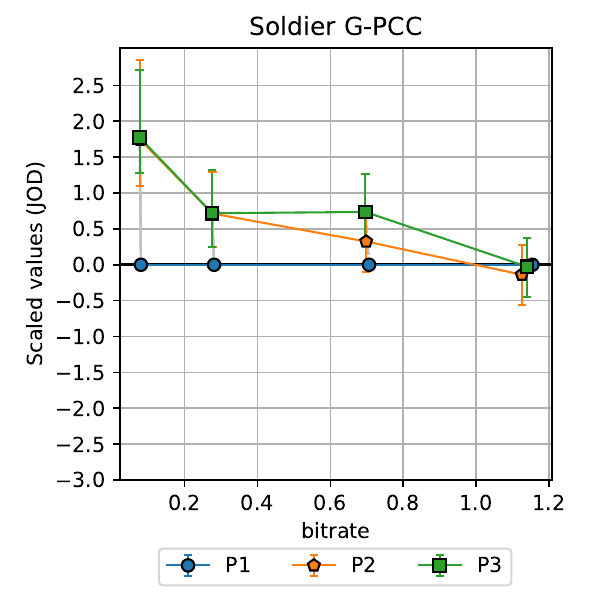}
    \includegraphics[width=0.29\linewidth]{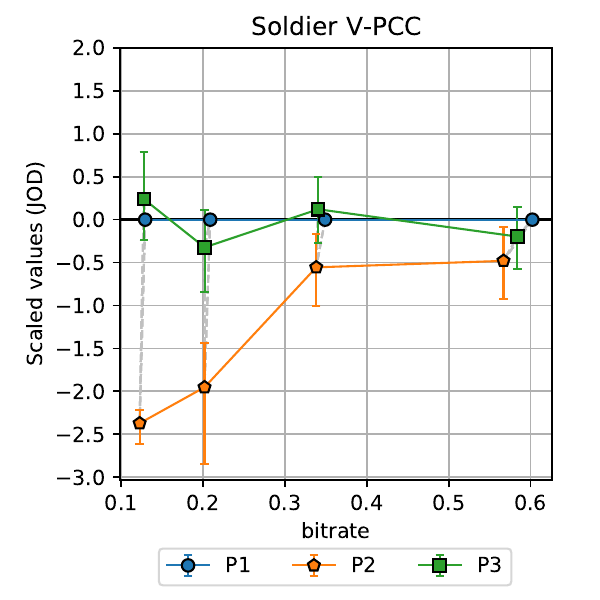}
    \includegraphics[width=0.29\linewidth]{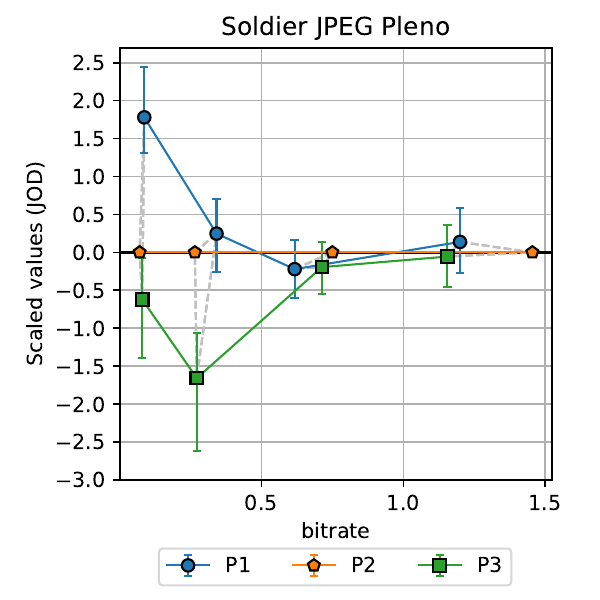}
    \end{minipage}

    \begin{minipage}[b]{\linewidth}
    \centering
    \includegraphics[width=0.29\linewidth]{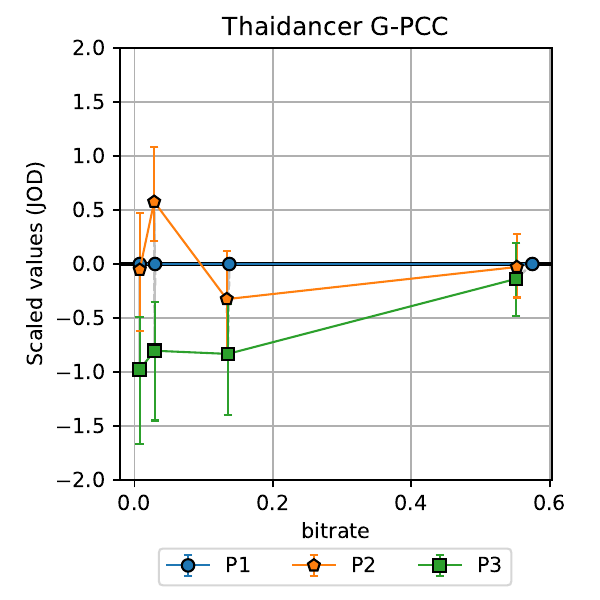}
    \includegraphics[width=0.29\linewidth]{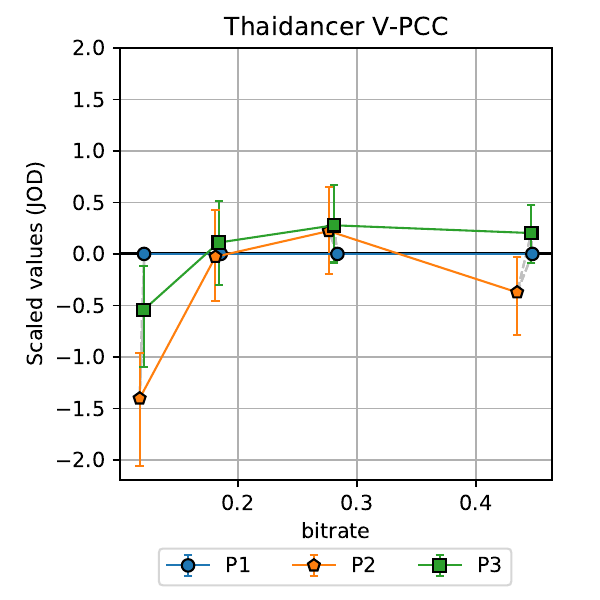}
    \includegraphics[width=0.29\linewidth]{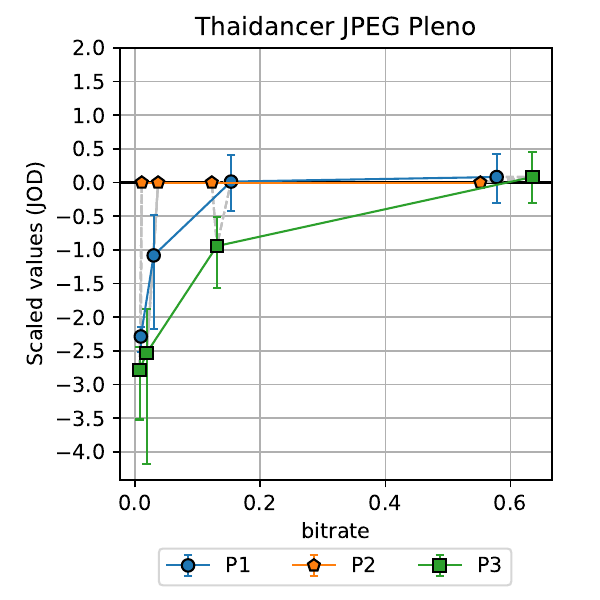}
    \end{minipage}
    
    \caption{Reconstructed JOD values against the bitrate for each image and codec.}
    \label{fig:pwc_results_jod}
\end{figure}

The reconstructed JOD values are reported in Figure \ref{fig:pwc_results_jod} against the bitrate. 
As underlined in Section \ref{sec:data_processing}, the displayed JOD values only reflect relationships between the subjective quality of point clouds compressed with the same codec and rate, i.e. it is not possible to derive the number of JOD between stimuli at different rates. 
The plots in Figure \ref{fig:pwc_results_jod} highlight the points that can be compared in terms of JOD by connecting them with light-grey lines. 
The conclusions derived from these plots are similar to what can be observed from Table \ref{tab:preference_per_codec}, but additional nuances regarding the influence of the content and rate can also be observed. 
For G-PCC, the advantage of P2 over P1 and P3 is mainly present at R1 and R2, while all configurations achieve very similar JOD values at R4. 
On the other hand, P3 is only the best-performing configuration for \emph{Soldier}. 
These plots suggest that a potential improvement in performance can be achieved in relation to the CTC parameters. 
However, the measured improvement is not high, going over 1 JOD only for \emph{Soldier} at R1. 
Moreover, the results of this study only suggest the existence of better rate allocation methods for G-PCC without proposing an optimal scheme. 
Regarding V-PCC, the plots indicate again that the alternative allocation methods are not able to outperform the CTC by a meaningful margin in any evaluated condition. 
P3 has a very similar performance as P1, with a reduction in performance only at R1 especially for \emph{Bouquet} but also for \emph{StMichael} and \emph{Thaidancer} to a smaller extent. 
On the other hand, P2 severely underperforms P1, especially for \emph{Soldier} but also for \emph{Thaidancer} at R1, suggesting that it is generally not beneficial to use higher $aqp$ than the CTC for the same target bitrates. 

\textcolor{black}{For JPEG Pleno, 
due to the scarcity of available configurations for some bitrates, some rate allocation strategies produce decoded point clouds with \emph{strictly better} quality than others at the same rate point, as displayed in Figure \ref{fig:allocation_relationship}. 
In these cases, it is reasonable to assume that the quality of the second cannot be significantly better than the first under normal circumstances, although a slightly higher score may be achieved due to statistical noise if the difference between the strategies is not perceptible. 
Any comparison of the scores between two strategies must therefore take into account whether there is a real \emph{trade-off} between geometry and color or not.} 

The plots in Figure \ref{fig:pwc_results_jod} for JPEG Pleno show that, if P2 is set as the baseline, there are 11 cases where the JOD difference is higher than 0.5, 10 of which happen at R1 and R2. 
For only two of those cases, the stimulus compressed at P2 has lower quality, namely for \emph{StMichael} and \emph{Soldier} compressed at R1. 
In the remaining cases, P2 has better quality than its pair, being 3 times against P1 and 6 times against P3. 
In only 4 of those cases, P2 is compressed with \emph{strictly better} configurations, all of them for \emph{Thaidancer} at R1 and R2. 
In the remaining cases, there is a \emph{trade-off} between geometry and color quality, where P2 has better geometry quality and worse color quality than P1, or better color quality and worse geometry quality than P3. 
These comparisons suggest that, \textcolor{black}{whenever two configurations allowing for a \emph{trade-off} between color and geometry result in similar bitrates}, assigning a higher value to the color compression parameter is usually better than choosing better geometry quality. 
However, the results of this study cannot point to one rate allocation strategy as being the best-performing, mainly due to the lack of granularity allowed by the current JPEG Pleno VM in regard to the configuration parameters. 

For all the evaluated codecs, these plots also show that the JOD differences are higher at lower bitrates. 
This may indicate that these codecs achieve nearly lossless visual quality at higher rates, as suggested by the visual results in Figure \ref{fig:soldier_gpcc}. 
However, these results may also be caused by the employed protocol where the stimuli are displayed side-by-side and therefore do not allow to discern subtle artifacts. 
Indeed, recent studies for image quality assessment \cite{testolina2023performance} have shown that side-by-side protocols are not able to detect distortion that can easily be spotted in the Flicker test \cite{aic2}, where the compressed image is repeatedly interleaved with the original one in the same area of the screen. 
Moreover, boosting techniques \cite{men2021subjective} have been proposed aiming at assessment at the high to nearly visually lossless range, which could be adapted to point clouds in future studies. 

\subsection{Joint bitrate allocation evaluation}
\label{sec:joint_analysis}

\textcolor{black}{
The separate analysis of the DSIS and PWC experiments in Sections \ref{sec:dsis} and \ref{sec:pwc} reveal several observations regarding the performance of different rate allocation strategies that can be summarised as follows:}

\begin{itemize}
    \item \textcolor{black}{The \textbf{DSIS} scores show that, in most cases, the difference in MOS values between allocation strategies is small, with a large overlap between confidence intervals.}
    \item \textcolor{black}{The largest differences are observed for JPEG Pleno. For some solid point clouds, P3 is outperformed by P1 and P2, while P1 trails the performance of the other strategies for the sparser models.}
    \item \textcolor{black}{For the \textbf{PWC} experiment, while higher quality differences are observed for lower bitrates, the scores from the three allocation strategies converged to similar values at higher rates.}
    \item \textcolor{black}{For G-PCC, the differences are small but there is a slight tendency for P2 to outperform P1 at lower rates.}
    \item \textcolor{black}{For V-PCC, P1 consistently achieves high performance when compared to other strategies.}
    \item \textcolor{black}{For JPEG Pleno, a closer analysis indicates that whenever there is a \emph{trade-off} between color and geometry qualities between two strategies, allocating a higher portion of the bitstream to color is beneficial in most cases.}
\end{itemize}

\begin{figure}
    \centering

    \begin{minipage}{0.95\linewidth}
        \hspace{50pt}
        \includegraphics[width=0.37\linewidth]{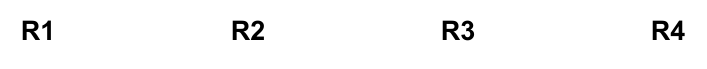}
        \hspace{19pt}
        \includegraphics[width=0.37\linewidth]{Figures--Profile_relationships-Rates.pdf}    
    \end{minipage}
    
    \begin{minipage}[b]{0.1\linewidth}
        \includegraphics[width=\linewidth]{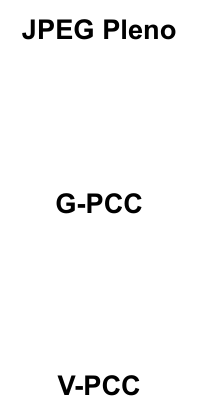}    
        \vspace{0.01pt}
    \end{minipage}
    \subfloat[\emph{Bouquet}]{
        \includegraphics[width=0.4\linewidth]{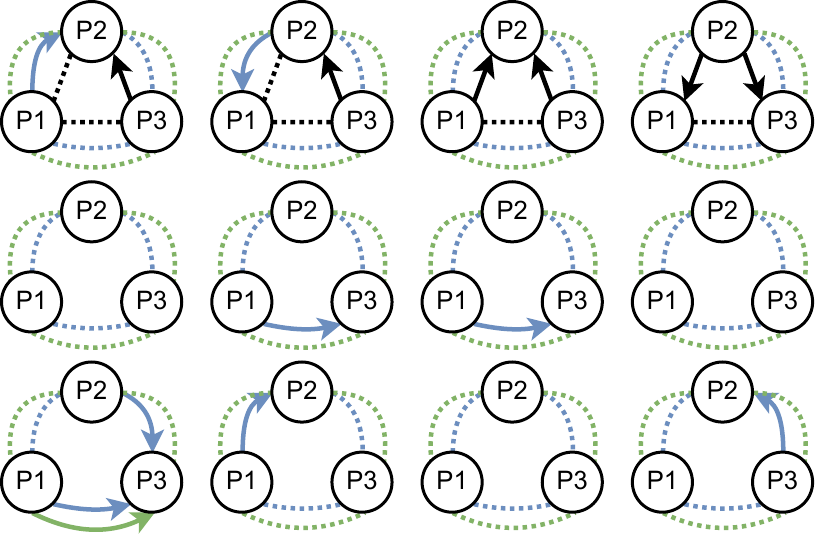}
    }
    \vrule
    \subfloat[\emph{StMichael}]{
        \includegraphics[width=0.4\linewidth]{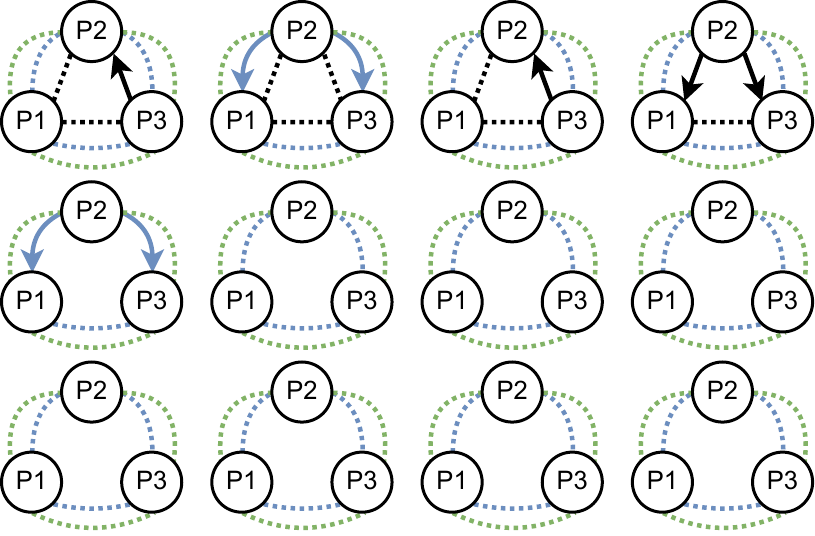}
    }

    \begin{minipage}[b]{0.1\linewidth}
        \includegraphics[width=\linewidth]{Figures--Profile_relationships-codecs.pdf}    
        \vspace{0.01pt}
    \end{minipage}
    \subfloat[\emph{Soldier}]{
    \includegraphics[width=0.4\linewidth]{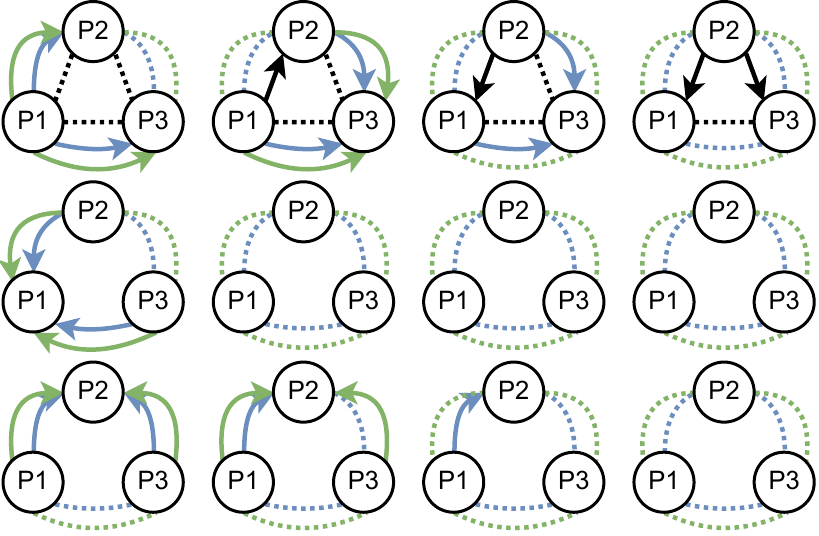}
    }
    \vrule
    \subfloat[\emph{Thaidancer}]{
    \includegraphics[width=0.4\linewidth]{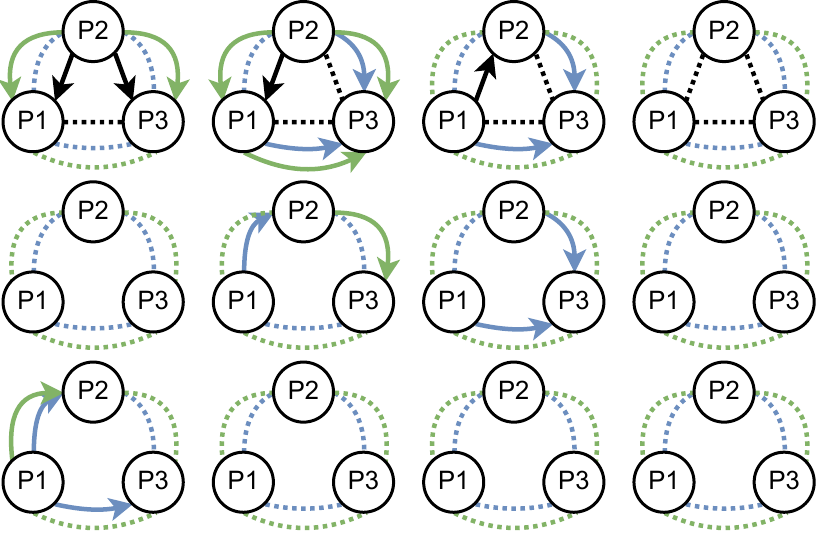}
    }

    \begin{minipage}[b]{0.1\linewidth}
        \includegraphics[width=\linewidth]{Figures--Profile_relationships-codecs.pdf}    
        \vspace{0.01pt}
    \end{minipage}
    \subfloat[\emph{Boxer}]{
    \includegraphics[width=0.4\linewidth]{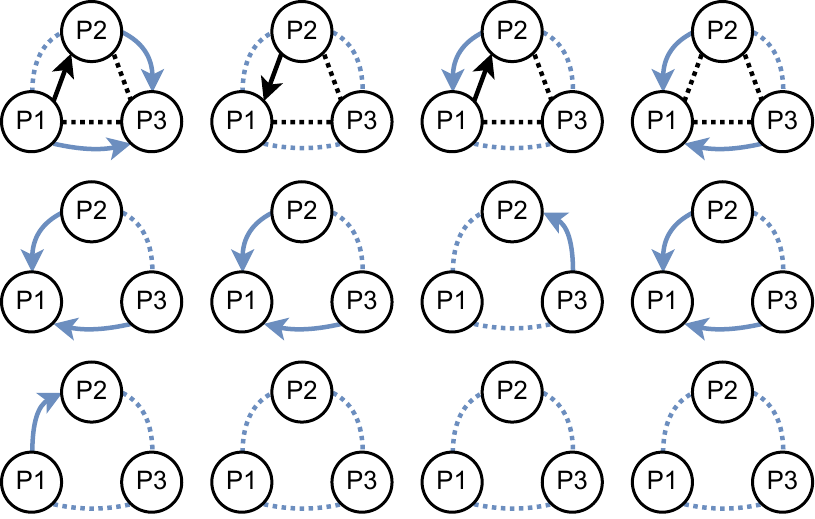}
    }
    \vrule
    \subfloat[\emph{House\_without\_roof}]{
    \includegraphics[width=0.4\linewidth]{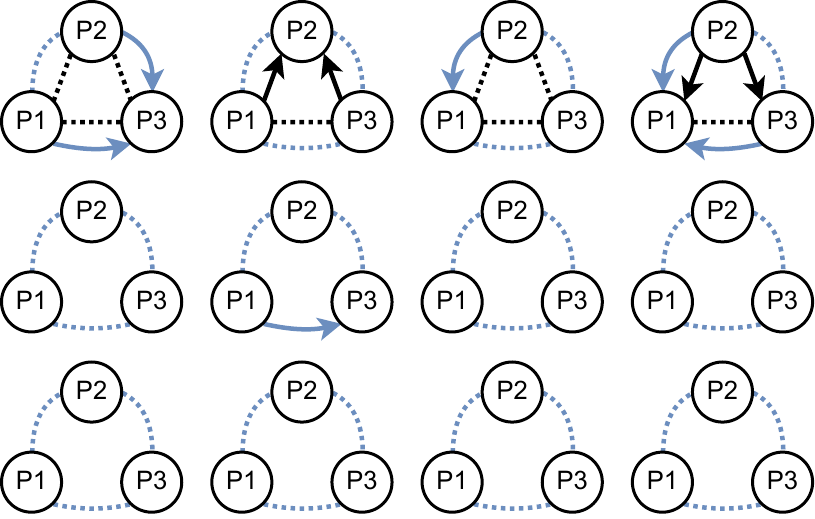}
    }

    \vspace{15pt}
    \includegraphics[width=0.95\linewidth]{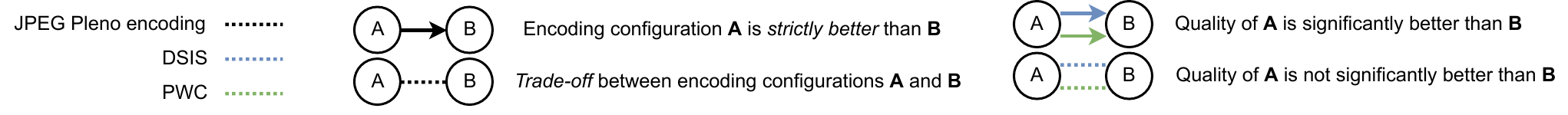}

    \caption{\textcolor{black}{Comparative results between rate allocation strategies for each codec and point cloud including results from both DSIS and PWC experiments.}}
    \label{fig:allocation_results}

\end{figure}

\textcolor{black}{
Ideally, the results obtained in both experiments should help inform which is the best rate allocation for each point cloud and rate point. 
However, this decision is not straightforward since in many cases it is not possible to affirm whether the score difference is due to statistical noise or an actual difference in perceived quality between stimuli. 
For this reason, an additional analysis is conducted with the results from the two experiments, which evaluates, for every possible pair of allocation strategies of each rate, point cloud content, and codec, whether one strategy provides better subjective quality with statistical significance. 
For the DSIS scores, the individual scores for two stimuli generated with different rate allocation strategies are considered as samples from two different distributions, and the one-tailed Welch t-test is applied to determine whether one population mean is higher than the other with a p-value of 0.05. 
For the PWC experiment, the difference between the JOD values between the same pair of stimuli is used, and a rate allocation strategy is considered to provide better quality if its JOD is higher than its counterpart with a difference equal to or larger than 1 JOD. 
The results are displayed in Figure \ref{fig:allocation_results}, where a solid arrow indicates that one rate allocation strategy achieves superior subjective quality than the other according to a given experiment. 
Otherwise, a dotted line is used. 
For JPEG Pleno, the relationship between allocation strategies defined in Figure \ref{fig:allocation_relationship} is also added to the diagrams of Figure \ref{fig:allocation_results}, where two configurations defined according to a \emph{trade-off} between geometry and color are connected with a dotted line, and a solid arrow is used whenever one configuration is \emph{strictly better} than the other. 
}

\textcolor{black}{
When considering only the four point clouds for which the two experiments were conducted, it is not possible to define a winning rate allocation strategy in several cases. 
Among the 48 different combinations of rate, codec, and point cloud content, no significant difference in quality is found in any comparison for 26 of them. 
In particular, this proportion is higher for R4, for which 11 out of the 12 triplets present the same performance across the three strategies, as well as for the \emph{StMichael} point cloud, where the same happens in 10 cases. 
The threshold of 1 JOD used in this analysis for the PWC experiment is observed to be particularly strict, determining only 18 superiority relationships between allocation strategies. 
In comparison, the DSIS result indicates that one strategy produces higher quality than the other on 35 occasions for the four solid point clouds. 
The two experiments do not produce any divergent results, and superiority relationships indicated by the PWC experiment are also indicated by the DSIS scores on 13 occasions. 
Since the results from both experiments do not diverge for the solid point clouds, it is reasonable to use only the DSIS results for \emph{Boxer} and \emph{House\_without\_roof} where PWC scores are not available. }

\textcolor{black}{
For JPEG Pleno, P3 is never observed to deliver better quality compared to P1 or P2 for any solid point cloud, even when its encoding configuration is \emph{strictly better}. 
For instance, when \emph{Bouquet} is encoded at R1, P1 is observed to have better quality than P2 according to the DSIS experiment, while it is P3 that has \emph{strictly better} coding configurations when compared to P2. 
In this case, it would be reasonable to select P1 as having the best quality among the three strategies. 
For R1 and R2, P1 or P2 are superior for most solid point clouds, while only for \emph{Thaidancer} this superiority can be attributed to a \emph{strictly better} coding configuration. 
However, for higher rates, the scores do not show strong indications of difference in quality between rate allocation strategies. 
The result of the analysis is different for \emph{Boxer} and \emph{House\_without\_roof}, where P2 is found to have consistent performance across bitrates. 
For both point clouds, P1 is preferred to P3 for R1 and the opposite relationship is observed for R4. 
The results for sparser point clouds indicate that higher differences in performance can be observed for sparser point clouds at higher rates, mainly due to the fact that P1 is not able to achieve high MOS scores even at R4. 
Overall, the joint analysis of the results supports that, for JPEG Pleno, assigning either a balanced bitrate between color and geometry or a larger portion to color is beneficial for solid point clouds. 
However, for models with lower point density, the subjective perception can be more affected by geometry quality at high bitrates. }

\textcolor{black}{
For G-PCC, P3 provides better quality than other strategies for all rates in \emph{Boxer}, but rarely has any advantage over other point clouds. 
P1 demonstrates superiority for \emph{Bouquet}, \emph{Thaidancer}, and \emph{House\_without\_roof} at mid-range bitrates R2 and R3, but is outperformed at R1 for \emph{StMichael}, \emph{Soldier} and \emph{Boxer}. 
In general, the joint results allow for the existence of rate allocations better than the CTC without a clear indication of how this allocation should be defined. 
However, for V-PCC, P1 is usually the best strategy whenever a significant difference is observed. 
For the alternative strategies, direct comparisons generally attribute an advantage to P3, although significant differences are observed only for \emph{Bouquet} and \emph{Soldier}. 
Therefore, further study would be needed to investigate whether it is possible to improve the subjective visual quality of V-PCC only by changing the rate allocation when compared to the CTC. 
}

\section{Conclusions}

This paper presents a comprehensive evaluation of the performance of the \textcolor{black}{JPEG Pleno and MPEG} point cloud compression standards. 
A dataset containing distorted point clouds was built with the goal of not only including different bitrates but also different sets of configuration parameters for the same target rate for each codec, using objective quality metrics to guide the process. 
This dataset was then subjectively evaluated in two experiments following different protocols: the DSIS experiment provided an absolute scale of comparison for all stimuli, while differences between stimuli compressed with the same codec and rate could be better evaluated following the PWC protocol. 
The results from the first experiment indicate that V-PCC outperforms the other codecs by a small margin for most point clouds, but also that its performance can be heavily affected if some configuration parameters are not properly selected. 
JPEG Pleno displayed good performance for solid point clouds, but results were highly dependent on the point density, with reduced quality for sparser point clouds. 
On the other hand, G-PCC was usually outperformed by the other two codecs but had the most stable behavior across different point clouds. 
This analysis also demonstrated that none of the evaluated objective quality metrics was able to accurately rank the performance of the codecs for the point clouds of this dataset. 
The PWC experiment revealed that alternative trade-offs between geometry and color quality can potentially provide better subjective quality than the CTC for G-PCC. 
However, the evaluated alternative rate allocation schemes are not optimal, leaving room for improvement in future studies. 
On the other hand, the proposed strategies for V-PCC were not able to consistently outperform the CTC at similar bitrates, reinforcing the conclusions suggested by the objective evaluation that producing more efficient sets of parameters than the CTC would be challenging. 
For JPEG Pleno, the analysis indicates that allowing a higher proportion of the bitstream to the color representation may be advantageous for increasing subjective quality \textcolor{black}{when encoding solid point clouds, while a more accurate representation of the geometry is important for models with lower point density at higher bitrates.}
Future works aiming at the evaluation of the performance of objective quality metrics across different types of compression artifacts and different trade-offs between color and geometry quality may benefit from the results of this paper as well as the openly provided dataset and scores. 
Moreover, different rate allocation schemes may be developed based on these results in order to optimize the performance of point cloud compression standards. 
Finally, the insights presented in this paper can also serve as a basis for modifications to the compression standards themselves, i.e. by identifying different patterns in the reaction of subjects to specific artifacts and using them to improve rate-distortion performance.



\begin{backmatter}

\section*{Acknowledgements} 
The authors acknowledge the participation of Filip Mikovíny in the development of the platform used for the interactive subjective evaluation of point clouds used in this study. 
\section*{Funding} 
The authors would like to acknowledge support from the Swiss National Scientific Research project entitled ``Compression of Visual information for Humans and Machines (CoViHM)” under grant number 200020\_207918.
\section*{Competing interests} 
The authors declare that they have no competing interests.
\section*{Ethics approval} 
Not applicable
\section*{Consent to participate} 
Not applicable
\section*{Consent for publication} 
Not applicable
\section*{Availability of data and materials} 
The datasets generated and/or analyzed during the current study are available in the MJ-PCCD repository, \url{https://www.epfl.ch/labs/mmspg/downloads/mj-pccd/}
\section*{Code availability} 
Not applicable
\section*{Authors' contributions} 
All authors participated in the design of the experiment, generation of the dataset, conducting the subjective assessment sessions, and writing of the manuscript. All authors read and approved the final manuscript.





\bibliographystyle{sn-nature}

\end{backmatter}
\end{document}